\documentclass{aa}
\usepackage[varg]{txfonts}

\usepackage{graphicx}
\def\ms{\mbox{$M_{\star}$}}

\def\msun {$M_{\sun}$}

\def\re{$R_{eff}$}

\def\tq{$t_q$}
\def\tqlb{$t_{\rm q,LB}$}
\def\tqin{$t_{\rm max}$}
\def\Dtq{$\Delta t_q$}

\usepackage[colorlinks, citecolor=blue]{hyperref}
\bibpunct[; ]{(}{)}{;}{a}{}{;}

\begin{document}

\title{SDSS-IV MaNGA: Global and local stellar population properties of elliptical galaxies}

\titlerunning{Global and local stellar population properties of ellipticals}

\author{I.~Lacerna\inst{\ref{uda},\ref{mas}} \and
H.~Ibarra-Medel\inst{\ref{illinois}} \and
V.~Avila-Reese\inst{\ref{unam}} \and
H.~M.~Hern\'andez-Toledo\inst{\ref{unam}} \and
J.~A.~V\'azquez-Mata\inst{\ref{unam}} \and
S.~F.~S\'anchez\inst{\ref{unam}}
}  

\institute{
Instituto de Astronom\'ia y Ciencias Planetarias, Universidad de Atacama, Copayapu 485, Copiap\'o, Chile\\
\email{ivan.lacerna@uda.cl} \label{uda} 
\and 
Instituto Milenio de Astrof\'isica, Av. Vicu\~na Mackenna 4860, Macul, Santiago, Chile\label{mas}
\and
University of Illinois Urbana-Champaign, Department of Astronomy, 1002 W Green St, Urbana, Illinois, 61801, United States\label{illinois}
\and 
Instituto de Astronom\'{\i}a, Universidad Nacional Aut\'onoma de M\'exico, A.P. 70-264, 04510 M\'exico D. F., M\'exico \label{unam} 
}

\abstract
%Context
{
We study the spatially resolved properties of 343 elliptical galaxies with the 
Mapping Nearby Galaxies at the Apache Point Observatory (MaNGA) survey. 
}
%Aims
{
Our goal is to understand the fundamental processes of formation and quenching of elliptical galaxies.
}
%Methods
{
We used the DESI Legacy Imaging Surveys for accurate morphological classification.
Based on integrated spectroscopic properties and colors, we 
classified seven classes of elliptical galaxies. 
We inferred the stellar age and metallicity gradients out to a 1.5 effective radius (\re) of classical "red and dead," recently quenched, and blue star-forming ellipticals (CLEs, RQEs, and BSFs), corresponding to
73\%, 10\%, and 4\% of the sample, respectively. Additionally, we reconstructed their global and radial histories of star formation and mass growth. 
}
%Results
{
The mass- and luminosity-weighted age gradients of CLEs are nearly flat or mildly negative, with small differences between both ages. The respective metallicity gradients are negative ($\nabla\log$[$Z_{mw}$] = $-0.11_{-0.08}^{+0.07}$ dex/\re{} and $\nabla\log$[$Z_{lw}$] = $-0.11_{-0.07}^{+0.06}$ dex/\re, respectively), being flatter 
as the mass is smaller.
The more massive CLEs formed stars
earlier and quenched faster than the less massive ones. The CLEs show a weak inside-out growth and a clear inside-out quenching.
They finished their quenching globally $3.8\pm 1.2$ Gyr ago on average, with quenching time-scales of $3.4\pm0.8$ Gyr.
At \ms < $10^{11}$ \msun, the age and $Z$ gradients of the RQEs and BSFs are flatter than those of the CLEs, but with larger scatters. They show very weak inside-out growth and quenching, which is slow and not even completed at $z\sim0$ for the BSFs. Instead, the massive RQEs show an outside-in quenching and positive gradients in the luminosity-weighted age 
and stellar metallicities. 
The RQEs of all masses quenched 1.2 $\pm$ 0.9 Gyr ago on average.
}
%Conclusions
{
Our results for the CLEs are consistent with a two-phase scenario where their inner parts formed by an early and coeval dissipative collapse with a consequent burst of star formation and further quenching, whereas the outer parts continued their assembly, likely by dry mergers. We also discuss some evolutionary scenarios for the RQE and BSF galaxies that would agree with the generic results. 
}


\maketitle          

\section{Introduction} 
\label{S1}

Elliptical (E) galaxies are the earliest-type galaxies in the Hubble Tuning Fork diagram \citep{Hubble1936}. For many decades, they have been described as galaxies that formed either in an intense early burst of star formation after a violent dissipative monolithic collapse \citep[e.g.,][]{Larson1974,Oemler1974, White1978} or assembled lately by dry 
mergers of fully formed galaxies \citep[e.g.,][]{Toomre1977, Dressler1980, Hernquist1993,Tutukov+2007, Kormendy+2009}. In more recent years, these scenarios were modified in light of the hierarchical clustering paradigm and the new observations. 
Ellipticals are proposed to form mainly in early collapsing high mass-density peaks, where the gaseous disks suffer wet mergers or strong gas feeding by cold streams and form stars intensively from the dissipated gas. After this violent process of star formation and gas consumption, feedback mechanisms and environmental conditions both avoid further gas accretion and quench star formation; a fraction of elliptical galaxies can continue to significantly grow later by dry mergers \citep[see for reviews, e.g.,][and more references therein]{Mo+2010, Kormendy2016}.  
This scenario motivated the concept of ``red and dead'' galaxies to be used to refer to Es \citep[e.g.,][]{vanDokkum2005,Bamford+2009,Darg+2010,Smith+2010,Crossett+2014,DavisBureau2016,Boardman+2017}. The previous scenario offers a good representation for massive E galaxies. They dominate the morphological mix of galaxies for stellar masses higher than $\sim10^{10.8}$ \msun\
\citep[][]{Gadotti2009,NairAbraham2010,FisherDrory2011,Cibinel+2013,Avila-Reese+2014,Thanjavur+2016}.

With new instruments and more dedicated observations in this century, it has been possible to learn new information. In particular, the number 
of local early-type (i.e., E and lenticular) galaxies 
with integrated properties different from the classical ``red and dead'' massive ones increases as the stellar mass is smaller and the environment becomes less dense
\citep[e.g.,][]{Kuntschner+2002,Bernardi+2006,Lee+2006, Schawinski+2007, Schawinski+2009, Kannappan+2009, Suh+2010, Thomas+2010, McIntosh+2014, Schawinski+2014, GZ2015, Vulcani+2015, Lacerna+2016, SpectorBrosch2017,AF+2018,Lacerna+2018}. Two examples of these early-type galaxies are the recently quenched ellipticals (RQEs) 
and the blue, star-forming ellipticals (BSFs). 
\citet[][]{McIntosh+2014} define the RQEs as
mostly blue E galaxies at $z \leq 0.08$ with light-weighted stellar ages shorter than 3--4 Gyr and lacking detectable emission lines from star formation. They suggest them as candidates for "first-generation" Es, which formed in a relatively recent spiral-spiral major merger.
BSFs are blue galaxies with high star formation rates and light-weighted stellar ages lower than 1--2 Gyr, indicating recent processes of
star formation in them 
\citep[][]{Lacerna+2016,Lacerna+2018}. 

Projects with integral field spectroscopy 
\citep[IFS; see for a recent review][]{Sanchez2020}
based on integral field units (IFUs),
such as SAURON \citep{Bacon+2001,deZeeuw+2002,Sarzi+2006,Emsellem+2007,Shapiro+2010}, ATLAS$^{\verb|3D|}$ \citep{Cappellari+2011_I,Cappellari+2011_VII,Young+2014}, CALIFA \citep{Sanchez+2012,GonzalezDelgado+2015,Aquino+2018}, SAMI \citep{Croom+2012,Bassett+2017,Barone+2018}, 
MASSIVE \citep{Ma+2014,Veale+2018},
and MaNGA \citep{Bundy+2015,Blanton+2017,Wake+2017}, have revolutionized our understanding of early-type galaxies (E+S0) by resolving stellar populations and minimizing the effects of aperture size and slit losses.
 For example, \cite{McDermid+2015} applied models of single stellar populations and nonparametric star formation histories in early-type galaxies of ATLAS$^{\verb|3D|}$ and found that massive E+S0 galaxies formed 50\% of their stars at $z\sim 3$. In contrast, low-mass E+S0 galaxies reach the same fraction of stars at $z\sim 0.6$.
\cite{Ibarra-Medel+2016} studied the stellar mass growth histories of galaxies in MaNGA and found that the inside-out formation mode is more pronounced, on average, in late-type galaxies than in early-type galaxies, independently of mass.
\citet[][see also \citealt{Garcia-Benito+2017,LopezFernandez+2018,Bitsakis+2019}]{GonzalezDelgado+2017} studied the spatially resolved star formation histories (SFHs) in CALIFA and found that the star formation rate (SFR) in outer regions of E+S0 galaxies has undergone an extended phase of growth in mass between $z = 2$ and $0.4$ 
(see also \citealt{Sanchez+2019} for the SFHs of E+S0 galaxies in MaNGA). 
Additionally, massive E galaxies show evidence of early and fast quenching of star formation, whereas less massive E galaxies show extended SFHs with slow quenching. 
The information provided by IFS has also been used to constrain the origin of the gas content and the assembly history of E+S0 galaxies in simulations \citep[][and references therein]{Lagos+2014,Naab+2014}.

In this paper, we use data from MaNGA, which is
so far the largest IFS galaxy survey. 
Unlike other works on early-type galaxies in general,
here, we focus on the spatially resolved properties of only E galaxies,
that is to say not including evident S0 galaxies (see also \citealt{DominguezSanchez+2019} and \citealt{Bernardi+2019} with MaNGA), 
with a particular focus on classical (red and quenched) E galaxies,  CLEs. The post-processing catalog from the DESI Legacy Imaging Surveys \citep{Dey+2019} allowed us to look for disky features and sharp edges in E galaxies, and then purify our sample to purely E galaxies %when 
as much as
possible. On the other hand, given the spectral information provided by MaNGA, 
we can accurately select
these kinds of galaxies, excluding those with signatures of active star formation (SF) or active galactic nuclei (AGN), as well as those with blue colors. 
Here, we use the spatially-resolved spectral analysis of the MaNGA CLEs to find constraints on the local and global SFH and stellar mass growth of these galaxies.
In addition, we pay attention to two other, particular, much less frequent E classes; the RQE and BSF galaxies.  
It is essential to determine the following: the mechanisms behind the recent star formation quenching of RQEs; and whether BSF galaxies are classic E galaxies that were rejuvenated by recent cosmological gas accretion or wet minor mergers or, alternatively, whether they are post-starburst galaxies with a recent morphological transformation.

We aim to carry out our explorations in terms of stellar mass and galaxy structural parameters. 
We study the profiles of specific SFR (sSFR), stellar ages, and stellar metallicities of classical E, RQE, and BSF galaxies, and 
we reconstructed their cumulative temporal distributions of stellar mass and SF at two different radial regions using the 
IFUs of MaNGA.
We present the statistics and general results on the 
populations for the classical E, RQE, and BSF galaxies.
In a companion paper, we will study the different properties presented here in terms of the environment and individual-based cases for the low-frequent populations of RQE and BSF galaxies. 

The outline of the paper is as follows.
Section \ref{S2} presents the description of MaNGA.
This section also describes the visual morphological classification of E galaxies and their integrated and resolved physical properties used throughout the paper. 
In Sect. \ref{S3}, we classify E galaxies based on integrated properties, such as color, equivalent width of H$\alpha$, light-weighted stellar age, emission lines of the BPT diagram, and total stellar mass. From this selection, we pay special attention to classical Es, RQEs, and BSFs and
study their 
radial profiles and gradients in stellar age and metallicity
in Sect. \ref{S4}.  
We estimate their cumulative mass temporal distributions and SFHs, both globally and radially, in Sect. \ref{S5}, and determine the long-term
quenching epoch and time-scale of the Es when it applies.
The implications of our results are discussed in Sect. \ref{Sdiscussion} and the conclusions are given in Sect. \ref{Sconclusions}.

Throughout this paper the reduced Hubble constant, $h$, is defined as $H_0 = 71$ $h$ km s$^{-1}$ Mpc$^{-1}$, unless another value of $h$ is specified, with the following dependencies: stellar mass in $h^{-2}~\rm M_{\sun}$, physical scale in $h^{-1}~\rm{ Mpc}$, and age and look back time in $h^{-1}~\rm{ yr}$. When a cosmological model should be assumed, we use a flat model with $\Omega_\Lambda =0.73$;  
the age for this model is 13.67 Gyr.
Table \ref{tabla0} lists %all 
the acronyms used in this paper.

%%% TABLA acronyms
\begin{table}
\caption{List of acronyms used in this paper.
}
\centering
\begin{tabular}{c c c}
\hline
\hline
AGB & Asymptotic Giant Branch \\
AGN & Active Galactic Nuclei \\
BPT & Baldwin, Phillips \& Terlevich diagram \\
BSF & Blue, Star-Forming (elliptical) \\
CLE & Classical Elliptical \\
CMTD & Cumulative Mass Temporal Distribution \\
E   & Elliptical \\
E+S0 & Early-type galaxies \\
EW  & Equivalent Width \\
FWHM & Full Width at Half Maximum \\
IFS & Integral Field Spectroscopy \\
IFU & Integral Field Unit \\
IMF & Initial Mass Function \\
lw  & luminosity-weighted \\
mw  & mass-weighted \\
PSF & Point Spread Function \\
\re & Effective Radius \\
RG  & Retired Galaxy \\
RQE & Recently Quenched Elliptical \\
S/N & Signal-to-Noise ratio \\
S0  & lenticular \\
SF  & Star Formation \\
SFG & Star-Forming Galaxy \\
SFH & Star Formation History \\
SFR & Star Formation Rate \\
SN  & Supernova \\
sSFH & Specific Star Formation History \\
sSFR & Specific Star Formation Rate \\
SSP & Simple Stellar Population \\
UND & Undetermined \\
$Z$ & stellar metallicity \\
% \hline
% ALHAMBRA & Advanced Large, Homogeneous Area Medium Band Redshift Astronomical Survey \\
% CALIFA & Calar Alto Legacy Integral Field spectroscopy Area survey \\
% DAP & Data Analysis Pipeline \\
% DESI & Dark Energy Spectroscopic Instrument \\
% DR  & Data Release \\
% DRP & Data Reduction Pipeline \\
% EAGLE & Evolution and Assembly of GaLaxies and their Environments \\
% FIREFLY & Fitting IteRativEly For Likelihood analYsis \\
% IPAC & Infrared Processing and Analysis Center \\
% MaNGA & Mapping Nearby Galaxies at the Apache Point Observatory \\
% MPA-JHU & Max Plank Institute for Astrophysics and the Johns Hopkins University \\
% MPL & MaNGA Product Launch \\
% MUSE & Multi Unit Spectroscopic Explorer \\
% NASA & National Aeronautics and Space Administration \\
% NSA & NASA-Sloan Atlas \\
% SAMI & Sydney – Australian Astronomical observatory Multi-Object Integral field spectrograph galaxy survey \\
% SAURON & Spectroscopic Areal Unit for Research on Optical Nebulae \\
% SDSS & Sloan Digital Sky Survey \\
% VIMOS & VIsible Multi-Object Spectrograph \\
\hline
\end{tabular}
\label{tabla0}
\end{table}

%%%%%%%%%%%%%%%%%%%%%%%%%%%%%%%%%
\section{Data} 
\label{S2}
%%%%%%%%%%%%%%%%%%%%%%%%%%%%%%%%%

\subsection{Sample} 

 Mapping Nearby Galaxies at the Apache Point Observatory (MaNGA) is an IFS survey of galaxies in the redshift range of 0.01 $<$ $z$ $<$ 0.15 on the SDSS 2.5 m telescope \citep{Gunn+2006}.
Each spectra has a wavelength coverage of 3600--10,300 \AA, an instrumental resolution of $\sim$ 65 km s$^{-1}$, and a median spatial resolution of 2.54 arcsec FWHM 
\citep[1.8 kpc at the median redshift of 0.037,][]{Drory+2015,Law+2015,Law+2016DRP}.
We use the MaNGA Product Launch-7 (MPL-7) from the Sloan Digital Sky Survey (SDSS) Data Release 15 \citep{DR15+2019},
which contains 4621 galaxies.

\subsection{Stellar population analysis}
\label{S_ssp}

We made use of the \verb|Pipe3D| pipeline 
\citep{Sanchez+2016_p21,Sanchez+2016_p171,Sanchez+2018_AGN}
to perform the spatially resolved stellar population analysis of the data cubes. We note that 
\verb|Pipe3D| is based on the stellar population synthesis code \verb|FIT3D| \citep{Sanchez+2016_p21,Sanchez+2016_p171}, which uses the \verb|GRANADA| and \verb|MILES| simple stellar population (SSP) libraries in the current implementation to model and subtract the stellar spectrum and fit the emission lines. The \verb|GRANADA| and \verb|MILES| (gsd156) stellar library is a combination of the empirical stellar library of \citet{Vazdekis:2010aa} and the theoretical stellar libraries of \citet{Gonzalez-Delgado+2005} and \citet{Gonzalez-Delgado+2010}. 
The SSPs use a \citet{Salpeter+55} initial mass function and
cover 39 stellar ages (from 1 Myr to 14.1 Gyr\footnote{In practice, we use results until 13 Gyr.}) and four metallicities (Z=0.002, 0.008, 0.02, and 0.03). It is important to note that  \verb|FIT3D| performs the SSP fitting within the wavelength of $3500\AA$ to $7000\AA$.  In addition, \verb|FIT3D| considers the effects on dust extinction during the stellar population synthesis using a \citet{Cardelli+1989} extinction law. \verb|FIT3D| provides the SSP decomposition of the modeled stellar spectra of each spatially resolved region of the galaxies. 
With this decomposition, we can estimate the respective SFHs and the cumulative mass temporal distributions \citep[or mass growth histories; e.g.,][]{ Ibarra-Medel+2016,Ibarra-Medel+2019,Sanchez+2019}. 
In order to estimate the contemporary (local and integrated) SFR values, 
we smoothed the SFHs within a period of 30 Myr. 
These quantities are independent of SFR estimations that use the H$\alpha$ emission line flux. For star-forming galaxies, both estimates are %actually 
very similar, showing that the SFR calculated in the last 30 Myr is related to massive young stars, contributing to the H$\alpha$ emission. For retired galaxies, the SFR calculated from the H$\alpha$ flux is likely overestimated because this flux is dominated by effects of post-AGB stars rather than of young stars \citep[see, e.g.,][]{Bitsakis+2019,Sanchez+2019}. However, in these 
galaxies, the SFR derived from the stellar population synthesis with Pipe3D should also be considered as an upper limit.
To fit the emission lines, \verb|Pipe3D| subtracts the modeled stellar spectra and fits a series of Gaussian profiles for each line \citep{Sanchez+2016_p171}. With those fits, \verb|Pipe3D| provides the total flux, dispersion, and the kinematics of each emission line. The quality of the emission line fits of \verb|Pipe3D| (and other codes) was tested using MUSE data \citep[][]{Sanchez-Menguiano+2018,Bellocchi+2019,Lopez-Coba+2020}. 
With the use of the adjacent continuum windows (from the modeled stellar spectra) for each emission line, \verb|Pipe3D| estimates the stellar continuum flux and provides the equivalent width of each emission line.

We estimated the integrated (within a 1.5 effective radius, \re) and radially resolved (within a given radial region) 
luminosity- and mass-weighted stellar ages, age$_{lw}$ and age$_{mw}$, respectively, using the following definitions:
\begin{equation}
\log(age_{mw})=\sum_j^{n_{ssp}}\log(age_{ssp,j})m_{ssp,j}/\sum_j^{n_{ssp}}m_{ssp,j},%\\
\label{eq_ageMW}
\end{equation}
\begin{equation}
\log(age_{lw})=\sum_j^{n_{ssp}}\log(age_{ssp,j})L_{ssp,j}/\sum_j^{n_{ssp}}L_{ssp,j},
\label{eq_ageLW}
\end{equation}
where $age_{ssp,j}$ is the single stellar population age with index $j$, luminosity $L_{ssp,j}$, and mass $m_{ssp,j}$;
$n_{ssp}$ is the total number of SSPs within 1.5 $R_{eff}$ or the given radial region. 
In the same way, we calculated the respective mass- and luminosity-weighted stellar metallicities, $Z_{mw}$ and $Z_{lw}$.
We also calculated the integrated and azimuthally-averaged radial surface-density values of the sSFR based on the stellar population synthesis method. We also calculated the following integrated quantities within 1.5 $R_{eff}$: H$\alpha$ equivalent width and the emission line fluxes of H$\alpha$, H$\beta$, [OIII]$\lambda$5007, and [NII]$\lambda$6584. The latter quantities were used to study the BPT diagram \citep{BPT1981, VeilleuxOsterbrock1987, Kewley+2001}.

\subsection{Morphology}
\label{sec_morpho}

A detailed description of the procedures in our visual classification of MPL-7 galaxies can be found in V\'azquez-Mata et al. \citep[in prep. See also][]{Hernandez-Toledo+2008, Hernandez-Toledo+2010}. 
Briefly, the classification of each MaNGA galaxy was carried out by 
visually inspecting
post-processed images from the SDSS and the DESI Legacy Imaging Surveys \citep{Dey+2019} in various stages. 

A first input morphological evaluation was achieved by visually inspecting all the $gri$ color images available in the SDSS image server. In a second stage, we used the Montage software packages from NASA/IPAC (Infrared 
Processing and Analysis Center) to re-assemble the $r$-band SDSS images corresponding to each MaNGA galaxy. These new background-subtracted and gradient-removed 
images were used to generate mosaics containing
gray logarithmic-scaled $r$-band images and filter-enhanced $r$-band images (after applying Gaussian filters of 
various sizes that later were subtracted from the original images). 

The DESI Legacy Imaging Surveys\footnote{http://legacysurvey.org/} have delivered optical images for an important fraction of the 
extragalactic sky that is visible from the Northern Hemisphere in three optical bands ($g$, $r$, and $z$), providing 
a survey of nearly uniform depth (5$\sigma$ depths of $g$ =24.0, $r$ =23.4, and $z$ =22.5 AB mag for a fiducial DESI point source). 
The DESI collaboration generated a post-processing catalog for the Legacy Surveys using an approach to estimate the light distribution, namely, 
source shapes and brightness properties. Each source was modeled 
using the Tractor package \citep[for more details, see][]{Dey+2019} using parametric light profiles, including delta functions (for point sources), 
de Vaucouleurs ($R/$\re)$^{1/4}$ law, the exponential disk, or a combination of the last two.  
The best fit model was 
determined by convolving each model with the specific point spread function (PSF) for each 
exposure, fitting each image, and minimizing the residuals for 
all images. 
Similar to the SDSS images, the DESI legacy
images for all MaNGA galaxies in the $r$-band were retrieved and 
digitally-processed to generate new mosaics containing filter-enhanced images, 
the $grz$ color images, and the corresponding residual images.

Early-type galaxies were classified according to their bulge dominance. 
Bulge-dominated galaxies have bright centers with a gradual fall-off in brightness 
at all radii, with the outer regions having an extended envelope without any sharp edges.  
On the other hand, disk-dominated galaxies have a sharp outer edge where the light drops off drastically, showing a flatter profile at intermediate radii. The disk can be either featureless or 
weakly featured (smooth disks) or with more defined or strong features (non-smooth disks).

Since a relevant goal for the present work is to isolate between early-type E and S0 candidates as much as possible, 
we carried out this separation by combining our previous experience \citep{Hernandez-Toledo+2008,Hernandez-Toledo+2010}
with a well-defined, purely light-based approach by 
\citet{Cheng+2011}, in which
red sequence galaxies are accommodated according to their light smoothness distribution into bulges, smooth disks, and non-smooth disks. Then, we assigned these groups to the Hubble types E, S0, and S0a, respectively. 
We explicitly notice that our bulge-dominated category is not composed of pure
bulges. In our experience with the mosaic images, many apparent bulge-dominated 
galaxies have characteristics indicating the presence of both bulge and disk-like 
components, similar to galaxies showing a disk embedded within a more extended classical bulge (c.f., 
\citealt{Cappellari+2011_VII}).
In our classification, bulges with centrally 
concentrated disks or inner disks were classified as Es. On the other hand, galaxies, where the 
apparent disk (either face-on or edge-on) is occupying a significant (or dominant) fraction of the 
extended bulge, were classified as lenticulars, resembling the classification grid by \cite{GrahamA+2019}
for early-type galaxies. 
At present, only the most evident S0 cases were eliminated from our sample.
In this way, we identify 539 E galaxies, which is $\approx 12\%$  of the MaNGA MPL-7 sample.

We emphasize that both our filter-enhanced and PSF-convolved residual images from the DESI 
post-processing catalog  were helpful to visually identify disky features and sharp edges in E 
galaxy candidates and to look for disk-features (broad 
or tight arm-like features) in S0 candidates,
including flattened edge-on disks. 
The depth of the DESI $r$-band images translates into a median surface brightness limit of $\sim$ 27.5 mag arcsec$^{-2}$ for a 3$\sigma$ detection of a 100 arcsec$^{-2}$ low surface brightness feature 
\citep{Hood+2018}.
In spite of the higher depth of the DESI images, a major limitation in our 
visual classification is the difficulty of classifying small or compact galaxies that show no visible structure or that cannot be resolved. Furthermore, the visual nature still allows for possible contamination of the E sample by face-on S0s. 
\citet{DominguezSanchez+2020} recently found evidence that (fast-rotating) E galaxies are not simply S0 galaxies that are viewed face-on.
In our experience, a more quantitative approach from an isophotal analysis, as shown in %Hernandez-Toledo (2008;2010) 
\citet{Hernandez-Toledo+2008,Hernandez-Toledo+2010},
in combination with a detailed 2D parametric decomposition of the DESI images are useful to identify possible face-on S0 galaxies.

\subsection{Color, stellar mass, redshift, structural, and kinematic properties} 

Color measurements were taken from the SDSS 
database with extinction corrected $modelMag$ magnitudes  (\verb|dered| parameter in SkyServer\footnote{See https://skyserver.sdss.org/dr15/}). 
This magnitude is defined as the better of two magnitude fits: a pure de Vaucouleurs profile and a pure exponential profile.  

The stellar mass and redshift are obtained from the MaNGA data reduction pipeline
\citep[DRP;][]{Law+2016DRP}, which includes galaxy properties from the NASA-Sloan Atlas (NSA\footnote{See http://nsatlas.org/}). For the stellar mass, we used \verb|NSA_Sersic_MASS| that is calculated from a K-correction fit for S\'ersic fluxes 
(two-dimensional S\'ersic fit flux using $r$-band structural parameters). We find a good agreement between this definition of stellar mass and the stellar mass estimated with the stellar mass-to-light ratio of \cite{Bell+2003}, considering a \cite{Kroupa2001} initial mass function (IMF)
and including an extra correction 
to the Petrosian magnitudes of -0.1 mag since these magnitudes underestimate the total flux for E galaxies \citep{Bell+2003,McIntosh+2014}.

We use the S\'ersic index and the concentration  as two structural properties of E galaxies. The S\'ersic index ($n$) is a two-dimensional, single-component S\'ersic fit in the $r$-band and
obtained from the NSA catalog (\verb|v1_0_1|). The concentration is defined as
$R_{90}/R_{50}$, where $R_{90}$ and $R_{50}$ are the azimuthally-averaged 90\% and 50\% light radius in Petrosian magnitudes in the $r$-band, respectively. They are also obtained from the NSA catalog.
We also calculated the ellipticity as $\epsilon = 1 - b/a$, in which the axis ratio $b/a$ is a two-dimensional, single-component Sersic fit in the $r$-band from the NSA catalog.
For the kinematics, we calculated $\lambda_{R_e}$, the specific angular momentum for the stellar populations within 1 $R_{eff}$, using the \verb|Pipe3D| pipeline.

We establish an upper limit of $z = 0.08$ in our sample 
as a compromise for having a considerable number of galaxies with reliable visual morphological classifications of E galaxies, avoid an excess of very red, massive galaxies that increase significantly in number at larger redshifts in MaNGA, and obtain a reasonable normalization of the cumulative mass temporal distributions
at a look back time of $\sim$1 Gyr
\citep[e.g.,][]{Bamford+2009,Hernandez-Toledo+2010,Lacerna+2014,Ibarra-Medel+2016, Sanchez+2017}. 
We obtain 345 E galaxies out to $z = 0.08$.
Finally, we adopt a lower limit in the stellar mass of 10$^9$ \msun\ to have a reliable sample.
There are 343 E galaxies that satisfy this condition, with a median $z$ of 0.044.
This is our final sample of E galaxies.

%%%%%%%%%%%%%%%%%%%%%%%%%%%%%%%%%
\section{Classification of elliptical galaxies based on their integrated properties} 
\label{S3}
%%%%%%%%%%%%%%%%%%%%%%%%%%%%%%%%%

%%% Fig BPT
\begin{figure}
\includegraphics[width=9.1cm]{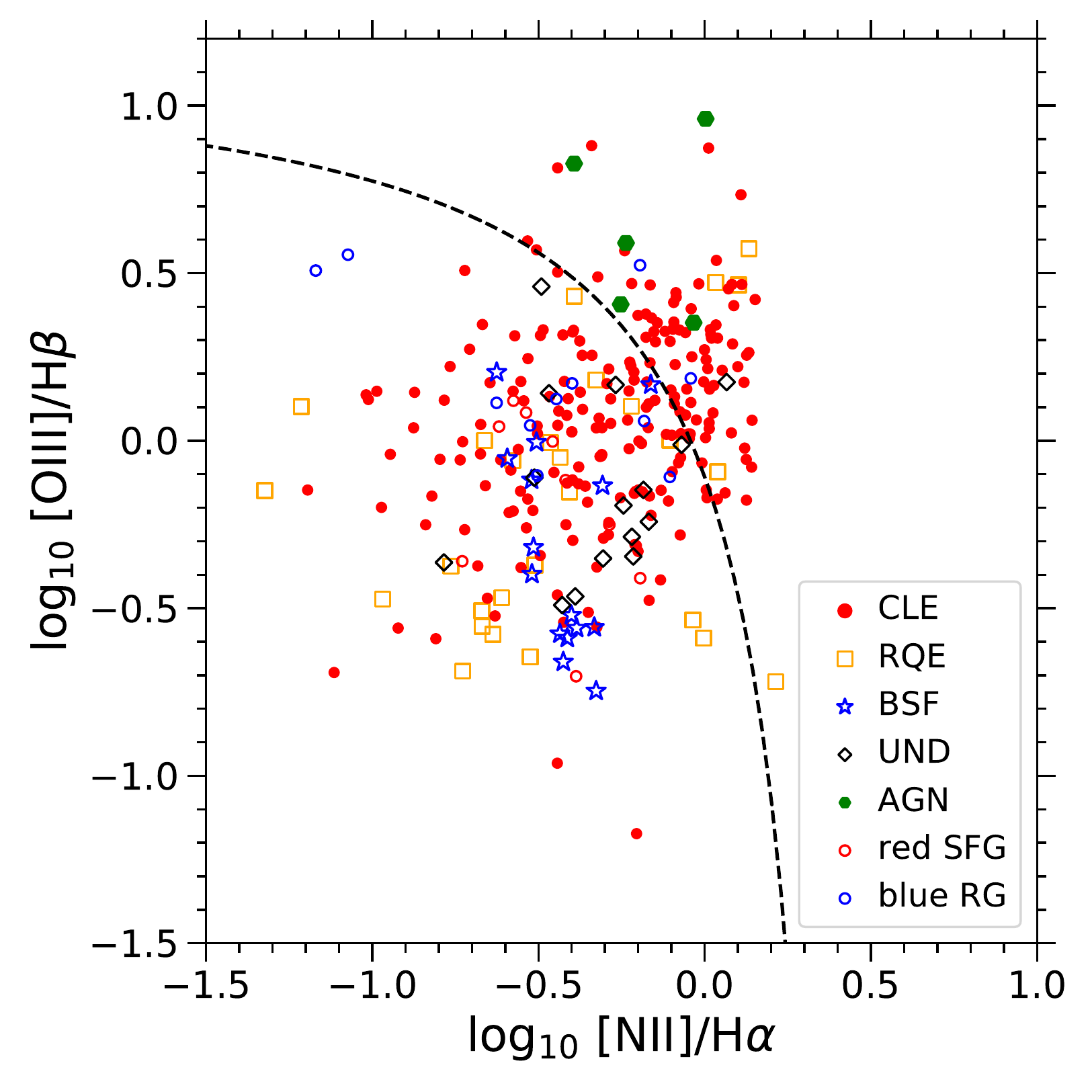}
\caption{
BPT diagram of elliptical galaxies from MaNGA.
The dashed line is given by \citet[][]{Kewley+2001}.
The symbols correspond to classical elliptical (CLE) galaxies in red filled circles,
recently quenched elliptical (RQE) galaxies in orange open squares, 
blue and star-forming (BSF) elliptical galaxies in blue open stars, undetermined (UND) elliptical galaxies in black open diamonds, AGN elliptical galaxies in green filled hexagons, red star-forming ellipticals in red open circles, and blue retired galaxies (RG) in blue open circles. See the details of their classification in Sect. \ref{S3}.
}
\label{figBPT}
\end{figure} 

Following \cite{Sanchez+2014} and \citet[][and more references therein]{Cano-Diaz+2016,Cano-Diaz+2019},
we define the retired galaxies (RGs) as those galaxies with an equivalent width (EW) of H$\alpha$ $<$ 3 \AA. There are 299 RGs in our sample of E galaxies.
Here, the RQE galaxies are defined as galaxies with an 
age$_{lw}$ $\leq$ 4 Gyr and either RGs or with 3 $\leq$ EW(H$\alpha$)/\AA < 6 lying above the relation of \citet[][]{Kewley+2001} in the BPT diagram (Fig. \ref{figBPT}). We identify 35 RQEs, which are all RGs with young stellar populations.
The other galaxies with 3 $\leq$ EW(H$\alpha$) < 6 \AA\ 
and that are not classified as RQEs because they lie below the \citet[][]{Kewley+2001} line, are nominally defined as undetermined (UND) E galaxies. 
There are 
15 UND galaxies.
Following \citet{Cano-Diaz+2019}, we define star-forming galaxies (SFGs) and galaxies with active galactic nuclei (AGN) as those systems that fall below and above the relation of \citet[][dashed line]{Kewley+2001}, respectively, and whose EW(H$\alpha$) is $\geq$ 6 \AA. 
There are 24 SFGs and 5 AGNs.

%%% TABLA E types
\begin{table*}
\caption{Classification of E galaxies.
}
\centering
\begin{tabular}{c c c c}
\hline
\hline
type  &  N (\%)  &  criteria \\
\hline
CLE &       251 (73.2\%) & age$_{lw}$ $>$ 4 Gyr, EW(H$\alpha$) $<$ 3 \AA\ and red     \\
RQE &        35 (10.2\%) & age$_{lw}$ $\leq$ 4 Gyr and EW(H$\alpha$) $<$ 3 \AA \\ 
BSF &        15 (4.4\%) & EW(H$\alpha$) $\geq$ 6 \AA, below \citet[][]{Kewley+2001} line and blue \\
AGN &         5 (1.4\%) & EW(H$\alpha$) $\geq$ 6 \AA\ and above \citet[][]{Kewley+2001} line \\
UND &        15 (4.4\%) & 3 $<$ EW(H$\alpha$) $\leq$ 6 \AA \\ 
blue RG &    13 (3.8\%) & age$_{lw}$ $>$ 4 Gyr, EW(H$\alpha$) $<$ 3 and blue \\
red SFG &     9 (2.6\%) & EW(H$\alpha$) $\geq$ 6 \AA, below \citet[][]{Kewley+2001} line and red \\
\hline
\end{tabular}

\tablefoot{Columns are the type of E galaxy, the number and percentage of each type, and the criteria to classify them.  
}
\label{tabla1}
\end{table*}

\begin{figure}
\includegraphics[width=9.cm]{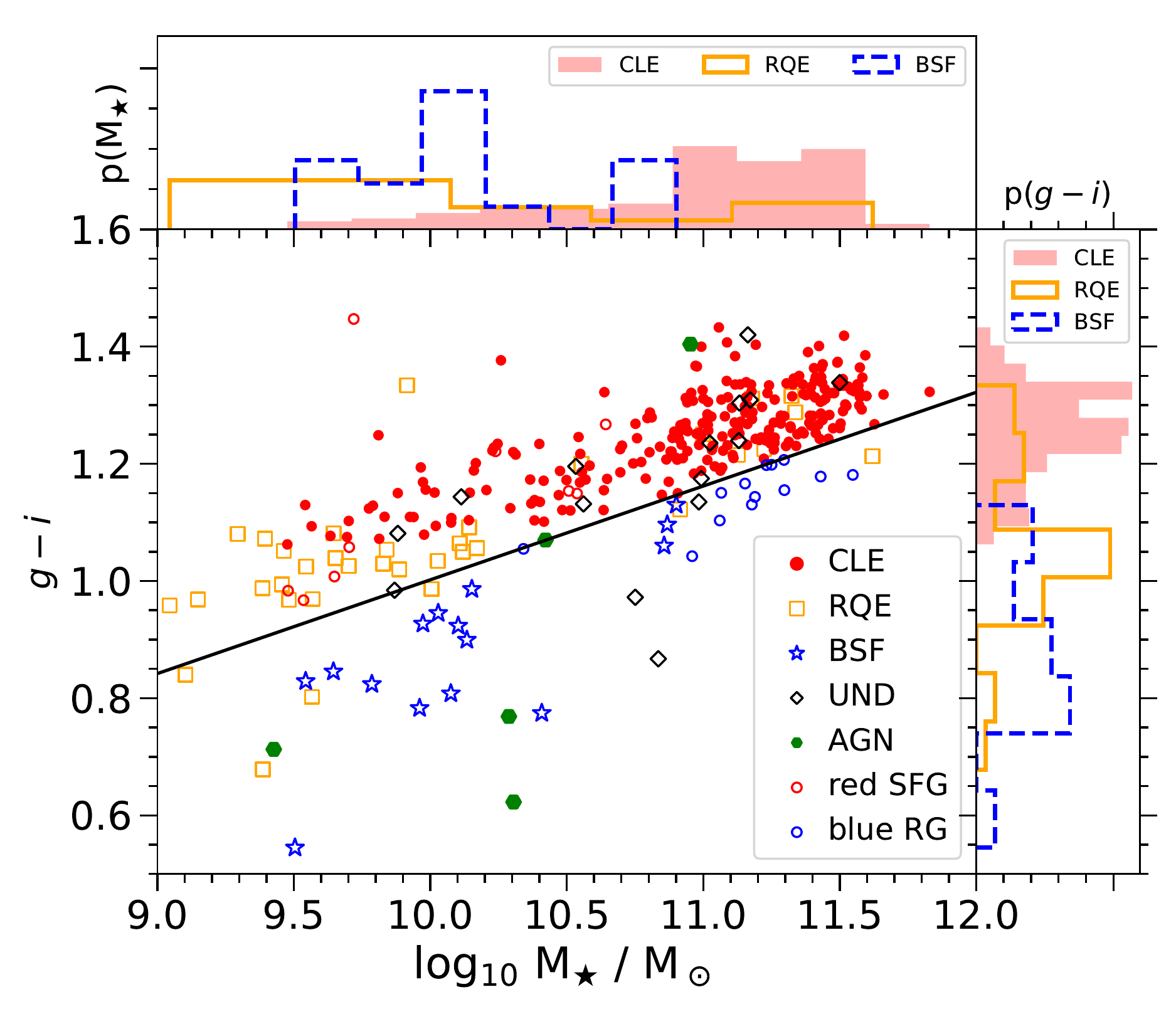}
\caption{
As a function of stellar mass, $g-i$ color  
for E galaxies from MaNGA. 
The black solid line separates red and blue galaxies.
The symbols for the galaxies are the same as in  Fig. \ref{figBPT}.
Top and right panels: normalized density distributions of stellar mass and $g - i$ color, respectively, for CLE 
galaxies (red solid histogram), RQE galaxies (orange open histogram), and BSF galaxies (blue dashed histogram).
The integral of each histogram sums to unity.
}
\label{fig_gi_Ms_0.08}
\end{figure}

Fig. \ref{fig_gi_Ms_0.08} shows the $g-i$ color as a function of stellar mass. 
We use the relation found by \citet{Lacerna+2014} to separate between red and blue galaxies. The color is a representation of the aging of galaxies.
RQEs are typically in the transition between red and blue galaxies.
Therefore, our RQEs are a reliable sample of galaxies that stopped the star formation activity recently.

We find 251 RGs with red colors and which are not classified as RQEs (i.e., with age$_{lw}$ > 4 Gyr). We refer to this population as classical elliptical (CLE) galaxies. 
There are 13 other RGs with blue colors and age$_{lw}$ > 4 Gyr. From the sample of SFGs, we identify 15 blue galaxies, which we refer to as blue and SF (BSF) galaxies.
There are 9 other SFGs with red colors. The numbers and classification criteria of the E galaxies are summarized in Table \ref{tabla1}.

%%%fig color (fig2)

The stellar mass (\ms) distributions of BSFs and RQEs are different compared with the distribution of CLEs, as shown in the top panel of Fig. \ref{fig_gi_Ms_0.08}. CLEs are typically more massive with a median \ms\ of $10^{11.1}$ \msun, whereas the median \ms\ is $10^{10.1}$ \msun\ for BSFs and $10^{9.9}$ \msun\ for RQEs.  The latter presents a nearly flat mass distribution below $10^{10}$ \msun.
As mentioned above, the percentage of E galaxies with different properties than CLEs increases for lower masses 
\citep[see also][]{Lacerna+2016}. 
The CLEs correspond to 86\% of the galaxies with \ms\ $\geq$ $10^{11}$ \msun, whereas RQEs are only $\sim$4\%, and no BSF is in this massive bin. For the range $10^{10.4}$ $\leq$ \ms\ $<$ $10^{11}$ \msun, CLEs represent 78\%, RQEs represent $\sim$4\%, and BSFs represent $\sim$5\%. The change in the fractions is more important for lower masses, \ms\ $<$ $10^{10.4}$ \msun, where the CLEs represent only 44\%, RQEs represent 29\%, and BSFs represent 12\%.
There are only RQE galaxies at \ms\ $\leq$ $10^{9.4}$ \msun. 
The overall fractions reported in Table \ref{tabla1} should be considered with caution, given that our MaNGA sample of Es is not from a complete volume survey. The fractions reported 
for each stellar mass bin are less affected by the volume incompleteness. 

%%%Fig RQE (fig3)
\begin{figure}
\includegraphics[width=8.8cm]{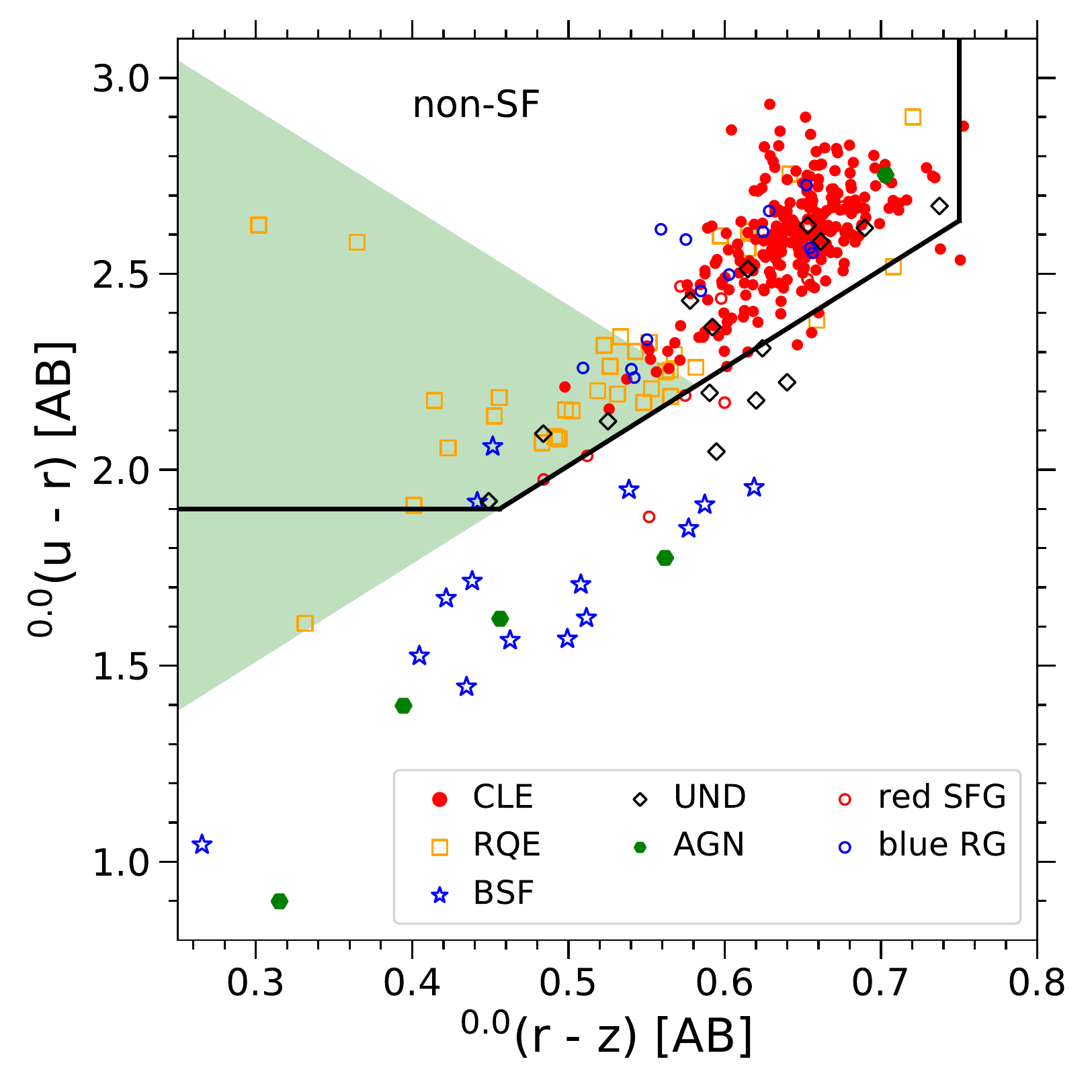}
\caption{Color--color diagram. 
The colors are in AB magnitudes and $K$-corrected at $z=0$.
The symbols for the galaxies are the same as  in Fig. \ref{figBPT}.
The thick black lines enclose the non-star-forming region of \citet{McIntosh+2014} based on \citet{Holden+2012}. 
The green shaded region corresponds to an empirical relation to select RQE galaxies defined by \citet{McIntosh+2014}.}
\label{fig_RQE}
\end{figure} 
%%%

%%%fig age (fig4)
\begin{figure}
\includegraphics[width=9.cm]{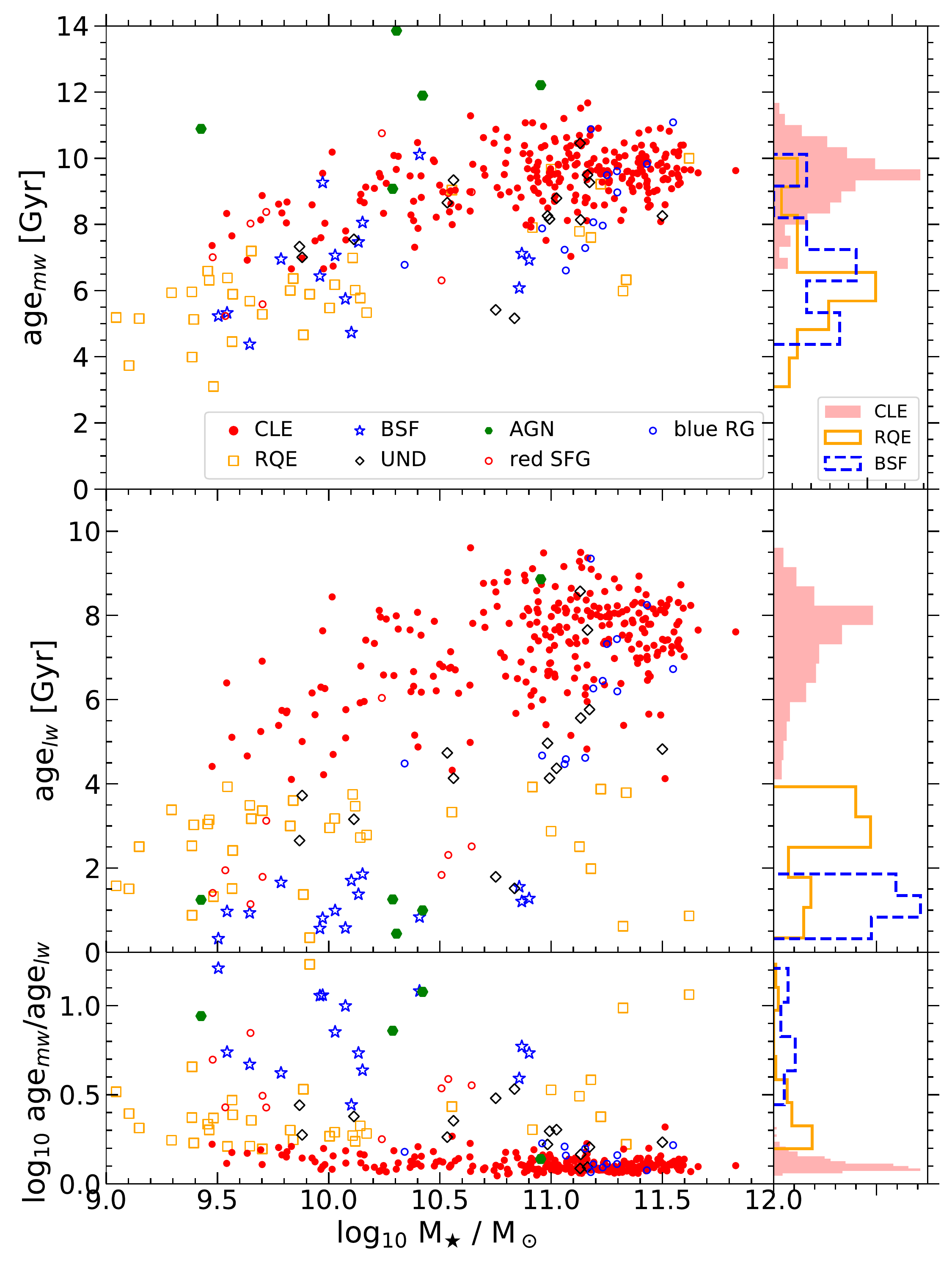}
\caption{Global mass-weighted stellar ages (top-left panel) and global luminosity-weighted stellar ages (middle-left panel) as functions of the stellar mass of E galaxies.
The symbols for the galaxies are the same as  in Fig. \ref{figBPT}.
The relative difference between both ages is shown in the bottom-left panel.
Top-right, middle-right, and bottom-right panels: normalized density distributions of mass-weighted ages,  luminosity-weighted ages, and the relative difference between both ages for CLE 
galaxies (red solid histogram), RQE galaxies (orange open histogram), and BSF galaxies (blue dashed histogram), respectively.
The integral of each histogram sums to unity. 
}
\label{fig_age}
\end{figure}

%%%fig Z integrated (fig5)
\begin{figure}
\includegraphics[width=9.cm]{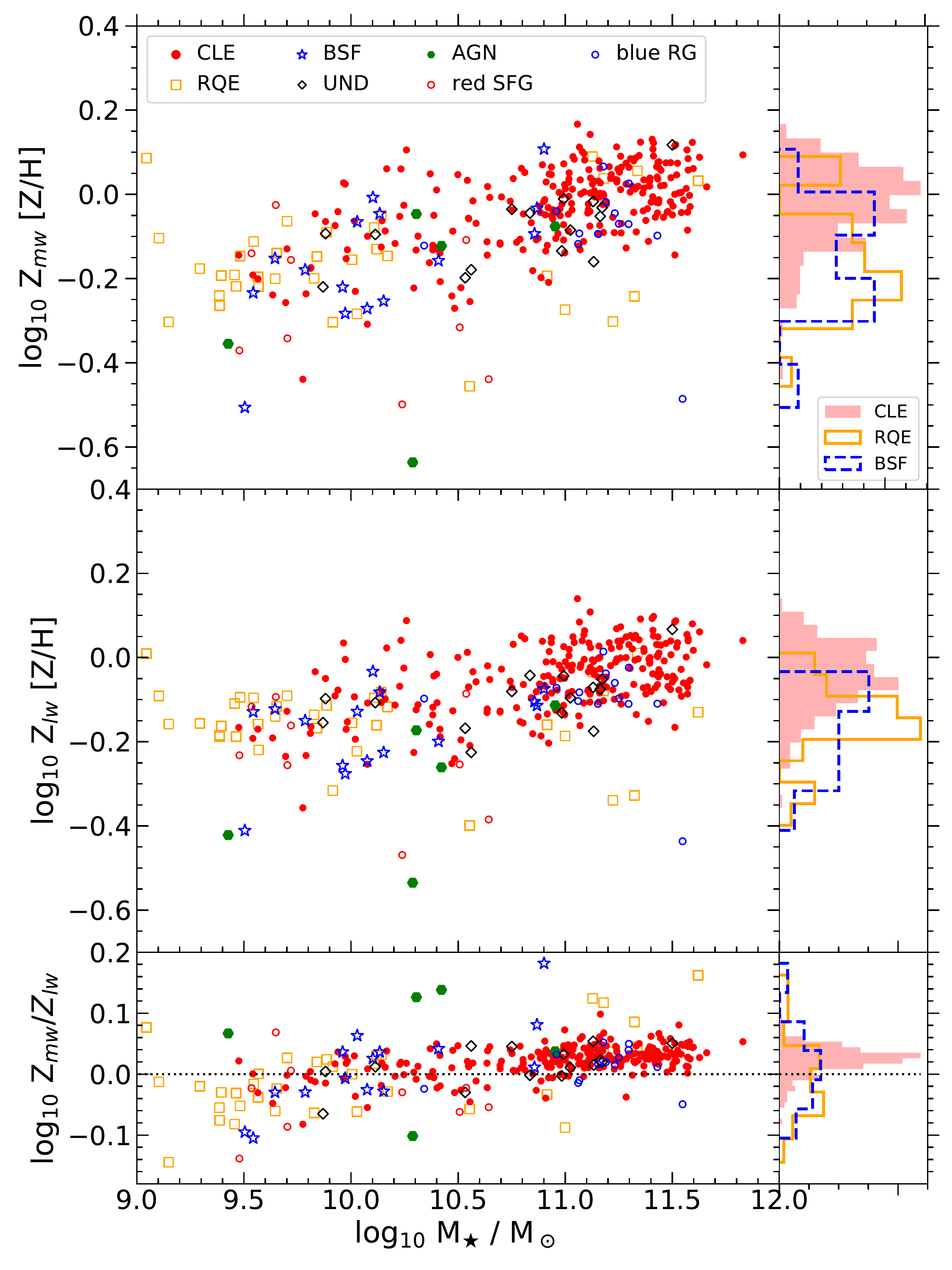}
\caption{Global mass-weighted stellar metallicities (top-left panel) and global luminosity-weighted stellar metallicities (middle-left panel) as functions of the stellar mass of E galaxies.
The symbols for the galaxies are the same as in Fig. \ref{figBPT}.
The relative difference between both metallicities (in dex units)  
is shown in the bottom-left panel.
Top-right, middle-right, and bottom-right panels: normalized density distributions of mass-weighted $Z$, luminosity-weighted $Z$, and the relative difference between both $Z$ for CLE galaxies (red solid histogram), RQE galaxies (orange open histogram), and BSF galaxies (blue dashed histogram), respectively. 
The integral of each histogram sums to unity. The dotted line in the bottom panels is the difference in $Z$, which equals 0.
}
\label{fig_Z}
\end{figure}

The $g-i$ color is also different between CLEs and BSFs by definition. The median $g-i$ color is 1.26 for the former and 0.90 for the latter. 
The RQEs are in the middle (median $g-i$ color of 1.05), with an overlap of the BSF and CLE color distributions.
The definition of \citet{McIntosh+2014} strongly motivates our classification of RQE galaxies.
In the following, we mention some differences. First, they use light-weighted median stellar ages $\leq$ 3 Gyr derived by \citet{Gallazzi+2005}, whereas this upper limit is slightly relaxed to 4 Gyr in our case by using integrated values of age$_{lw}$ within 1.5 \re\ as described in Sect. \ref{S_ssp}. Second, they follow the method of \citet{PeekGraves2010} to define spectroscopically quiescent galaxies within an ellipse of EW(H$\alpha$) and EW([OII]) using fluxes from 
the Max Plank Institute for Astrophysics and the Johns Hopkins University \citep[MPA-JHU,][]{Kauffmann+2003May,Brinchmann+2004,Tremonti+2004,Salim+2007} catalog. Additionally, they consider quiescent galaxies as %to 
those with
an S/N < 3 in H$\alpha$ flux or with EW(H$\alpha$) $\leq$ 2 \AA. In our case, we use the definition of RGs, that is, EW(H$\alpha$) < 3 \AA\ integrated within 1.5 \re, as the criterion for quiescent galaxies. Finally, RQE candidates are inside the non-star-forming region of the $urz$ diagram according to \citet{McIntosh+2014}. 
We do not use this condition because we consider that our condition for RGs is good enough to select non-star-forming galaxies. Fig. \ref{fig_RQE} shows the $urz$ diagram for our sample of E galaxies. Nearly all the RQEs using our selection criteria are located inside the modified non-star-forming region of \citet{McIntosh+2014}. We also include the empirical selection of RQE galaxies proposed by them using only the $urz$ diagram. We find 
that 25 (71\%) of the RQEs using our selection criteria are located within this region. This fraction is smaller than the 82\% of RQEs identified by \citet{McIntosh+2014} with their empirical method, but it is expected given the different criteria mentioned above. 
Inside this empirical region, we also find six (2\%) CLEs, 
two (13\%) BSFs, 
three (20\%) UNDs, 
three (23\%) blue RGs, one (11\%) red SFG, and no AGN. 
In summary, the \citet{McIntosh+2014} empirical selection can identify a high fraction of RQEs and it includes mild pollution from other Es, especially from blue RGs and UNDs.

%%%Fig structural properties (fig6)
\begin{figure*}
\includegraphics[width=9.2cm]{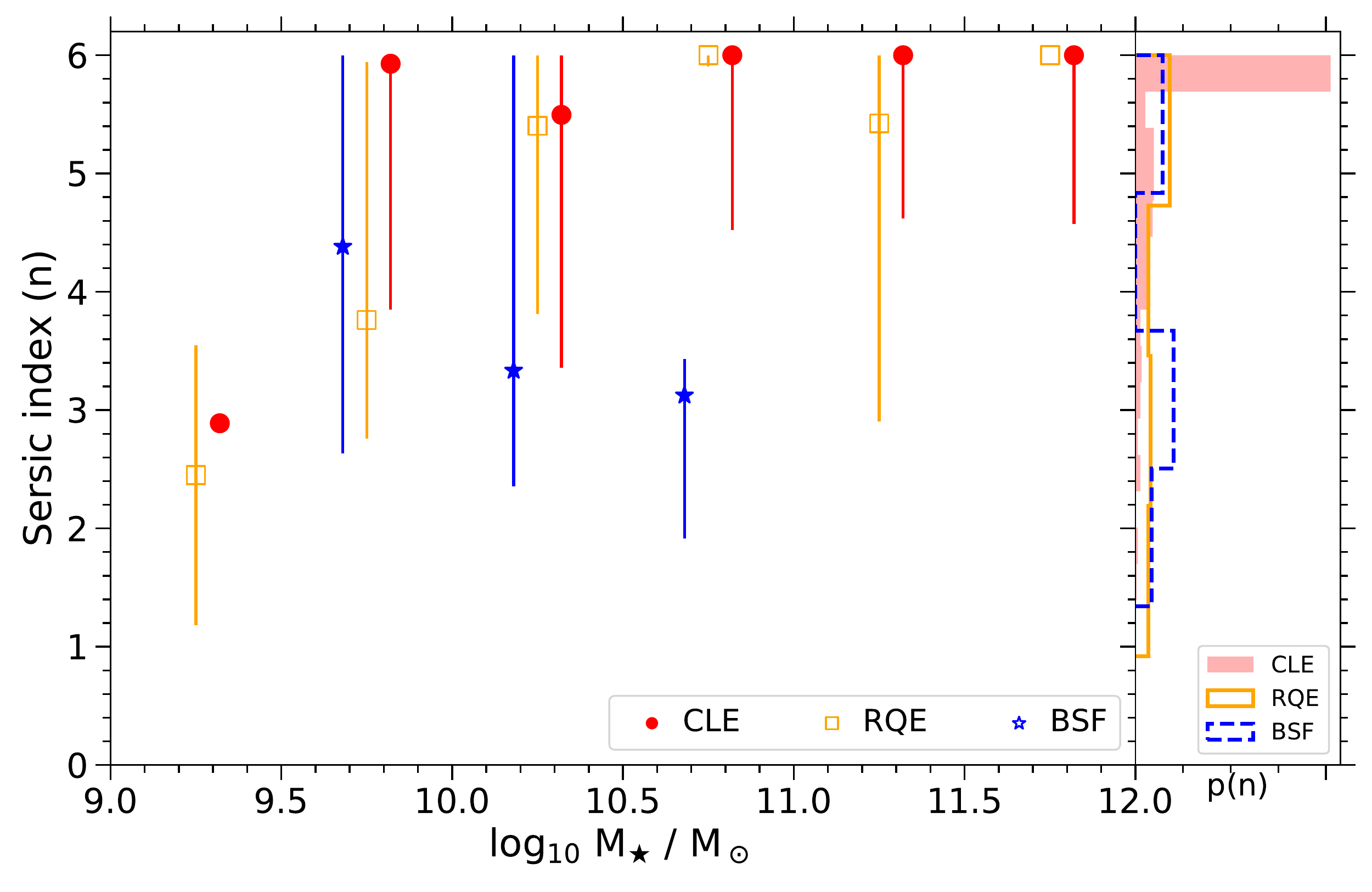}
\includegraphics[width=9.2cm]{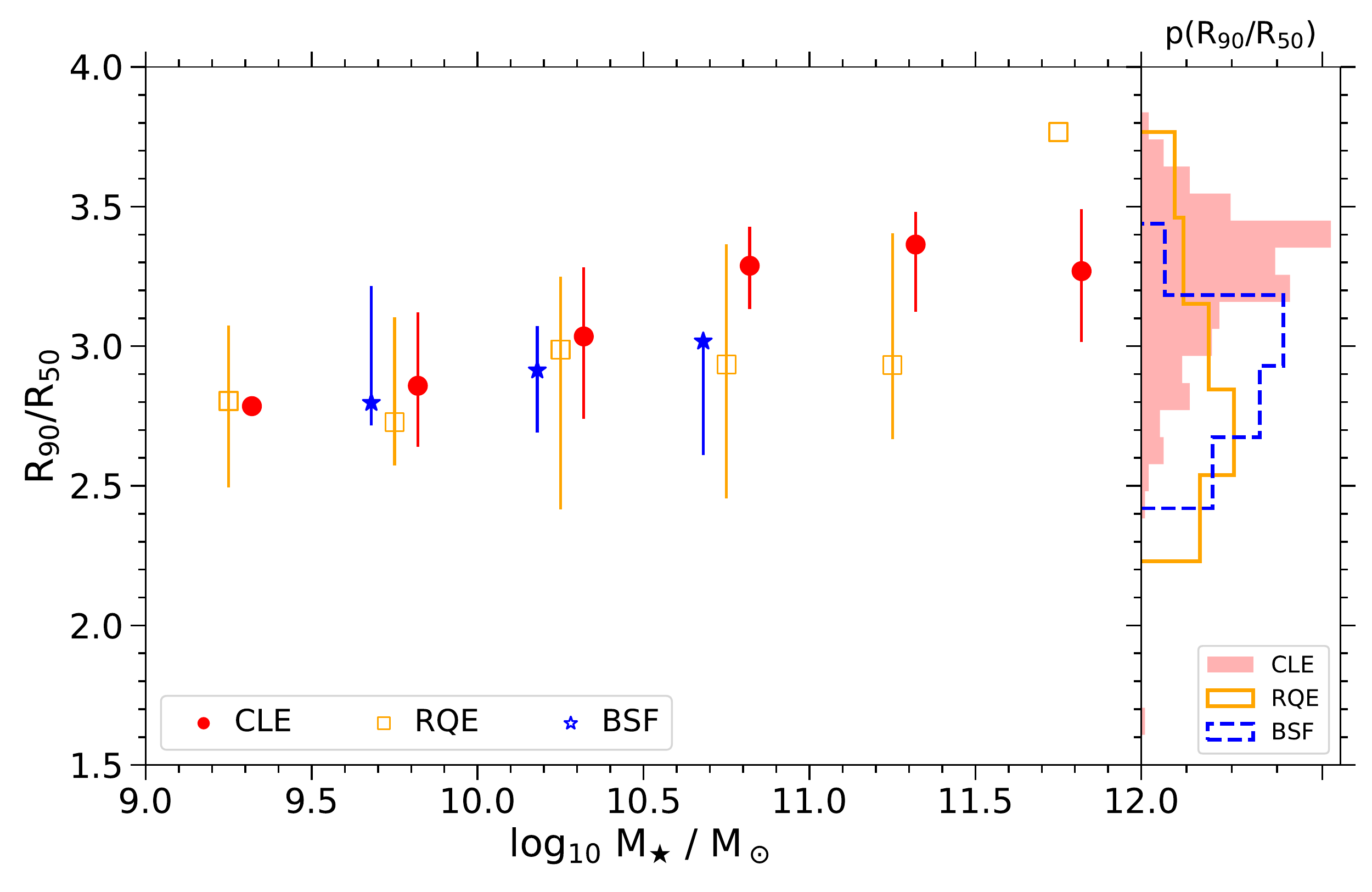}
\caption{Two left panels: Median S\'ersic index as a function of stellar mass and normalized density distributions of the S\'ersic index (n), respectively. Two right panels: Median concentration ($R_{90}/R_{50}$) as a function of stellar mass and normalized density distributions of the concentration.
The symbols of the medians are red filled circles for CLEs, orange open squares for RQEs, and blue stars for BSFs. The error bars correspond to the 16th and 84th percentiles.
The distributions for CLE, RQE, and BSF galaxies are shown in red solid, orange open, and blue dashed histograms, respectively. The integral of each histogram sums to unity.
}
\label{fig_struct}
\end{figure*}

One of our aims is to explore whether the RQE galaxies, as defined here, are a distinct class of Es with respect to the CLEs, or if they are just the expected tails 
in the color or age distribution of the CLE galaxies.  Based on our results, we discuss this in Sect. \ref{SD_RQE}.

Figure \ref{fig_age} shows the integrated (within 1.5 $R_{eff}$) mass- and light-weighted stellar ages. 
The mass-weighted ages of CLEs are old, with a median of 9.5 Gyr, and a narrow distribution; CLEs less massive than $\sim 3\times 10^{10}$ \msun\ are younger the less massive they are. The mass-weighted ages of the RQEs and BSFs are significantly younger and present much broader distributions than the CLEs, with medians of 6.0 and 6.9 Gyr, respectively, and there is not a clear trend with \ms. Regarding the luminosity-weighted ages, the CLEs have a median age$_{lw}$ of 7.5 Gyr. The RQEs are younger than 4 Gyr by definition, with a median age$_{lw}$ of 3.0 Gyr. 
The BSFs present the youngest ages, with a median age$_{lw}$ of 1.0 Gyr, 
which suggests recent episodes of star formation. 
 We note that the luminosity-weighted age is more sensitive to small fractions of recent generations of stars, which significantly contribute  in luminosity but little in mass, while the mass-weighted age represents the average epoch 
 when the bulk of the stars formed. Therefore, the (age$_{mw}-$age$_{lw}$) difference measures how coeval or dispersed in time the SFH of a galaxy was,
 and it has been proposed as an estimator of the star formation timescale \citep{Plauchu-Frayn+2012}.
 If the difference is big, it is because either the SFH is very extended or at least two different populations compose the current stellar populations; the one that is dominant in mass formed early and the other one formed much later. 
 According to the bottom panels of Fig. \ref{fig_age}, most of the CLEs have small (age$_{mw}-$age$_{lw}$) differences (around 2 Gyr on average and $\sim 2.5$ Gyr for the less massive galaxies) 
 with median relative differences of 0.1 dex, 
 suggesting that the star formation in CLEs was early and nearly coeval. This is not the case for the RQE and BSF galaxies. For RQEs, the differences are typically larger than  2 Gyr (median relative differences of 0.3 dex for RQE galaxies more massive than 10$^{9.4}$ \msun) and with a large scatter, especially for the more massive galaxies. The BSFs show the largest differences in (age$_{mw}-$age$_{lw}$), from $\sim$ 3 Gyr to 9 Gyr with median relative differences of 0.7 dex, a clear signal of very recent episodes of star formation.
 Due to the uncertainties
 in the method explained in Sect. \ref{Sect:caveats}, 
 the differences between both ages are lower limits.

Figure \ref{fig_Z} shows the integrated (within 1.5 $R_{eff}$) mass- and luminosity-weighted stellar metallicities. The CLEs are the most metal-rich E galaxies on average, and they present a trend of lower metallicities as \ms\ is smaller. The relative (Z$_{mw}/$Z$_{lw}$) differences are small in these galaxies, around $0.025$ dex on average for $\ms\ga 10^{10.4}$ \msun and around 0 dex for the less massive galaxies, showing that their stellar populations are rather homogeneous not only in age but also in metallicity. 
The trend with mass could be because younger populations in more massive galaxies came from late minor mergers with slightly younger, less metallic galaxies.
This effect with the mass seems to have been more pronounced for the most massive RQEs, which show the
largest (Z$_{mw}/$Z$_{lw}$) positive differences. 
Interestingly enough, for most of the less massive RQEs, Z$_{lw}$ is slightly higher than Z$_{mw}$, suggesting that the young stellar populations formed from slightly more enriched gas. 
This (Z$_{mw}/$Z$_{lw}$) difference is also present for the less massive BSFs. Due to the uncertainties in the method explained in Sect. \ref{Sect:caveats}, the differences between both metallicities are lower limits.

%%%Fig rotators
\begin{figure}
\includegraphics[width=8.8cm]{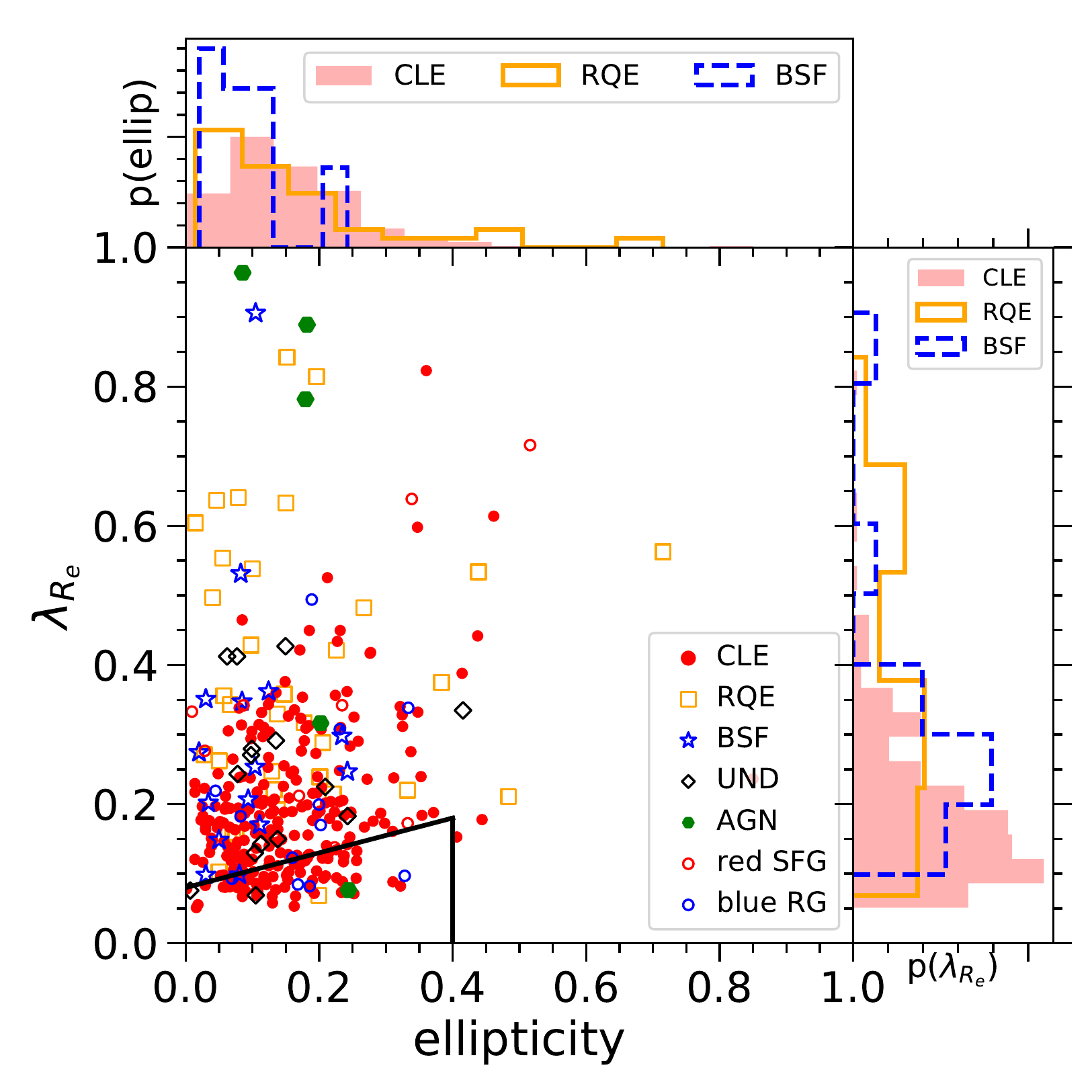}
\caption{Stellar spin parameter as a function of ellipticity. 
The symbols for the galaxies are the same as in Fig. \ref{figBPT}.
The thick black lines enclose the non-regular rotators from \citet{Cappellari2016}.
Top and right panels: normalized density distributions of ellipticity and $\lambda_{R_e}$ for CLE galaxies (red solid histogram), RQE galaxies (orange open histogram), and BSF galaxies (blue dashed histogram),  respectively. The integral of each histogram sums to unity.
}
\label{fig_rot}
\end{figure}

We show two structural properties of the E galaxies in Fig. \ref{fig_struct}: the S\'ersic index ($n$) and the $r$-band concentration ($R_{90}/R_{50}$). The CLEs have a strong peak at the largest value of $n=6$. The RQE galaxies have a wider distribution of the S\'ersic index. Their median values of $n$ increases with stellar mass and correspond to bulge-dominated systems ($n > 2.5$). 
The BSFs also show a wide distribution of the S\'ersic index, but the median values of $n$ decrease with stellar mass. They also exhibit values of bulge-dominated systems. 
On the other hand, the median concentration is roughly independent of stellar mass up to $10^{10.5}$ \msun\ for the three subsamples of Es and tends to increase for larger masses for the CLEs. They are typically more concentrated than BSFs and RQEs with an overall median $R_{90}/R_{50}$ of 3.3 compared with medians of 3.0 and 2.9 for BSFs and RQEs, respectively, except the most massive RQE with $R_{90}/R_{50}$ = 3.8. The median values at all the masses are larger than 2.6, which is a usual limit to define early-type morphologies in the literature (e.g., see  \citealt{Strateva+2001,McIntosh+2014}).

%%%fig ageMW and ageLW gradients (fig8)
\begin{figure*}
\includegraphics[width=18.4cm]{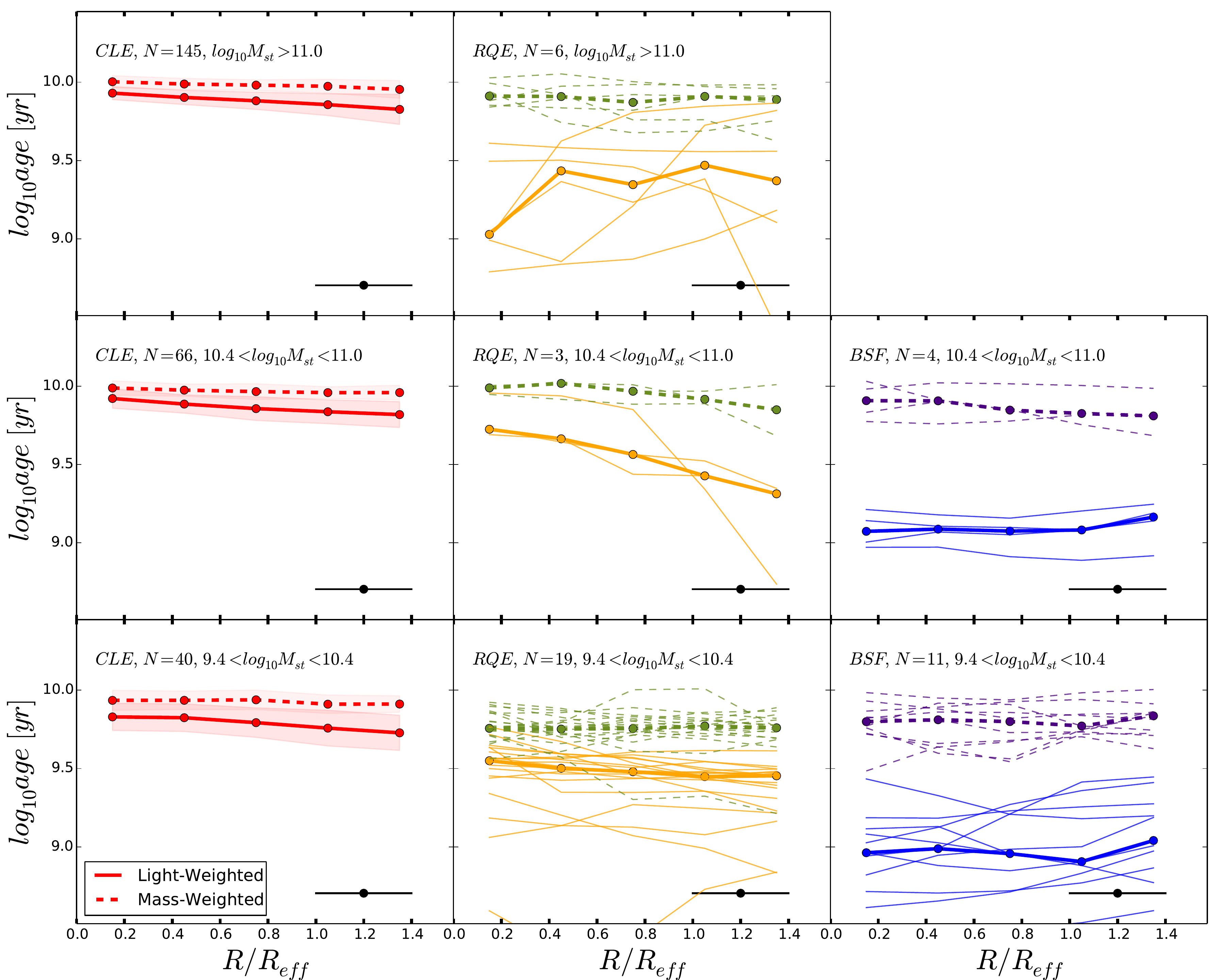}
\caption{Median mass-weighted and luminosity-weighted stellar ages profiles (dots connected with thick dashed and thick solid lines, respectively). 
The left column is for CLEs, the central column is for RQEs, and the right column is for BSFs. Each row is a stellar-mass bin, which increases from the bottom to the top.
The shaded regions in the left panels correspond to the 
16th and 84th percentiles (darker for the luminosity-weighted quantities).
Given the relatively low numbers of RQE and BSF galaxies, we show the individual profiles as thin lines.
The number of galaxies of each type in a given mass bin is shown in each panel. 
The horizontal error bar is the typical PSF size that represents the minimum radial resolution of the IFU.
}
\label{fig_ageMW_ageLW_grad}
\end{figure*}

Figure \ref{fig_rot} shows the $\lambda_{R_e}$$-$$\epsilon$ diagram \citep{Emsellem+2007, Emsellem+2011} for the kinematic classification of our samples of morphological E galaxies. 
See, for example, \cite{Graham+2018}, \cite{Smethurst+2018}, \cite{Bernardi+2019}, and \cite{Tabor+2019}
for information about this kinematic classification for MaNGA galaxies. 
The galaxies enclosed by the black lines correspond to non-regular rotators (i.e., slow rotators),
according to \cite{Cappellari2016}. 
We find that nearly all the Es have low values of $\epsilon$ with a median value of 0.14.
There is a wide distribution of $\lambda_{R_e}$. The peak is lower than $\lambda_{R_e} = 0.2$  for the CLEs, with a median of 0.17.
The RQEs are the subsample of Es that show the highest values, that is, more extended tails, of both $\epsilon$ and $\lambda_{R_e}$, 
with median values of ($\epsilon$,$\lambda_{R_e}$) = (0.14,0.34).
Some signatures of a late morphological transformation may still be present in their kinematics.
In the case of the BSFs, they 
have median values of ($\epsilon$,$\lambda_{R_e}$) = (0.08,0.25).
Almost all the BSFs present a disky central structure (see Appendix \ref{S_Ap_FineStructure}), which may explain the relatively high values of $\lambda_{R_e}$.

%%%fig metallicity gradients (fig9)
\begin{figure*}
\includegraphics[width=18.4cm]{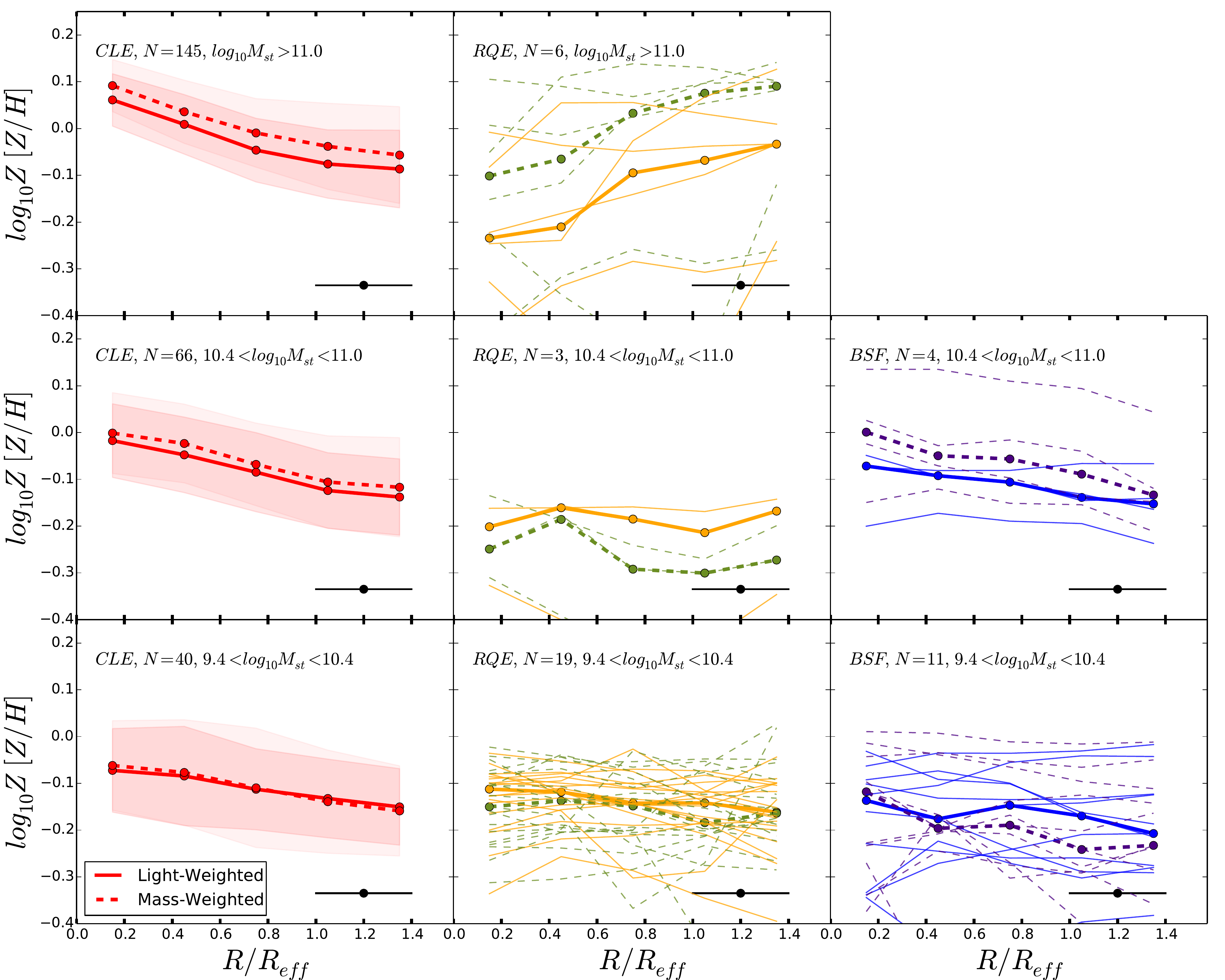}
\caption{Median mass-weighted and luminosity-weighted stellar metallicity profiles (dots connected with thick dashed and thick solid lines, respectively). The left column is for CLEs, the central column is for RQEs, and the right column is for BSFs. Each row is a stellar-mass bin, which increases from the bottom to the top. 
The shaded regions in the left panels correspond to the 16th and 84th percentiles (darker for the luminosity-weighted quantities). Given the relatively low numbers of RQE and BSF galaxies, we show the individual profiles as thin lines.
The number of galaxies of each type in a given mass bin is shown in each panel. 
The horizontal error bar is the typical PSF size that represents the minimum radial resolution of the IFU.
}
\label{fig_Z_grad}
\end{figure*}

%%%%%%%%%%%%%%%%%%%%%%%%%%%%%%%
\section{Radial profiles}
\label{S4}
%%%%%%%%%%%%%%%%%%%%%%%%%%%%%%%

This section presents information related to radial trends of the age, stellar metallicity, and sSFR obtained from the IFU spatially-resolved observations of the E galaxies defined as CLEs, RQEs, and BSFs. We first present the ``stacked'' radial profiles in three mass bins (Sect. \ref{Median-profiles}). 
 In Sect. \ref{S_grad} we present the scatter plots of the individual age and metallicity gradients for each galaxy as functions of the stellar mass. In these plots, rather than average radial trends of the population, we see the gradient that each galaxy has and can then evaluate the fraction of galaxies that do or do not follow the average (stacked) trends.

\subsection{The stacked radial profiles}
\label{Median-profiles}

\begin{figure*}
\includegraphics[width=18.4cm]{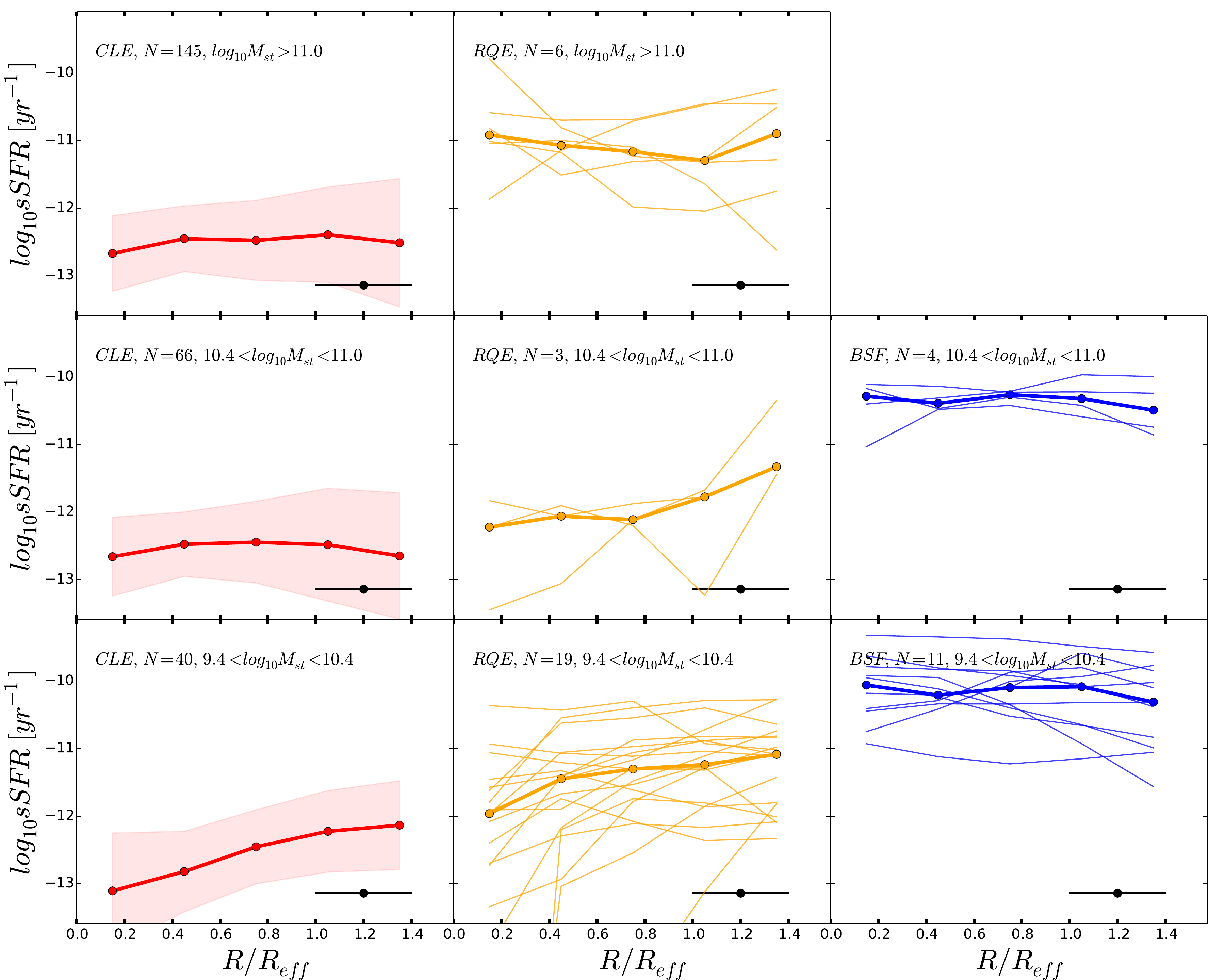}
\caption{Median sSFR profiles in the last 30 Myr obtained from the SFHs (dots connected with thick solid lines). 
The sSFR values below $\sim 10^{-12}$ yr$^{-1}$ are very uncertain and can be considered as upper limits. 
The left column is for CLEs, the central column is for RQEs, and the right column is for BSFs. Each row is a stellar-mass bin, which increases from the bottom to the top.
The shaded regions in the left panels correspond to the 16th and 84th percentiles.
Given the relatively low numbers of RQE and BSF galaxies, we show the individual profiles as thin lines.
The number of galaxies of each type in a given mass bin is shown in each panel.
The horizontal error bar is the typical PSF, which represents the minimum radial resolution of the IFU.
}
\label{fig_sSFR_grad}
\end{figure*}

To generate the ``stacked'' radial profiles, we first calculated the individual radial profiles for the given quantity in the same way as described in \cite{Ibarra-Medel+2019}. We azimuthally averaged the 2D maps of each galaxy at five radial bins in terms of \re\footnote{When calculating the azimuthally-averaged radial profiles, we only used five radial bins in units of \re{} of the given galaxy because the typical PSF size of MaNGA observations does not allow for better resolving galaxies (see the horizontal error bars in panels of Figs. \ref{fig_ageMW_ageLW_grad}--\ref{fig_sSFR_grad}). In any case, we also calculated the radial profiles using twelve radial bins, and we find that both the individual and ``stacked'' profiles are, in most  cases, very similar to those using five radial bins (see Appendix \ref{S_Ap_12bins}).}. 
We estimated the best linear fit of the stacked profiles by randomly resampling them each one thousand times. The distribution of the resampling profiles follows the standard deviation and mean values of each stellar mass bin and radial bin. Then, we performed a linear fit for each resampled profile following the 2D version of the algorithm of \citet{Sheth+2012}. This method allowed us to obtain a distribution in the slopes for each “stacked” profile. The final gradient is the value that has the 50th percentile of this distribution (see Table \ref{tab1}). The reported errors of the gradients in the table are the standard deviations of their respective distributions. The gradients are in units of dex/\re{} and are defined as:
\begin{equation}
\nabla \log(Y) = d\log(Y)/dR ,
\label{gradient}
\end{equation}
where $Y$ is either age$_{mw}$, age$_{lw}$, $Z_{mw}$, or $Z_{lw}$, and $R$ is the  
galacto-centric radius in units of \re. 

The ages and stellar metallicities per spaxel were calculated by weighting the respective SSP distributions by stellar mass or luminosity (e.g., Eqs. \ref{eq_ageMW} and \ref{eq_ageLW}). We then used the weighted azimuthally-averaged profiles (using the stellar mass or flux maps) to obtain a profile that is not biased by local outliers (like very young regions) in a  given radial region. For the sSFR radial profiles, we used the arithmetic azimuthal average for each radial bin. We calculated the standard deviation of each value per radial bin (we used this information to estimate the individual gradients below in Sect. \ref{S_grad}). Figs. \ref{fig_ageMW_ageLW_grad}--\ref{fig_sSFR_grad} show the “stacked” radial profiles with the median values and the 16th and 84th percentiles of our quantities for the five radial bins in three stellar mass bins.

Figure \ref{fig_ageMW_ageLW_grad} shows the stacked (median) age$_{lw}$ and age$_{mw}$ radial profiles for CLEs, RQEs, and BSFs. 
It also shows the individual profiles of the RQE and BSF galaxies.
We consider a lower mass limit of \ms = 10$^{9.4}$ \msun\ in the lowest mass bin because there are only RQE galaxies at lower masses.
By fitting a straight line to the stacked age profiles, the gradients in units of dex/\re{} (see Eq. \ref{gradient}) can be calculated at each mass bin. 
Table \ref{tab1} shows these gradients for the CLE, RQE, and BSF galaxies for each mass bin.
The median age$_{lw}$ is slightly older in the central part of the CLE population compared to their outskirts for all the masses. A similar behavior is seen in the case of RQEs with $\ms < 10^{11}$ \msun, although with an overall younger age$_{lw}$ than CLEs of the same mass (median age$_{lw}$ between 10$^{9.3}$ and 10$^{9.7}$ yr 
for the former and $>$ 10$^{9.7}$ yr for the latter), which is expected by the selection criteria of the RQE galaxies. 
The massive RQEs show a different median trend in which the central part is younger than the outskirts. The BSF population also shows a trend of slightly increasing age$_{lw}$ with radius, 
although the profile is flatter on average 
at the $10^{10.4}$ < \ms/\msun < $10^{11}$ mass bin. The age$_{lw}$ values of BSFs are significantly younger than those of CLEs with the same mass at all the radii.
Finally, the stacked age$_{mw}$ profiles are flatter than the age$_{lw}$ profiles, especially for the RQEs (see Table \ref{tab1}). The radial age$_{mw}$ profiles of both RQEs and BSFs do not differ among them in a significant way.

In summary, the age$_{lw}$ and age$_{mw}$ stacked (median) profiles of the CLE galaxies show a mild decrease with radius at all mass bins, that is to say the central parts tend to be slightly older than the external parts, which suggests a weak inside-out formation scenario for the CLE population. The age$_{lw}$ radial profiles decrease slightly stronger than the age$_{mw}$ ones, 
with average differences that are not larger than 0.2 dex.
The age$_{lw}$ and age$_{mw}$ 
stacked profiles are more diverse for the RQEs and BSFs. 
In particular, the stacked age$_{lw}$ profile of RQEs that are more massive than 10$^{11}$ \msun{} 
tends to show a positive gradient, which could suggest an outside-in growth mode, although with large individual uncertainties.
The big differences among the mass- and luminosity-weighted age profiles, 
which are typically larger than 0.2 dex (especially for the BSFs),
show the presence of a relatively high fraction of young stellar populations in the BSF and RQE galaxies.

Figure \ref{fig_Z_grad} shows the stacked (median) luminosity- and mass-weighted stellar metallicity (Z$_{lw}$ and Z$_{mw}$, respectively) profiles.  The metallicity of CLEs is, on average, higher for more massive galaxies, and it decreases with radius and is steeper the more massive the galaxies are. 
We note that the most massive CLEs tend to show some flattening in the Z$_{lw}$ and Z$_{mw}$ profiles at $R$ $\gtrsim$ 1.0 \re.
By fitting a straight line to the stacked profiles out to 1.5 \re, the gradients in units of dex/\re{} (see Eq. \ref{gradient}) can be calculated at each mass bin. Table \ref{tab1} shows these gradients for both the luminosity- and mass-weighted stellar metallicities. 
Compared to the CLEs, the metallicity profiles are flatter for the RQEs, and the most massive RQE galaxies
show even a positive gradient in both luminosity- and mass-weighted metallicities, although
with large individual uncertainties. 
The BSF population tends to have a negative gradient, with more pronounced mass-weighted stacked profiles than the luminosity-weighted ones at the two mass bins.

%%%fig sSFR gradients (fig10)

Finally, Fig. \ref{fig_sSFR_grad} shows the stacked (median) profiles 
of sSFR($r$)=$\Sigma_{\rm SFR}(r)/\Sigma_{*}(r)$, where the SFR was calculated in the last 30 Myr from the stellar population synthesis results.
The sSFR shows how efficient the current SFR is with respect to the average in the past. It provides a criterion to evaluate whether a galaxy (or a galaxy region) is actively forming stars, if it is quenching, or, alternatively, if it has  already quenched. For example, it is common in the literature to define a galaxy as quenched or retired when its sSFR is lower than a critical value defined in between $10^{-11}$ and $10^{-10.7}$ yr$^{-1}$. 

We find that the stacked sSFR profiles of low-mass CLE and RQE galaxies increase with radius, with the central part having values lower than the outer part by $\sim$1 dex. This result suggests a more effective quenching process inside rather than outside \citep{GonzalezDelgado+2017,Belfiore+2018,Ellison+2018,Sanchez+2018_AGN}. For the higher mass bins,
the median sSFR profile of the CLE population is roughly flat, though with a weak decrease toward the inner regions, suggesting a slightly more efficient quenching in these regions. For the massive RQE population, the sSFR profiles of most of the objects tend to be negative, suggesting an outside-in quenching. 
The BSFs exhibit higher values of sSFR at all the radii than CLEs and RQEs. In contrast to low- and intermediate-mass CLEs and RQEs, the sSFR profiles of BSF Es tend to be flat or slightly decrease with radius. 
We note that the uncertainty in the constrained sSFR values below 10$^{-12}$ yr$^{-1}$ is large, so  these results for CLE galaxies should be considered with caution.

\begin{table*}
\begin{center}
\caption{Gradients of the ``stacked'' profiles out to 1.5 \re{} in three stellar mass bins.
}\label{tab1}
\resizebox{\textwidth}{!}{
\begin{tabular*}{0.97\textwidth}{@{\extracolsep{\fill}} c c c c c }
\hline\hline
type  &  $\nabla\log$  & 9.4 < $\log$(\ms/\msun) < 10.4 & 10.4 $\le$ $\log$(\ms/\msun) < 11 & 11 $\le$ $\log$(\ms/\msun) < 12 \\ 
%  & $[dex/R_e]$ & $[dex/R_e]$ & $[dex/R_e]$ \\
\hline
CLE & age$_{lw}$ & $-0.10 \pm 0.06$ & $-0.08 \pm 0.05$ & $-0.6 \pm 0.05$ \\
CLE & age$_{mw}$ & $-0.01 \pm 0.04$ & $-0.02 \pm 0.03$ & $-0.03 \pm 0.03$ \\
CLE & Z$_{lw}$ & $-0.08 \pm 0.05$ & $-0.12 \pm 0.05$ & $-0.11 \pm 0.04$ \\
CLE & Z$_{mw}$ & $-0.11 \pm 0.06$ & $-0.11 \pm 0.06$ & $-0.13 \pm 0.05$ \\
\hline
RQE & age$_{lw}$ & $-0.09 \pm 0.13$ & $-0.13 \pm 0.02$ & $ 0.26 \pm 0.19$ \\
RQE & age$_{mw}$ & $-0.01 \pm 0.05$ & $-0.07 \pm 0.02$ & $-0.00 \pm 0.06$ \\
RQE & Z$_{lw}$ & $-0.03 \pm 0.03$ & $ 0.04 \pm 0.05$ & $ 0.19 \pm 0.10$ \\
RQE & Z$_{mw}$ & $-0.01 \pm 0.06$ & $-0.08 \pm 0.06$ & $ 0.18 \pm 0.13$ \\
\hline
BSF & age$_{lw}$ & $ 0.10 \pm 0.16$ & $ 0.00 \pm 0.05$ & \\
BSF & age$_{mw}$ & $-0.05 \pm 0.06$ & $-0.07 \pm 0.06$ & \\
BSF & Z$_{lw}$ & $-0.02 \pm 0.07$ & $-0.06 \pm 0.03$ & \\
BSF & Z$_{mw}$ & $-0.07 \pm 0.08$ & $-0.14 \pm 0.05$ & \\
\hline
\end{tabular*}}
\end{center}
\end{table*}

%%%%%%%%%%%%%
\subsection{Individual stellar age and stellar metallicity gradients}
\label{S_grad}
%%%%%%%%%%%%%

%%%fig age gradients
\begin{figure}
\includegraphics[width=9.cm]{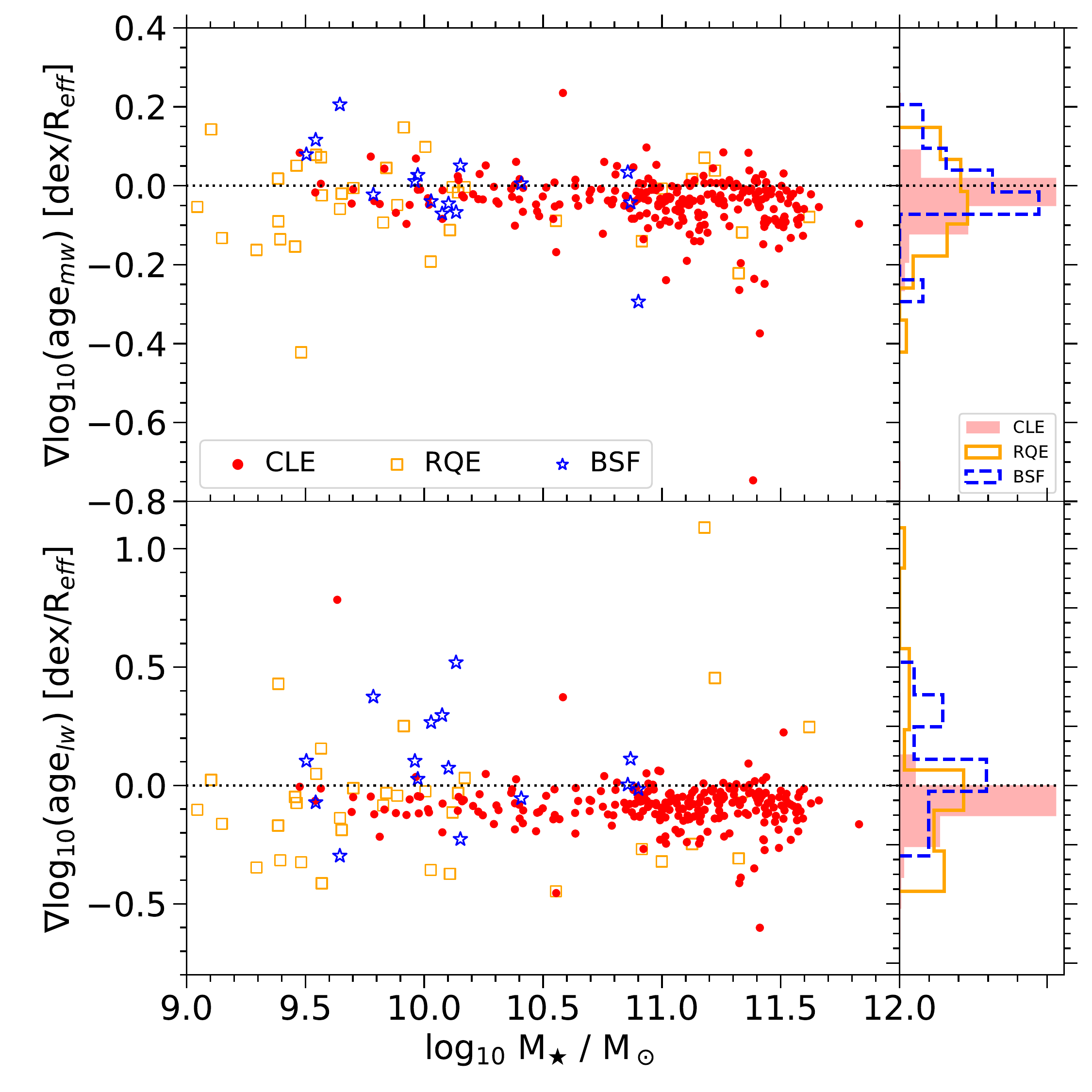}
\caption{Mass-weighted stellar age gradients (top-left panel) and luminosity-weighted stellar age gradients (bottom-left panel) as functions of the stellar mass of E galaxies.
The symbols are red filled circles for CLEs, orange open squares for RQEs, and blue stars for BSFs. 
Top-right and bottom-right panels: normalized density distributions of mass-weighted age gradients and luminosity-weighted age gradients, respectively, for CLE 
galaxies (red solid histogram), RQE galaxies (orange open histogram), and BSF galaxies (blue dashed histogram).
The integral of each histogram sums to unity. 
The dotted lines in the panels are gradients equal 0.}
\label{fig_gradAge_indiv}
\end{figure} 

%%%fig Z gradients
\begin{figure}
\includegraphics[width=9.cm]{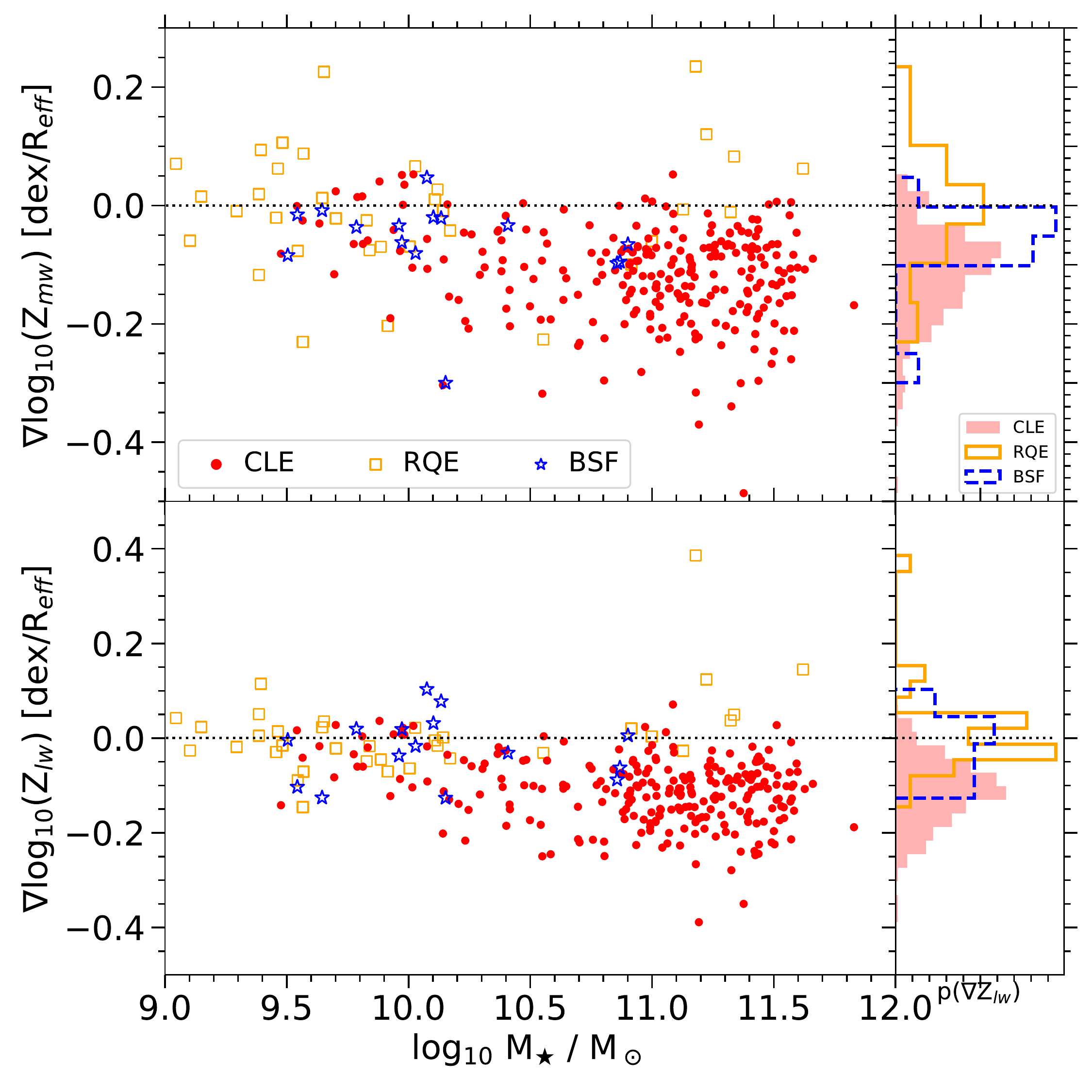}
\caption{Mass-weighted stellar metallicity gradients (top-left panel) and luminosity-weighted stellar metallicity gradients (bottom-left panel) as functions of the stellar mass of E galaxies.
The symbols are red filled circles for CLEs, orange open squares for RQEs, and blue stars for BSFs. 
Top-right and bottom-right panels: normalized density distributions of mass-weighted metallicity gradients and luminosity-weighted metallicity gradients, respectively, for CLE 
galaxies (red solid histogram), RQE galaxies (orange open histogram), and BSF galaxies (blue dashed histogram).
The integral of each histogram sums to unity. 
The dotted lines in the panels are gradients equal 0.
}
\label{fig_gradZ_indiv}
\end{figure}

We use the same methodology described in Sect. \ref{Median-profiles} but for the individual gradient values.
In Figs. \ref{fig_gradAge_indiv} and \ref{fig_gradZ_indiv}, we show the individual gradients in stellar age and metallicity as functions of stellar mass. 
Table \ref{tablaMedianGrad} presents the median values of the gradients and the 16th-84th percentiles for different mass bins and all the masses.

The CLE galaxies mostly show very mild negative or flat gradients in age$_{mw}$, whereas their gradients in age$_{lw}$ are more negative. Only 19\% and 8\% of the CLE sample have positive gradients in age$_{mw}$ and age$_{lw}$, respectively, confirming that the ``stacked'' age profiles in the mass bins shown in Fig. \ref{fig_ageMW_ageLW_grad} capture the general behavior of the whole population well. Also, in agreement with the ``stacked'' profiles, the individual metallicity gradients are more negative than the age gradients, and their scatter is also larger. For $Z_{mw}$ and $Z_{lw}$, only 6\% and 5\% of the CLE sample show positive gradients, respectively. 
There is also a clear trend of steeper gradients as the galaxies are more massive, both for $Z_{lw}$ and $Z_{mw}$. We discuss the implications of these results in Sect. \ref{SD_CLE}.

The RQE galaxies show evident larger scatters in their gradients compared to the CLEs. 
The medians of their mass- and light-weighted gradients tend to be mildly negative except for the most massive bin (\ms > 10$^{11}$ \msun), see Table \ref{tablaMedianGrad}. The gradients of age$_{lw}$, $Z_{lw}$, and $Z_{mw}$ tend to be positive for massive RQEs.
In the case of BSF galaxies, the scatter in age gradients is larger compared with that for CLEs, but the scatter in metallicity gradients is smaller or comparable between BSFs and CLEs. The median of the gradients in age and metallicity are slightly more positive than those of the CLEs of the same mass.
We discuss the implications of these results for RQEs and BSFs in Sects. \ref{SD_RQE} and \ref{SD_BSF}, respectively.

%%% TABLA individual gradients
\begin{table*}
\caption{Median mass- and luminosity-weighted gradients of stellar age and stellar metallicity out to 1.5 \re{} in units of dex/\re\ for different mass bins from Figs. \ref{fig_gradAge_indiv} and \ref{fig_gradZ_indiv}. 
The last column is for the full mass range. The errors correspond to the 16th and 84th percentile
range.
}
\centering
\begin{tabular}{c c c c c c c}
\hline
\hline
type  &  $\nabla\log$  & 9.4 < $\log$(\ms/\msun) < 10.4 & 10.4 $\le$ $\log$(\ms/\msun) < 11 & 11 $\le$ $\log$(\ms/\msun) < 12 & 9 < $\log$(\ms/\msun) < 12 \\
\hline
\rule{0pt}{2.5ex}
 CLE & age$_{mw}$  & $-0.02_{-0.03}^{+0.06}$ & $-0.03_{-0.04}^{+0.04}$ & $-0.04_{-0.06}^{+0.04}$ & $-0.03_{-0.05}^{+0.04}$\\ \rule{0pt}{2.5ex}
 CLE & age$_{lw}$  & $-0.08_{-0.05}^{+0.06}$ & $-0.08_{-0.05}^{+0.07}$ & $-0.08_{-0.10}^{+0.05}$ & $-0.08_{-0.07}^{+0.06}$\\ \rule{0pt}{2.5ex}
 CLE & $Z_{mw}$  & $-0.06_{-0.06}^{+0.07}$ & $-0.11_{-0.08}^{+0.05}$ & $-0.12_{-0.08}^{+0.06}$ & $-0.11_{-0.08}^{+0.07}$\\ \rule{0pt}{2.5ex}
 CLE & $Z_{lw}$  & $-0.05_{-0.07}^{+0.06}$ & $-0.11_{-0.08}^{+0.05}$ & $-0.12_{-0.07}^{+0.06}$ & $-0.11_{-0.07}^{+0.06}$\\
\hline
\rule{0pt}{2.5ex}
 RQE & age$_{mw}$  & $-0.02_{-0.10}^{+0.09}$ & $-0.09_{-0.04}^{+0.06}$ & $-0.03_{-0.11}^{+0.08}$ & $-0.02_{-0.11}^{+0.09}$\\ \rule{0pt}{2.5ex}
 RQE & age$_{lw}$  & $-0.05_{-0.28}^{+0.08}$ & $-0.32_{-0.09}^{+0.04}$ & 0.11$_{-0.37}^{+0.47}$ & $-0.08_{-0.24}^{+0.19}$\\ \rule{0pt}{2.5ex}
 RQE & $Z_{mw}$  & $-0.02_{-0.05}^{+0.09}$ & $-0.10_{-0.09}^{+0.03}$ & 0.07$_{-0.08}^{+0.07}$ & $-0.01_{-0.07}^{+0.09}$\\ \rule{0pt}{2.5ex}
 RQE & $Z_{lw}$  & $-0.02_{-0.05}^{+0.04}$ & 0.00$_{-0.02}^{+0.01}$ & 0.09$_{-0.06}^{+0.11}$ & 0.00$_{-0.04}^{+0.05}$\\ 
   \hline
 \rule{0pt}{2.5ex}
  BSF & age$_{mw}$  & 0.01$_{-0.07}^{+0.08}$ & $-0.02_{-0.15}^{+0.04}$ &  & 0.01$_{-0.07}^{+0.07}$\\ \rule{0pt}{2.5ex}
 BSF & age$_{lw}$  & 0.10$_{-0.24}^{+0.22}$ & $-0.01_{-0.03}^{+0.07}$ &  & 0.07$_{-0.14}^{+0.22}$\\ \rule{0pt}{2.5ex}
 BSF & $Z_{mw}$  & $-0.03_{-0.05}^{+0.02}$ & $-0.08_{-0.02}^{+0.03}$ &  & $-0.04_{-0.06}^{+0.02}$\\ \rule{0pt}{2.5ex}
 BSF & $Z_{lw}$  & $-0.00_{-0.11}^{+0.05}$ & $-0.05_{-0.03}^{+0.03}$ & 
 & $-0.02_{-0.08}^{+0.05}$\\
\hline
\end{tabular}
\label{tablaMedianGrad}
\end{table*}

%%%fig integrated MGH
\begin{figure*}
\includegraphics[width=18.4cm]{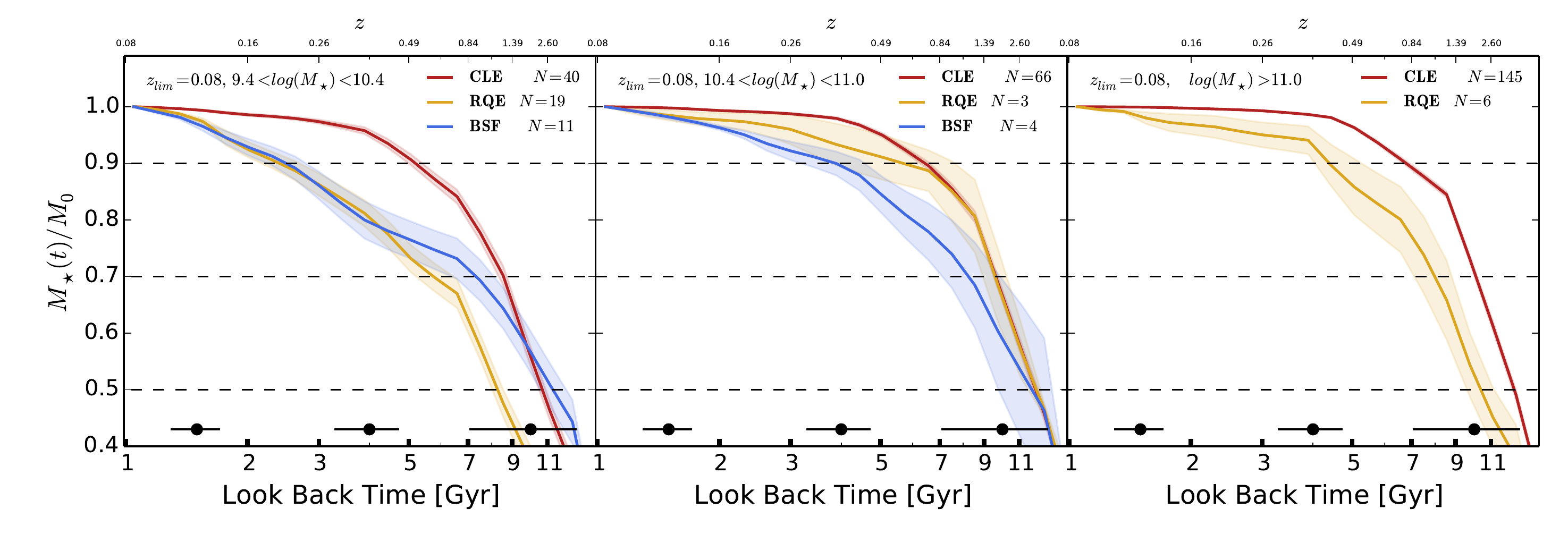}
\caption{Mean cumulative mass temporal distributions, CMTDs, within 1.5 \re{} in three stellar mass bins (mass increases to the right). The CMTDs are normalized to the mass attained at $z_{lim}=0.08$.
The red lines correspond to CLEs, the orange lines correspond to RQEs, and the blue lines correspond to BSFs.
The shaded region is 
the error of the mean (standard deviation divided by $\sqrt N$)  
of the respective distribution.
The number of galaxies of each type in a given mass bin is shown in each panel. The horizontal error bars show the average time resolutions of the SSP template stellar library. 
}
\label{fig_intMGH}
\end{figure*}

We also explore the individual gradients of age$_{mw}$, age$_{lw}$, $Z_{mw}$, and $Z_{lw}$ as functions of the integrated $g-i$ color, integrated mass-weighted and luminosity-weighted stellar ages, integrated mass-weighted and luminosity-weighted stellar metallicities, and $\lambda_{R_e}$. 
For the age$_{mw}$ and age$_{lw}$ gradients of CLEs, RQEs, and BSFs, we do not find any clear trend with the color, integrated galaxy ages and metallicities, and $\lambda_{R_e}$. We find mild dependences of the $Z_{mw}$ and $Z_{lw}$ gradients on the integrated luminosity-weighted age (more negative for older galaxies), the $Z_{lw}$ gradients 
on both integrated metallicities (more negative for richer galaxies), and the $g-i$ color (more negative for redder galaxies). Since we find a dependence of the metallicity gradients on the stellar mass, these weak dependences of the metallicity gradients on the integrated stellar ages, metallicities, and color can be understood from the correlation between these physical parameters and the stellar mass (Figs. \ref{fig_gi_Ms_0.08}--\ref{fig_Z}).

%%%%%%%%%%%%%%%%%%%%%%%%%%%%%%%%%
\section{Stellar population age distributions}
\label{S5}
%%%%%%%%%%%%%%%%%%%%%%%%%%%%%%%%%

%%%fig MGH_radial2
\begin{figure*}
\includegraphics[width=18.2cm]{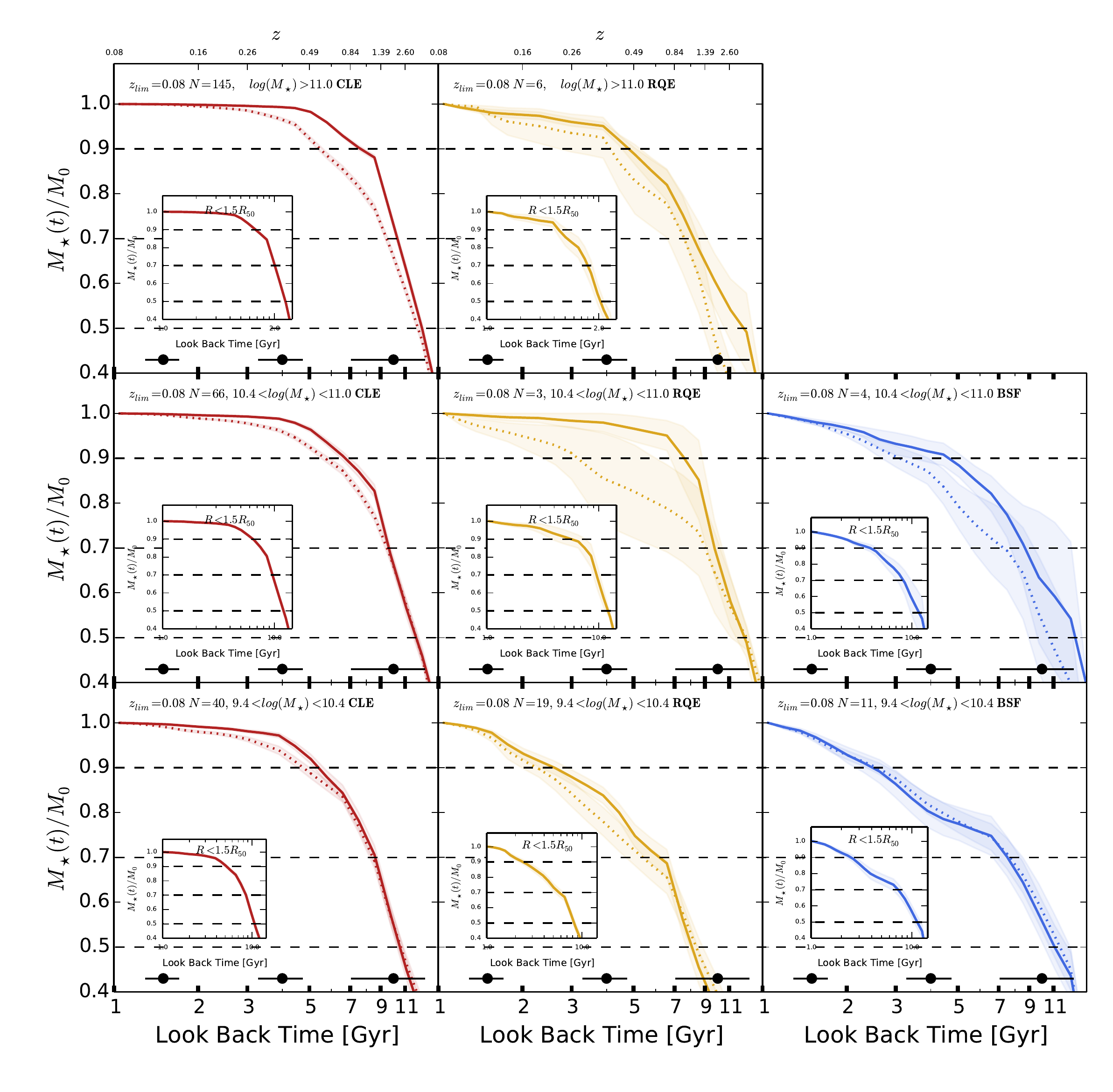}
\caption{Mean cumulative mass temporal distributions, CMTDs, for each E type in three mass bins.  
Each row is a stellar-mass bin, which increases from the bottom to the top.
The CMTDs are normalized to the respective masses at $z_{lim}=0.08$. The left column is for CLEs, the central column is for RQEs, and the right column is for BSFs.  In each panel, the solid line corresponds to the mean CMTD within 0.5 \re,
whereas the dotted line corresponds to the mean CMTD between 1 and 1.5 \re. 
The light and dark shaded regions are the errors of the mean (standard deviation divided by $\sqrt N$) of the distributions in the inner part and the outer part of the galaxies, respectively. The horizontal error bars show the average time resolution of the SSP template stellar library at the given epoch. 
The inset box shows the integrated mean normalized CMTDs within 1.5 \re.
}
\label{fig_MGH_radial2}
\end{figure*} 

The fossil record method applied to the IFU data allows us to obtain not only mean spatially-resolved stellar population ages but the whole age distribution of the fitted SSPs integrated and within regions of the galaxies. With this information, it is possible to construct several temporal quantities, integrated and spatially resolved, that allow us to explore the evolution of galaxies.  
Examples of these quantities are
the differential and cumulative SFHs, SFR$(t)$, and SFR$(<t)= \int_{0}^t$ SFR$(t')dt'$, respectively; 
the birthrate parameter \citep{Scalo1986,BM1998},
\begin{equation}
    b(t)\equiv\frac{\mathrm{SFR}(t)}{\langle \mathrm{SFR}(t)\rangle} =  \frac{\mathrm{SFR}(t)}{\frac{1}{t} \int_{0}^t \mathrm{SFR}(t')dt'} %$\ms(<t)=\int_{t_i}^t SFR(t')[1-R_{\rm ml}(t')]dt'$
\end{equation}
 and the cumulative mass temporal distribution, CMTD, 
 \begin{equation}
  \ms(<t)=\int_{t_i}^t \mathrm{SFR}(t')[1-R_{\rm ml}(t')]dt',
 \end{equation}
 where $R_{\rm ml}$ is the stellar mass-loss term. The CMTD is closely related to the cumulative SFH; the difference is that the stellar mass accumulated at each temporal step in the case of the CMTD takes into account the mass lost by stars as they evolve.

\subsection{Cumulative mass temporal distributions}
  
Figure \ref{fig_intMGH} shows the integrated (within 1.5 \re{}) mean CMTDs normalized to $M_0$ for the CLE, RQE, and BSF galaxies in three stellar mass bins. 
The normalized CMTDs are constructed in the same way as reported in \citet[][in that paper they were called normalized mass growth histories]{Ibarra-Medel+2016}, where $M_0$ is the stellar mass attained at the same cosmological epoch for each galaxy.  We define this epoch as the highest redshift of our sample, $z_{lim}=0.08$ (look back time $\sim 1$ Gyr; in this way, we assure that all studied galaxies have their normalized CMTD determined at this %initial 
time), that is, $M_0$ is the stellar mass at $z=0.08$. This mass is obtained by interpolating the CMTD around the time corresponding to $z=0.08$. The normalized CMTDs can be interpreted as the speed at which the stellar mass grows at different epochs \citep{Ibarra-Medel+2016}. 

For all masses, the CLEs accumulated higher fractions of their final stellar masses (defined at $z=0.08$) at earlier epochs on average, 
with higher mass growth rates in the past 
compared to RQEs and BSFs. At lower masses ($\ms < 10^{10.4}$ \msun), the CLEs 
accumulated  90\% of their final mass $\sim$ 5 Gyr ago, whereas it is between 2 and 3 Gyr ago for RQEs and BSFs. For masses $\ms > 10^{10.4}$ \msun, the CLEs accumulated  90\% of their mass $\sim$ 7 Gyr ago, whereas it is between 4 and 6 Gyr ago for BSFs and RQEs.
The mass growth of the CLEs tends to be earlier, with higher mass growth rates in the past, as the more massive they are (downsizing). In contrast, the RQEs show 
a changing trend with mass. They accumulate, on average, 90\% of their mass at later epochs 
($\sim$ 3 Gyr ago) at the lowest mass bin, later ($\sim$ 6 Gyr ago) at intermediate masses (10$^{10.4} < \ms < 10^{11}$ \msun), and between 4 and 5 Gyr ago in the most massive bin. We note that the trends for RQEs with \ms > 10$^{10.4}$ \msun\ have large scatters. 
However, in general, the RQEs have a systematic delay in their normalized CMTDs compared with the CLEs, as is seen in Fig. \ref{fig_intMGH}. 
The BSFs also present a clear delay in their CMTDs with respect to the CLEs. These results are consistent with the younger mean ages for the RQEs and BSFs with respect to the CLEs reported in Sect. \ref{S3}.

Figure \ref{fig_MGH_radial2} shows the mean normalized CMTDs for CLEs, RQEs, and BSFs in two radial bins; the inner part of the galaxies ($R$ $<$ 0.5 \re) and the external part of the galaxies (1.0 $<R/$\re\ $< 1.5$).  In this case, $M_0$ is the mass attained within each one of these two radial bins  at $z=0.08$. The results are shown for the same three mass bins as in the previous figures. The CLEs typically present a weak inside-out growth mode;  their inner stellar mass growth rates were slightly higher in the past than the outer ones, and as a result, the inner part accumulated 90\% of its mass at earlier times than the outer part. 
This trend is more evident for the more massive CLEs. We notice, however, that the inside-out growth mode is much more pronounced for late-type galaxies \citep[see, e.g.,][]{GonzalezDelgado+2015,Ibarra-Medel+2016,Garcia-Benito+2017,Avila-Reese+2018,LopezFernandez+2018}. The RQEs show less clear and more scattered trends with mass. The inside-out growth mode seems to occur in RQE galaxies, but it is much weaker and noisier for the most massive ones, $\ms > 10^{11}$ \msun. This different behavior in the CMTDs for massive RQEs is in agreement with the ``stacked'' age and metallicity radial profiles presented in Sect. \ref{S4} for the RQEs in the same mass bin. Regarding the BSFs, their average inner and outer normalized  CMTDs
are statistically similar, suggesting either a radial homogeneous stellar mass assembly or strong radial mixing processes. The more massive BSFs show more explicit evidence of an inside-out mode. However, given the small numbers of RQEs and BSFs, the mean trends shown in Fig. \ref{fig_MGH_radial2} should be taken with caution since they can be dominated by the extreme behavior of a few galaxies (see below).  

%%%Fig delta T
\begin{figure}
\includegraphics[width=8.8cm]{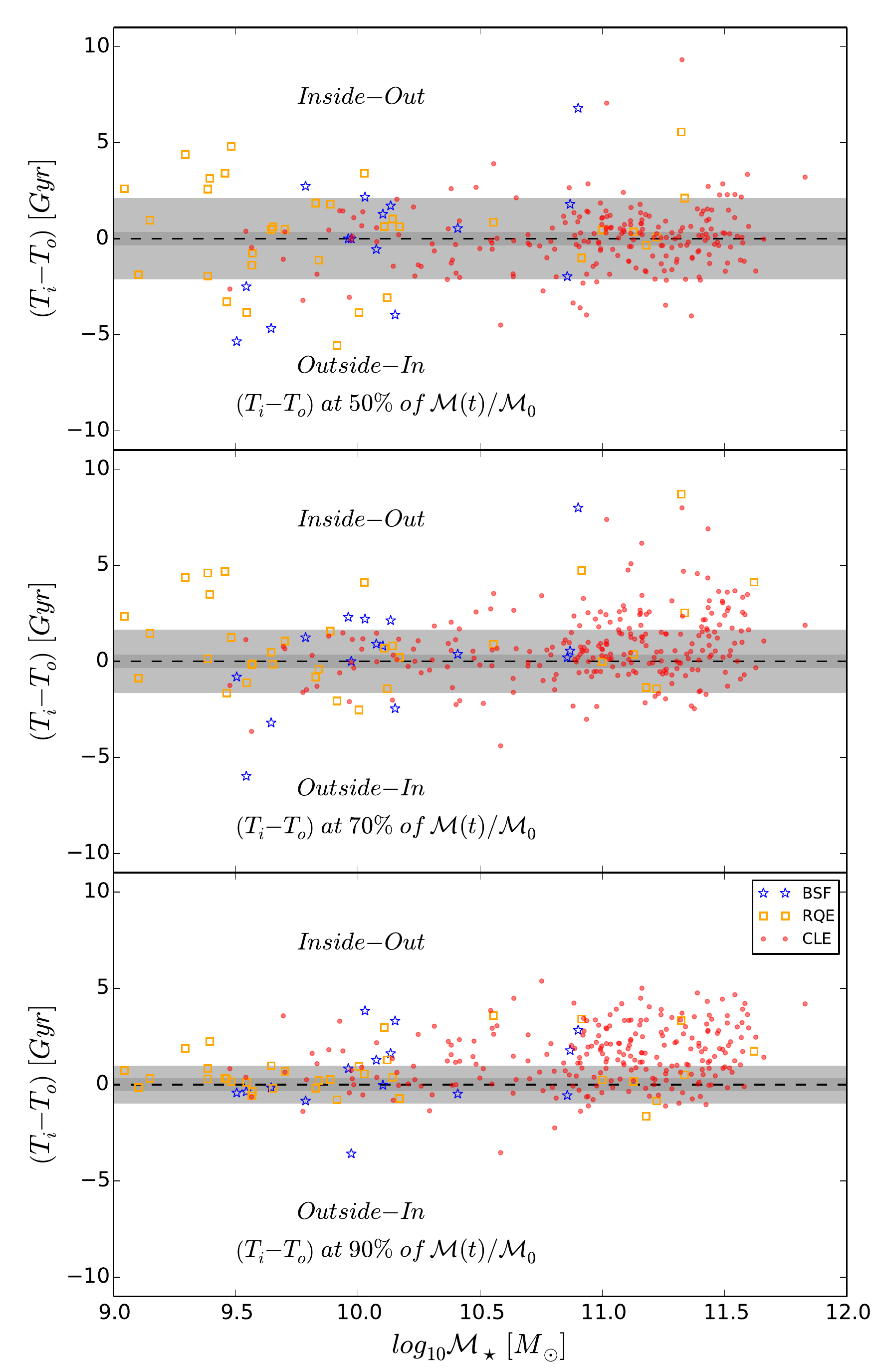}
\caption{Difference of look-back times between the internal part ($R$ < 0.5 \re) and the external part (1 < $R$ < 1.5 \re), ($T_i - T_0$), when each accumulates 50\%, 70\%, and 90\% of the mass 
 (from top to bottom panels) as functions of the total stellar mass.
The red dots, orange squares, and blue stars correspond to CLE, RQE, and BSF galaxies, respectively.  
Above (below) the dashed lines, the E galaxies would be assembled inside-out (outside-in). 
The shaded regions correspond to the
temporal resolution allowed by the SSP template library at the earliest ages of the galaxies in each panel. 
}
\label{fig_deltaT}
\end{figure} 

We notice that the differences in mean normalized CMTDs
of the external and inner parts of a given E galaxy can be smaller than the age resolution of the used SSP libraries for ages corresponding to look-back times higher than $\sim 8$ Gyr\footnote{As discussed in \citet[][see the references therein]{Ibarra-Medel+2016,Ibarra-Medel+2019}, the spectra of old stellar populations are very similar among them in such a way that the exact age determination of them is very uncertain. For example, the fossil record method applied to MaNGA IFU data with a reasonable signal-to-noise ratio (S/N) recovers the ages of stellar populations older than 10 Gyr with an uncertainty that is not better than $\pm (2.5-3)$ Gyr. Estimates of the uncertainty in age determination with the fossil record method at different stellar population ages are shown in Figs. \ref{fig_intMGH} and \ref{fig_MGH_radial2} as the horizontal error bars.}. 
Fig. \ref{fig_deltaT} shows this comparison. Here, we plotted for each galaxy the difference between the look back times of the normalized CMTDs 
corresponding to the internal and the external parts, $(T_i - T_o)$, when each attains 50\%, 70\%, and 90\% of the mass in the corresponding radial bin as functions of the galaxy \ms{}. When $(T_i - T_o)$ is larger (smaller) than 0, we have formally the inside-out (outside-in) growth mode. However, when $|T_i - T_o|$ is smaller than the uncertainty in the age determination, we cannot conclude anything more than that the radial growth mode is roughly uniform within this age uncertainty. The light gray bands in Fig. \ref{fig_deltaT} show our rough estimates of the age resolution allowed by the fossil record method at the  different mass fractions, which correspond to different epochs on average.  
The radial mass growth of CLEs passes from roughly homogeneous at epochs when they accumulated $\sim$50\% of their inner and outer masses (within the age uncertainty of the fossil record method) to clearly inside-out mass growth mode when they accumulated 90\% of their inner and outer masses, that is to say the inside-out mode becomes relevant at later epochs of the mass growth. 
The (late) inside-out mode tends to be stronger for more massive CLEs in all the panels.
 The RQE galaxies do not show systematic differences with \ms{} in the mass growth between the internal and external parts. Most %of the 
 intermediate- and low-mass RQEs show some evidence of the inside-out growth mode, though weaker than 
 the CLEs. In the case of the massive RQEs,  two (1/3)
 of them show evidence of an outside-in mode at least since times where 70\% of their inner and outer parts have formed,   while other two massive RQEs show a clear inside-out growth mode. The BSFs show a large scatter in this figure, though not too different from the CLEs of similar masses.

\subsection{Specific star formation histories}

%%%fig integrated specific SFH (fig16)
\begin{figure*}
\includegraphics[width=15.cm]{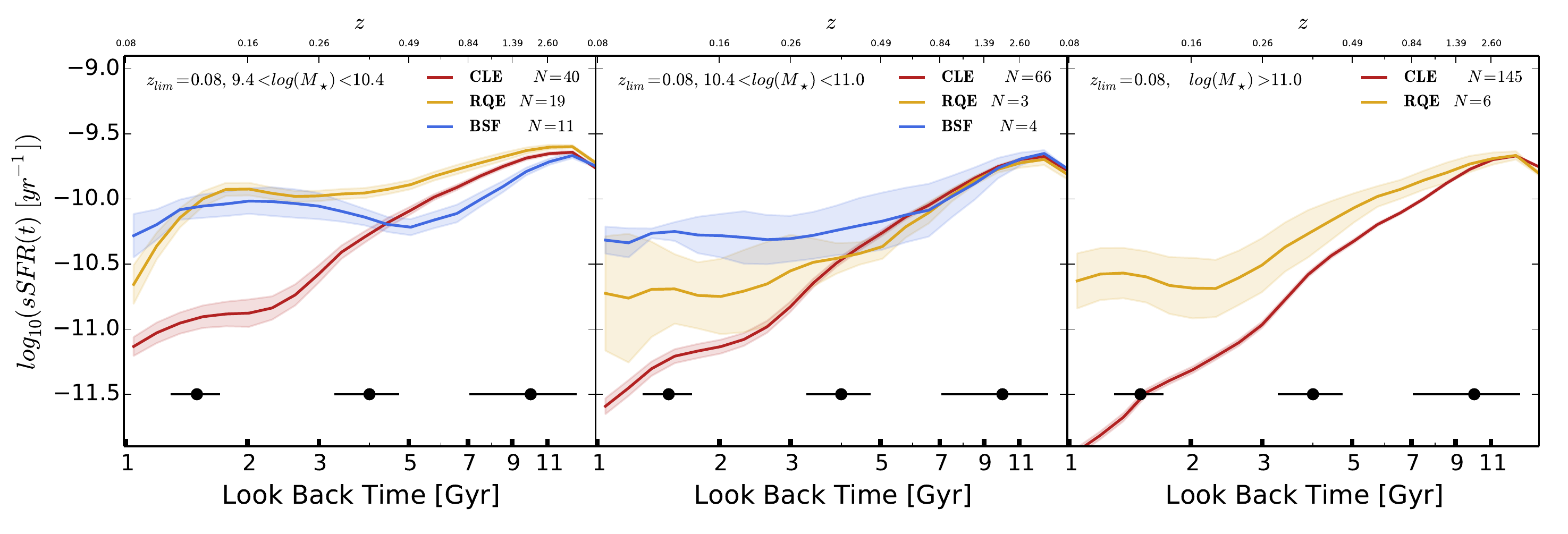}
\caption{Integrated mean specific SFRs as functions of look back time, i.e., integrated mean sSFHs, within 1.5 \re{} in three stellar mass bins (mass increases to the right). The sSFHs are normalized at $z_{lim}=0.08$. The red lines correspond to CLEs, the orange lines correspond to RQEs, and the blue lines correspond to BSFs. The shaded region is the error of the mean
of each sSFH. The number of galaxies of each type in a given mass bin is shown in each panel. 
The horizontal error bars show the average time resolution of the SSP template stellar library at the given epoch.
}
\label{fig_intSFH}
\end{figure*} 

%%%fig SFH_radial2 (fig 17)
\begin{figure*}
\includegraphics[width=15.8cm]{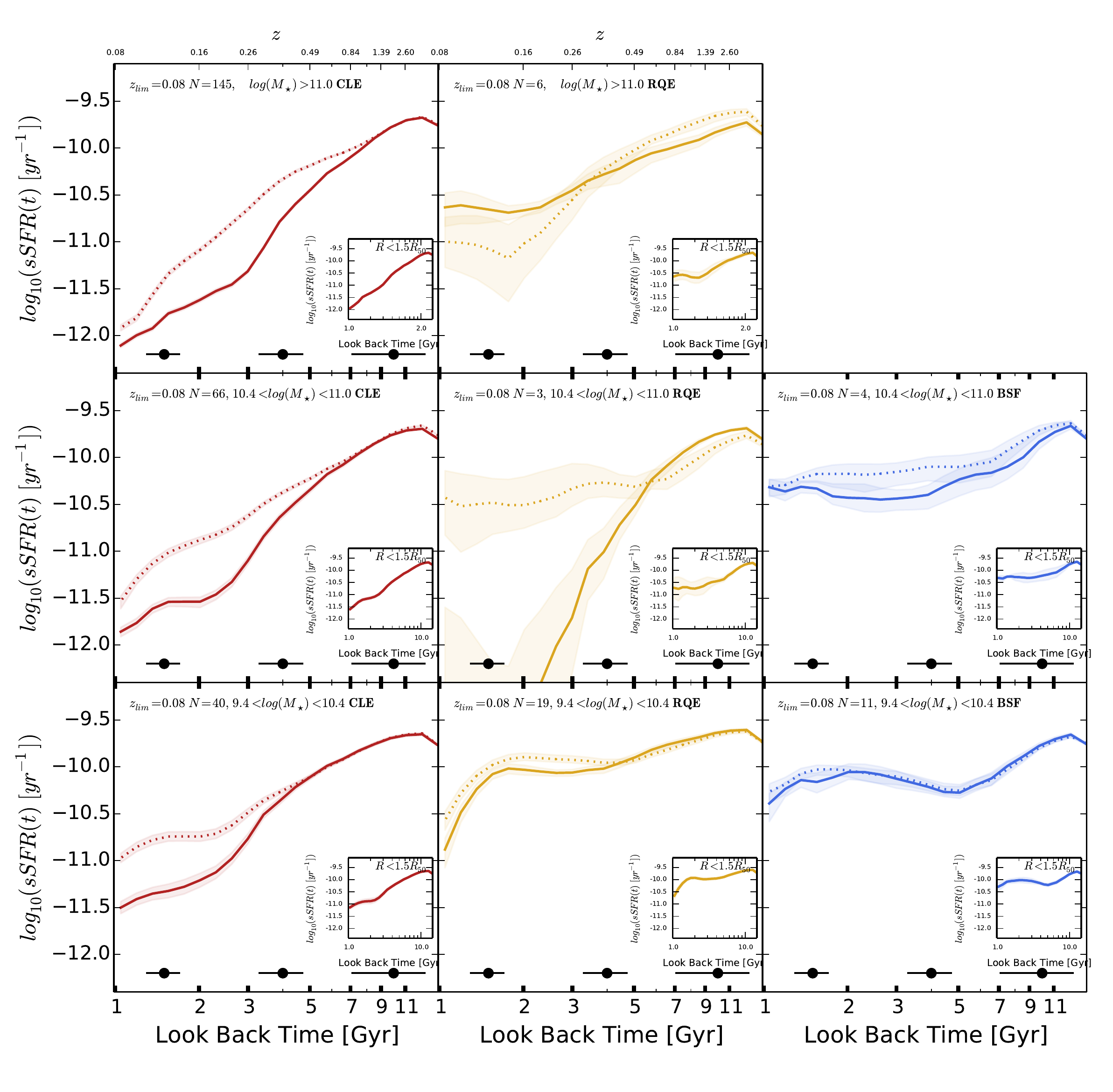}
\caption{Mean sSFHs for each E type in three mass bins, which increases from the bottom to the top.
The left column is for CLEs, the central column is for RQEs, and the right column is for BSFs. 
In each panel, the solid line corresponds to the mean sSFR within 0.5 \re, whereas the dotted line corresponds to the mean sSFR between 1.0 and 1.5 \re. 
The light shaded region and dark shaded region are the
errors of the mean
of the sSFRs in the inner part and the outer part of the galaxies, respectively. 
The horizontal error bars show the average time resolution of the SSP template stellar library at the given epoch. 
The inset box shows the integrated mean specific SFHs within 1.5 \re.
}
\label{fig_SFH_radial2}
\end{figure*}

Figure \ref{fig_intSFH} shows the mean integrated (within 1.5 \re{}) specific SFHs for CLEs, RQEs, and BSF Es in three stellar-mass bins\footnote{We smoothed the specific SFHs by a step filter with a width of 32 Myr to minimize the impact of numerical spikes.}. 
They are normalized to their values attained at $z_{lim}=0.08$. 
The mean specific SFHs of CLEs show a sharp decrease with time, which is
faster the more massive the galaxies are. 
This behavior is likely associated with quenching processes rather than natural aging due to gas consumption; see more details below.
For the RQEs, their mean specific SFHs are roughly similar to those of the CLEs at the largest look back times, but then they decrease more slowly. In particular, the mean specific SFHs of the RQEs remain roughly constant since $\sim 5$ to 2 Gyr ago for masses \ms\ < $10^{11}$ \msun\ and since $\sim 2.5$ to 1.5 Gyr ago for larger masses, and then they decrease more quickly. This result implies that the progenitors of the RQEs at these epochs were more efficient in growing their stellar mass by SF than the CLEs, and only recently suffered significant decline in SF. 
The specific SFHs of the BSFs show behavior similar to the RQEs but without sharply decreasing lately, as the RQEs.

Figure \ref {fig_SFH_radial2} shows the mean specific SFHs for CLEs, RQEs, and BSFs in two radial regions and three stellar mass bins. Similar to Fig. \ref{fig_MGH_radial2}, the inner part of the galaxies ($R$ $<$ 0.5 \re) and the external part of the galaxies (1.0 $<$ $R$ $<$ 1.5 \re) are both shown. The inner mean specific SFHs of the CLEs decrease faster than the outer ones, suggesting that CLEs suffer an inside-out quenching, which is more pronounced for the more massive ones. In the case of the 
low-mass RQE and BSF galaxies, the inner and outer mean specific SFHs 
have shown relatively constant SFHs since 5 Gyr down to the present for the BSFs and down to $\sim$ 2 Gyr ago for the RQEs. 
The BSFs of intermediate masses also show constant radial SFHs down to the present. At the same mass, the inner part of the RQEs shows a rapid decrease in the sSFR to lower redshifts, whereas the specific SFH of the outer parts is still constant. 
The most massive RQEs show slight evidence of outside-in quenching, with the outer parts starting to significantly decline their SF faster $\sim4$ Gyr ago, on average, than the inner parts.

%%%%%%%%%%%%%%%%%%%%%%%%%%%%%%%%%
\subsection{Epoch and time-scale of long-term quenching in SF} 
\label{SecQ}

In the literature, there is not a common and exact definition of SF long-term quenching in galaxies; for a few recent discussions, see \cite{Schawinski+2014}, \cite{Smethurst+2015}, \cite{Pacifici+2016}, and \cite{Carnall+2018}, for example.
As their gas reservoir is consumed, galaxies decrease their SFRs according to their internal metabolism, even the star-forming ones, which likely accrete gas by cosmological infall during long periods of their lives. Let us call this cycle of decreasing gas accretion and its consumption into stars due to the internal metabolism of galaxies nominally as aging. In a general way, by quenching, one understands a decline in the SFR significantly faster than normal aging, which is produced by some internal or external mechanisms. These mechanisms may deprive
galaxies of gas infall, promote its ejection from the galaxy, or avoid the cold gas to be transformed into stars
(for  reviews of different quenching mechanics, see, e.g., \citealt{Ibarra-Medel+2016}, \citealt{Smethurst+2017a}, and more references therein).
The different quenching mechanisms have different time scales of SF cessation.

We have mentioned above that the sSFR histories of CLEs show a pronounced 
decline. Now, we would like to evaluate how quickly this decline of SF for our sample of MaNGA E galaxies occurs. We anticipate that it is beyond the scope of this paper to explore possible processes and mechanisms of quenching. A first step for estimating the time-scale of quenching is to determine the epoch at which a given galaxy becomes quenched long-term. 
Several approaches are used in the literature to define the long-term quenching time, \tq, of galaxies. 
For example, \tq\ is defined as the cosmic time when the sSFR decreased below a given factor $f$ times the inverse of the Hubble time \citep[e.g.,][]{Firmani+2010,Pacifici+2016} or when the birthrate parameter $b$ decreased below a given factor $f'$ \citep{Carnall+2018}.  Both factors are related, given that sSFR$(t)\approx b(t)/[(1-R_{\rm ml})t)]$; the approximation is because it is assumed that $R_{\rm ml}$
is independent of time, which is roughly correct after $\sim 2$ Gyr of evolution for a given SSP. 
A second step is to define a characteristic time of maximal SFR from which the galaxy started to decline, on average, its SFR, \tqin. For galaxies with several stages of increase and decrease of the SFR, there is not an unambiguous time of maximum SFR. Operationally, here, we relate \tqin\ to the luminosity-weighted average age of the galaxy, age$_{lw}$\footnote{We pass from age to look back time adding to age$_{lw}$ the observation time of each galaxy, and from look back time to cosmic time using the Universe age of 13.67 Gyr for the cosmology assumed here.}, considering it as an average epoch at which the galaxy still had active star formation. 
The time difference \Dtq= \tq $-$ \tqin\ is then the quenching time-scale. 

To define the long-term quenching time \tq\ from the criterion $b(t)=f'$, 
we use the last of the times when this criterion was attained.
For our sample of E galaxies, we experimented with values of $f'$ from 0.8 down to 0.1.  For 
high values, there are galaxies for which \Dtq\ is negative, and many of our BSF Es appear to be quenched. For low values of $f'$, the quenching time of some Es, mainly RQEs, is not attained and the \Dtq\ values of most of the Es are too large. We have found that $f'=0.4$ is an optimal value because all of our BSF Es are confirmed to not be retired galaxies, and there are only a few RQEs that do not attain the criterion of quenching.

%%%Fig quenching (fig18)
\begin{figure}
\includegraphics[width=8.8cm]{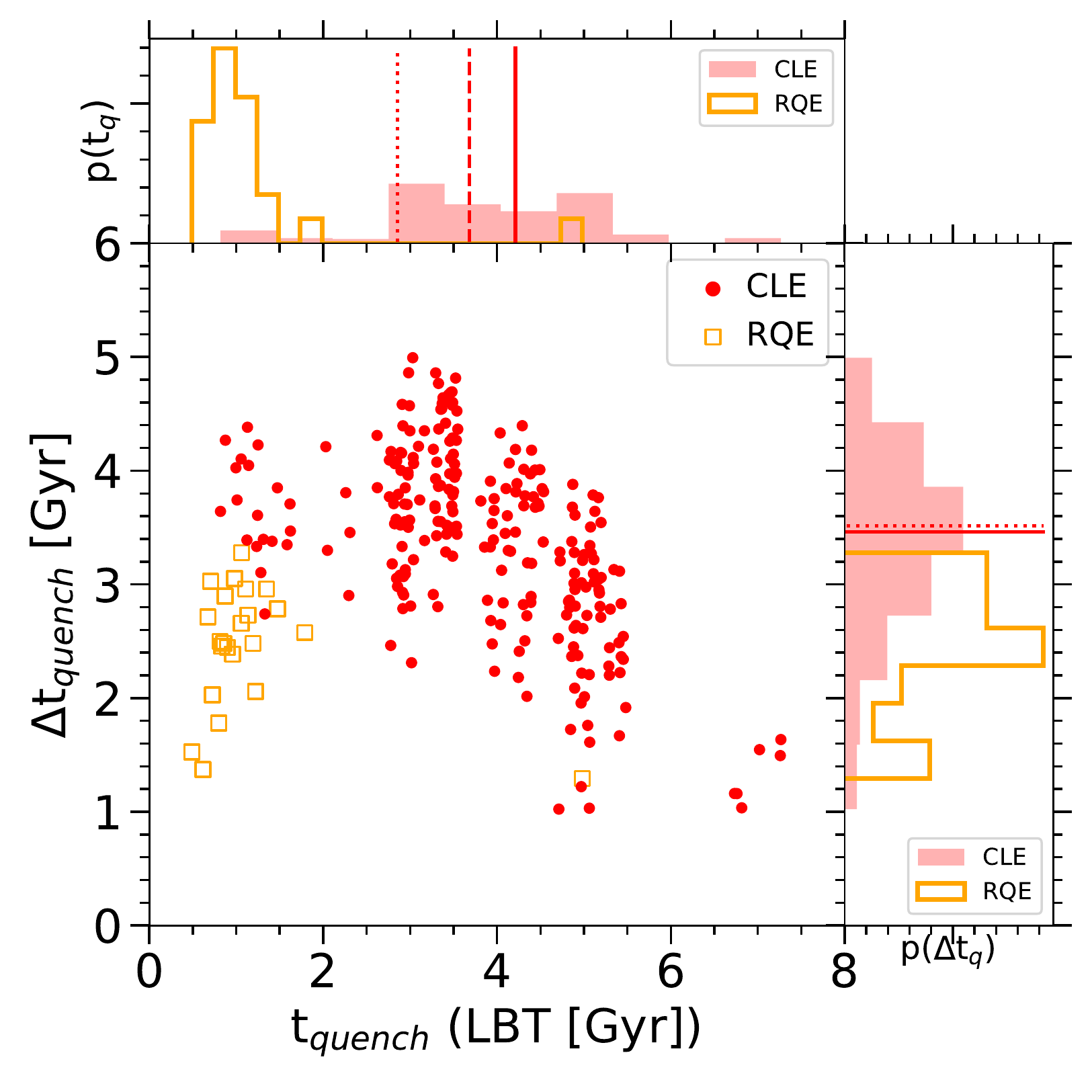}
\caption{
Long-term quenching look back time (\tqlb) and quenching time-scale (\Dtq) calculated from the nonparametric stellar population age distributions (see Sect. \ref{SecQ} for details). Red circles are for CLEs and orange squares are for RQEs. The upper and right panels show the normalized distributions of \tqlb\ and \Dtq, respectively,
for CLE galaxies (red solid histogram) and RQE galaxies (orange open histogram). The integral of each histogram sums to unity.
The vertical (horizontal) lines in the upper (right) panel correspond to median values of \tqlb\ (\Dtq) for CLE galaxies in three stellar mass bins: 9.4 < $\log$(\ms/\msun) < 10.4 (dotted), 10.4 $\le$ $\log$(\ms/\msun) < 11 (dashed), and 11 $\le$ $\log$(\ms/\msun) < 12 (solid).}
\label{fig:quenching}
\end{figure}

Figure \ref{fig:quenching} shows the quenching look back times, \tqlb, and 
quenching time-scales, \Dtq, calculated from the (nonparametric) stellar population age distributions of each one of our MaNGA E galaxies. The upper and right panels show the normalized histograms of \tqlb\ and \Dtq, respectively.  On average, CLEs quench $3.8\pm1.2$ Gyr ago and not earlier than 7.3 Gyr ago ($z\approx 1$). For values of $f'$ that are lower than 0.4, the quenching look back times result even later (lower redshifts). 
The RQEs, as expected, quench later than the CLEs, $1.2\pm 0.9$ Gyr ago on average.  

Regarding the long-term quenching time-scale \Dtq, the CLEs take $3.4\pm 0.8$ Gyr to quench on average, while RQEs quench faster, in $2.4\pm0.5$ Gyr on average. For the CLE population, we find some anticorrelation between \Dtq\ and \tqlb: The earlier the long-term quenching is, the faster it happens; for those CLEs that quenched $\sim7$ Gyr ago, \Dtq$=1$--$1.5$ Gyr. 
In contrast, RQEs quench late and relatively quickly, occupying a different region than CLEs in the \Dtq--\tqlb\ plane; this provides one more %time 
piece of evidence that the RQEs significantly segregate from the dominant population of CLEs. For the latter, %CLEs
there is a trend of \tqlb\ with \ms\ in the sense that more massive galaxies quench earlier. In the histograms of Fig. \ref{fig:quenching}, the median values of \tqlb\ and \Dtq\ corresponding to the low-, intermediate-, and high-mass bins used throughout this paper are indicated with the dotted, dashed, and solid lines, respectively. 

We also calculated the normalized long-term quenching time-scale, $\eta_q\equiv$ \Dtq/\tqin, which measures the galaxy quenching duration as a fraction of the age of the Universe when the quenching started (recall that we approximate this cosmic time to correspond to the luminosity-weighted age of the given galaxy). 
For CLEs, on average, $\eta_q=0.50\pm 0.15$, that is, these galaxies quenched roughly half the time they attained their most active SF phase.
For RQEs,  $\eta_q=0.23\pm 0.06$.
This result could be interpreted as a high quenching efficiency, but it 
could also be related to the efficiency of a possible post-merging quenching process rather than the quenching of an individual galaxy, in which case it is not easy to interpret $\eta_q$ for them. 

A typical function used to describe the SFH of galaxies is the so-called $\tau$-delayed model:
\begin{equation}
{\rm SFR}(t) = \frac{A(t-t_i)}{\tau^2}\exp{\frac{t-t_i}{\tau}},
\end{equation}
where $t_i$ is the onset time of SF
and $A$[\msun] is a normalization factor. 
The SFR grows linearly after $t_i$, it attains a maximum at $t_{\rm max} = t_i + \tau$,
and then it decreases exponentially.
In \citet{LopezFernandez+2018}, the authors used several parametric SFHs for their 
stellar population synthesis inversion
models applied to the IFS observations from the CALIFA survey \citep{Sanchez+2012}.  
For the $\tau$-delayed function and in the case of their E galaxy sample, they report look-back initial times of $t_{i,LB}=10.1\pm 5.1$ Gyr and $\tau= 1.4\pm1.4$.  For the $\tau$-delayed function, we can define the long-term quenching time-scale as \Dtq= \tq $-t_{\rm max}$. 
By using our definition of \tq\ as the time when $b(t)=0.4$, we find that  \Dtq\  strongly depends on $\tau$ and weakly on $t_i$, \Dtq$\approx(0.15t_i + 2.46)\tau + 0.02t_i + 0.37$. 
For the range of values of $t_i=3.35\pm0.5$ Gyr constrained in \citet{LopezFernandez+2018} for E galaxies, and for $\tau=1.4$ Gyr, \Dtq $\approx 4.5$--$4.7$ Gyr. These values are $\sim 1 $ Gyr larger than the average \Dtq $=3.4$ Gyr value found for CLEs here. The SFHs of individual galaxies are diverse, and in many cases, the $\tau$-delayed function does not describe them well. Despite this and the many differences between \citet{LopezFernandez+2018} and our fossil record method, it is encouraging that the differences in the obtained typical values of \Dtq\ for E galaxies are not large. 
Finally, we notice that when approaching the SFHs to the $\tau$-delayed model, values of $\tau\sim1$ Gyr imply long-term quenching times-scales (using our definition of \tq, when $b=0.4$) of 3.0--3.4 Gyr for reasonable $t_i$ values of 1--4 Gyr, respectively.

%%%%%%%%%%%%%%%%%%%%%%%%%%%%%%%%%
\section{Discussion}
\label{Sdiscussion}
%%%%%%%%%%%%%%%%%%%%%%%%%%%%%%%%%

In this section, we discuss the caveats of our methodology (Sect. \ref{Sect:caveats}), and summarize and discuss the interpretations of the results presented in the previous sections.  
The CLEs are the main population of E galaxies, and our large sample allows us to claim for generic evolutionary trends for them (Sect. \ref{SD_CLE}). This could not be the case for the RQE and BSF samples due to the diversity of their properties and evolutionary features as well as the low numbers. In any case, we propose and discuss some possible generic evolutionary trends for them in Sect. \ref{RQE-BSF}.

\subsection{Caveats of the methodology}
\label{Sect:caveats}

We note that \verb|Pipe3D| uses a brute-force minimization that explores all possible parameters that will achieve the best linear combination of the SSP stellar libraries, which best fits the observed spectrum. 
At the same time, \verb|Pipe3D| performs a Gaussian fitting of the observed emission lines \citep{Sanchez+2016_p171,Sanchez+2016_p21}.
The inversion methods, in general, have an intrinsic precision limit in determining the ages of the stellar populations that compose the spectrum (therefore, in the recovered SFH or CMTD) due to the degeneracy of spectral features with age, especially for old populations; see \citet{Sanchez+2016_p171} and \citet{Ibarra-Medel+2016} for an extensive discussion on this question, and more references therein. According to these papers, the average logarithmic age uncertainty of \verb|Pipe3D| is around 
0.1--0.15 dex. This uncertainty drives the peaks of the SFH to spread out and lower their amplitudes, with also consequences in the shape of the inferred normalized CMTDs. In addition to the age resolution effects, systematical uncertainties in the results from the inversion method are introduced by the SSP and dust extinction models and the choice of the IMF. \citet{Sanchez+2016_p171} and \citet{Ibarra-Medel+2016} extensively discuss the expected effects of these ingredients on the SFHs and CMTDs (their so-called mass growth histories).

\citet{Ibarra-Medel+2019} directly explore how well the integral and radial SFHs and CMTDs are recovered with \verb|Pipe3D|, including the effects of galactic extinction and observational settings, such as the S/N,
IFU spatial resolution, and galaxy inclination. They used post-processed state-of-the-art hydrodynamical cosmological simulations of well-resolved late-type and early-type galaxies. Regarding the inversion method, they find that the precision in the inferred SFHs (and CMTDs and sSFHs), as expected, decreases as the stellar age of the ``observed'' stellar populations is older: The inversion method tends to smooth the real SFHs more and more for ages older than $5$ Gyr. 
This effect diminishes the original inside-out trend at the radial level, especially for the late-type galaxies, which present a more complex SFH and a more prominent inside-out mode than the early-type ones. Nevertheless, \verb|Pipe3D| still manages to recover the qualitative behavior of the real radial SFHs and CMTDs at early epochs; the recovery is much more precise at late epochs. The main reason for these issues in recovering the SFHs is the limited age sampling of the used SSPs as the ages are older
\citep{Sanchez+2016_p21,Ibarra-Medel+2019}. 
For that reason and for the purposes of this paper, we used the  \verb|Pipe3D| age sampling resolution of the SSPs as a rough indicator of uncertainty of the inversion method in the recovery of the real SFHs or CMTDs at different epochs. 

\citet{Ibarra-Medel+2019} also tested the ability of the fossil record method to recover the age and metallicity radial profiles of the mock galaxies analyzed under the observational settings of MaNGA. They found that the mass- and luminosity-weighted profiles tend to be flattened by the fossil record method, particularly for late-type galaxies, the more inclined they are and when the IFU spatial resolution is poor. They note that young populations, even if their fraction in mass is small, significantly "contaminate" the given spectrum so that the inversion method tends to overestimate the fraction of these populations, with the consequent recovery of younger ages and lower stellar masses than the true ones. 
When young populations are present, the effects of this contamination affect the radial bins  that contain intrinsically larger fractions of old populations more, that is to say the innermost ones, thus assigning them lower ages and metallicities, which lowers the gradients. \citet{Ibarra-Medel+2019} warn against overinterpreting the age and metallicity gradients of galaxies with poorly sampled IFS data and high inclinations. For early-type galaxies, especially the CLEs, the above mentioned effects are much less severe than for late-type galaxies.

%%%
\subsection{The CLE galaxies}
\label{SD_CLE}
%%%

\subsubsection{Radial gradients: Comparisons with previous works and implications}

There are different radial mass growth 
modes of early-type galaxies discussed in the literature based on their calculated stellar age radial profiles. 
Most of the reported luminosity-weighted age slopes for E+S0 galaxies are distributed within the $[+0.1,-0.1]$ interval\footnote{In some papers, the gradient slopes are calculated per dex in age and per dex in physical radial units (kpc), and in other ones in per dex in age and in units of \re{}. The reported slopes are then in units of dex/dex or dex/\re{}, respectively.}, that is, around 0 (e.g., \citealt{Sanchez-Blazquez+2007}, \citealt{Reda+2007}, \citealt{Spolaor+2010}, \citealt{Rawle+2008,Rawle+2010} with VIMOS IFU; \citealt{Kuntschner+2010} with SAURON;  \citealt{Boardman+2017} with Mitchell Spectrograph IFU;
\citealt{Goddard+2017_466mass} with MaNGA; and \citealt{SanRoman+2018} with the ALHAMBRA survey, see more references therein), and also for only E galaxies (\citealt{Koleva+2011} with long-slit data and \citealt{MartinNavarro+2018} with CALIFA).

 The median mass- and luminosity-weighted age gradients we have found for CLEs here are $-0.03^{+0.04}_{-0.05}$ and $-0.08^{+0.06}_{-0.07}$ dex/\re{}, respectively, which suggest a weak inside-out formation. Among those works that are similar to ours, \citet{Goddard+2017_466mass} find that the mass-weighted median age gradients tend to be slightly positive for E+S0 galaxies in MaNGA, which suggests a weak outside-in formation.
\citet{DominguezSanchez+2019} find negative light-weighted age gradients for E galaxies in MaNGA, but the most massive galaxies show mildly positive age gradients. They find that the age gradients are weaker if there are IMF gradients as well.
Using the CALIFA survey, \citet{Zibetti+2020} report mildly positive luminosity-weighted age gradients for E+S0 galaxies, although the profiles tend to be U-shaped in the inner regions ($\lesssim 0.4$ \re{}).  
In contrast, \cite{GonzalezDelgado+2015} find mildly negative light-weighted mean age gradients for E galaxies in CALIFA, and \cite{ZhengZheng+2017} find mildly negative mass-weighted mean age gradients for Es in MaNGA, which are consistent with an inside-out formation scenario. 
\cite{Parikh+2019} also report mildly negative age gradients for E+S0 galaxies in MaNGA, but \cite{LiH+2018} and \cite{Ferreras+2019} find mildly positive age gradients for low-mass E+S0 galaxies in MaNGA and SAMI, respectively.

Different observational settings, spectral fitting codes, stellar population models, and methods to derive the stellar age gradients may explain the diversity of the results discussed above.
In Appendix \ref{S_Ap_compGoddard2017}, we show examples of the luminosity- and mass-weighted age profiles for individual MaNGA galaxies shown in \citet[][see their Fig. 7]{Goddard+2017_466mass}. 
Our luminosity-weighted age profiles are consistent with those of \cite{Goddard+2017_466mass}. However, our mass-weighted age profiles are also mildly negative or flat, which is opposite to the positive mass-weighted age gradients reported by them. 
At this point, it is difficult to make conclusive statements about the formation scenarios of early-type galaxies solely based on the stellar age profiles, given the mixed results in the literature using different approaches. Fortunately, more information can be obtained from the spatially-resolved IFS observations. It is also important to note that our study is particularly unique with respect to previous ones because we focus (i) only on morphologically well-defined E galaxies (S0 are not taken into account) and (ii) only on ``red and dead'' E galaxies (CLEs), separating those E galaxies with evidence of recent quenching, blue colors, or ongoing star formation.

For the metallicity radial profiles, the results in the literature are in agreement  that the vast majority of early-type galaxies present negative gradients.
However, the different works do not agree in how steep  these negative gradients are \citep[see for recent reviews][]{SanRoman+2018,Zibetti+2020}. The shallow slopes we find for our sample of CLE galaxies (see Tables \ref{tab1} and \ref{tablaMedianGrad}) 
are consistent with most of the recent works based on studies of large galaxy samples \citep[e.g.,][]{GonzalezDelgado+2015,Wilkinson+2017, Goddard+2017_466mass,ZhengZheng+2017,SanRoman+2018, LiH+2018}. Gas-rich dissipative collapse models \citep[][]{Larson1974}, including SN-driven winds \citep[e.g.,][]{Pipino+2010} or their revised versions in the hierarchical clustering cosmological scenario \citep[e.g.,][]{Cook+2016,Taylor+2017}, predict steep negative gradients in metallicity after the dissipative phase.
The steepness of the gradients can be lower by low SF efficiencies induced by SF feedback \citep{Pipino+2008,Pipino+2010} in less massive galaxies or by non-dissipative (dry) mergers \citep{Kobayashi2004,DiMatteo+2009,Hirschmann+2015,Cook+2016,Taylor+2017} in more massive galaxies, at least externally. If the mergers are gas-rich, central SF can regenerate the gradients \citep{Hopkins+2009b}, though the efficiency of this process is limited by processes that can suddenly quench SF. The final metallicity gradients depend on all of these processes and the fractions of stellar mass that formed or assembled in the different phases.

Regarding the dependence of the metallicity gradient on mass, previous observational results are in disagreement \citep[see for a review][]{Oyarzun+2019}.  We find that the gradient slopes tend to become flattened as the CLEs are less massive, which agrees with the proposal that SN-driven winds lower the SF efficiency in galaxies with weaker gravitational potentials. On the other hand, for the more massive CLEs, while the inner gradients remain relatively steep, at radii larger than $\sim 1$ \re\ the metallicity profiles tend to flatten (see the top-left panel of Fig. \ref{fig_Z_grad}), which agrees with the proposal of posterior dry mergers. The latter result is consistent with \cite{Oyarzun+2019}, who also find that the stellar metallicity profiles of massive E+S0 MaNGA galaxies fall with radius but flatten externally. 
In conclusion,  we find negative metallicity gradients with relatively shallow slopes, which support the scenario of early gas-rich dissipative collapse for CLEs combined with mass-dependent processes that tend to lower the steepness of the gradients.

As \cite{Oyarzun+2019} extensively discuss, some of the apparent differences in the measured metallicity gradients of early-type galaxies among many observational works may be due to the definition of metallicity gradients that is used as well as to the different metallicity tracers or stellar population fitting codes employed to calculate the stellar metallicity. 
We add to this discussion that differences may also arise from the way the samples of early-type galaxies are selected. Our work is probably 
more accurate in this sense since it exclusively refers to red, quenched, and morphologically classified pure E galaxies.

Finally, we find that the radial sSFR profiles decrease toward the center on average, suggesting inside-out quenching. 
Indeed, the average radial specific SFHs show that the quenching is faster in the inner part on average compared to the external one, see Fig. \ref{fig_SFH_radial2}. In the case of the CLEs that are less massive than $10^{10.4}$ \msun, some residual SF would be needed to maintain the sSFR relatively flat in the external parts during the last 2 or 3 Gyr. 
The inside-out quenching can be produced by the following: an early central compaction process that triggers bursts of SF with the consequent gas consumption and gas ejection by SN-driven feedback, thus not leaving too much gas for further SF \citep[this process also would help to reduce the negative metallicity gradient;][]{DekelBurkert2014,Tacchella+2016,Avila-Reese+2018}; 
the AGN feedback \citep[e.g.,][]{Bluck+2016,Bluck+2020,Guo+2019};
morphological quenching, that is, when the dynamically-formed hot spheroid stabilizes the inner gaseous disk avoiding further SF \citep[e.g.,][]{Martig+2009};
or a combination of the previous mechanisms over time \citep{LinL+2019}. 

\subsubsection{A two-phase formation scenario}

From the archaeological study of our well-defined morphological sample of 251 CLE galaxies, we find that they accumulated 70\% of their current stellar masses 8--11 Gyr ago with an evident ``downsizing'' trend, especially in what regards the formation of their last 10--20\% mass fraction.  As discussed above, the CLEs mostly show mild negative or zero light- and mass-weighted age gradients as well as relatively steep negative gradients in both light- and mass-weighted metallicities.
Additionally, the mass- and luminosity-weighted ages do not differ too much among them for both the integrated ages  
and for the ages along the galacto-centric radius. 
At earlier epochs, the inner and outer parts of CLEs formed their stellar populations roughly homogeneously; while at later epochs, the outer parts tend to form later than the inner ones, especially for the more massive galaxies. On the other hand, the inner parts tend to decrease their specific SFRs  
faster than the outer parts. All of these results suggest that the CLEs formed a significant fraction of their stars early ($>8$ Gyr) and nearly coeval in time and radius, followed by an inside-out growth and a quenching process that also proceeded inside out.
These trends become stronger the more massive the galaxies are.

From the archaeological analysis of the CLEs, we estimated that they finished their quenching globally (retired) on average $3.8\pm1.2$ Gyr ago and not earlier than 7.3 Gyr ago, with quenching time-scales (from a typical epoch of active SF to the long-term quenching time) of $3.4\pm0.8$ Gyr on average. Additionally, though with a large scatter, the trend is that CLEs that retired earlier did it faster. A possible explanation for this trend could be that CLEs generally form early by intense bursts of SF, followed by a rapid decline of SF. However, many of them can continue growing in stellar mass lately by (dry) accretion of other galaxies, whose stellar populations can be younger, that is to say these galaxies were forming stars after the central one already retired. The archaeological analysis of a galaxy assembled in this way shows a later epoch of SF quenching and a larger quenching time-scale than the early formed primary galaxy. 
The emergence of a quenched population of galaxies could result from our archaeological analysis too late compared to observations from cosmological surveys, which show 
 a non-negligible fraction of passive or quenched galaxies at $z\sim 2$ \citep[see, e.g.,][]{Muzzin+2013,Whitaker+2013,Pandya+2017}. 
However, we note the following. (i) Our galaxies may have several epochs of quenching (and SF increasing), but we define the long-term quenching time as the last of these epochs; therefore, one expects some fraction of our galaxies to be counted as quenched several times at epochs earlier than the assigned \tq. (ii) For accurate comparisons and to overcome the progenitor bias, it is important to use large volume-complete samples, and to use similar criteria in the fossil record and high redshift samples to define quenching times and galaxy quenched populations. The study of both questions mentioned above is beyond the scope of this paper. 
An analysis of long-term quenching times and time-scales in the inner and outer regions of the CLEs, and separating galaxies into centrals and satellites, may help explore this question of two-phase E galaxy formation. Such a study will be presented elsewhere.

In a general way, the results reported here for CLEs are consistent with the prediction from hydrodynamic simulations of the formation of spheroid-dominated galaxies \citep[e.g.,][]{Naab+2009,Oser+2010,Navarro-Gonzalez+2013,Rodriguez-Gomez+2016,Rosito+2018,Rosito+2019L3,Rosito+2019}.   
These simulations show that massive spheroid-dominated galaxies follow, in general, a two-phase formation scenario, which consists of an initial gas-rich (dissipational) collapse onto dark matter halos, where the bulk of the stellar mass is formed, and a late dry accretion of satellites and eventual (dry) major merger events.
The consistency between our results and those of simulations is evident even in some specific details. For example, according to  \cite{Rosito+2019L3}, for values of $\lambda_{R_e}$= 0.17, which is the median measured in our CLEs, the stellar mass-weighted ages should be older than 8 Gyr. This is consistent with the overall old stellar populations of our CLE galaxies from the $age_{mw}$ distribution (see Fig. \ref{fig_age}).

\subsection{The RQE and BSF galaxies}
\label{RQE-BSF}

In the following, we discuss the nature and evolution of the RQE and BSF galaxies. These two samples are about 15\% out of our whole sample of E galaxies and are mainly characterized by showing evidence of late events of SF. This is in line with \citet{Thomas+2010}, who from a SDSS sub-sample of early-type galaxies, have reported that $\sim$10\% of them present young ages (<2.5 Gyr), being they are more frequent at lower masses and lower-density environments. They interpreted these early-type galaxies as the result of rejuvenation processes \citep[see also][]{Shapiro+2010}. We caution that our attempt to find general evolutionary trends for the RQE and BSF samples is flawed by low numbers as well as for the possibility that these galaxies rather than homogeneous subpopulations are just objects that deviate from the main trends due to peculiar and diverse evolutionary conditions. In forthcoming papers, we will present an individual spectroscopic analysis of each RQE and BSF galaxy, including kinematics, as well as detailed photometric and environmental analysis. This detailed analysis will allow us to confirm or modify the generic interpretations we propose here tentatively for RQEs and BSF galaxies.

%%%
\subsubsection{RQE galaxies}
\label{SD_RQE}
%%%

Determining whether RQEs do in fact have different properties and evolutionary paths with respect to CLEs, as defined here, is necessary.
Our results suggest that, indeed, this is the case, reassuring that our selection criteria were able to pick a class of physically different E objects.
The RQEs are significantly younger than the CLEs on average,  even at a given \ms, see Fig. \ref{fig_age}; on average, the differences are by $\sim 5$ Gyr using $age_{lw}$, which is expected given the selection criterion with $age_{lw}$, and by 3--4 Gyr using $age_{mw}$.
At low masses of 9.4 < log(\ms/\msun) < 10, the RQEs are younger  than CLEs with differences of $age_{mw}$ $\sim$ 2 Gyr on average.
The gradients in $age_{mw}$ of the RQEs are, on average, flat but there is a large scatter around 0, see Fig. \ref{fig_gradAge_indiv}. The gradients in $age_{lw}$ are even more scattered and clearly differ between RQE galaxies that are more and less massive than $\sim 10^{11}$ \msun; for the former, the gradients tend to be positive while for the latter, they tend to be negative (see Fig. \ref{fig_gradAge_indiv}, and Tables \ref{tab1} and \ref{tablaMedianGrad}). 
We recovered 71\% of the RQEs within the empirical region of the $urz$ diagram that can capture the youngest portion of quiescent blue elliptical galaxies according to \citet[][see Fig. \ref{fig_RQE}]{McIntosh+2014}. In fact, only 2\% of the CLEs are contained inside this region. At low masses of 9.4 < log(\ms/\msun) < 10, we recovered 92\% of the RQE galaxies, whereas 28\% of the CLEs are located within this region. 
Furthermore, the progenitor(s) of RQEs was (were) most of the time in an active phase of SF (at least one of them if they suffered a late merger), which is the opposite for the progenitor of CLEs, see Fig. \ref{fig_intSFH}. Finally, the RQEs of all masses not only quenched recently (\tq$=1.2\pm0.9$ Gyr ago), but notably faster than the CLEs, even when comparing them to those CLEs that also have late long-term quenching times (see Fig. \ref{fig:quenching}). For instance, low-mass RQE galaxies show a median \tq$\sim$ 1 Gyr and \Dtq = 2.5 Gyr, whereas those are 2.9 Gyr and 3.5 Gyr for CLE galaxies of the same mass, respectively.
All of these facts suggest that RQEs are not a simple statistical extension of the properties of CLEs, including the low-mass range in which they overlap, and they may support the distinction of RQEs and CLEs as two different classes of E galaxies. 
The definition of RQE galaxies is strongly motivated by \citet{McIntosh+2014}, but we admit it is sort of arbitrary, and the numbers reported above might change with other definitions that classify them. There is the possibility that the RQE galaxies actually correspond to a continuous transition of CLE galaxies at low masses, in which the boundary is not totally clear.

The luminosity-weighted ages of RQEs are significantly lower than the mass-weighted ones at all masses. However,
while for RQEs with $\ms<10^{11}$ \msun\ the difference increases outward, for most of the more massive ones, it increases inward (see Fig. \ref{fig_ageMW_ageLW_grad}). A difference with mass is also seen in the stellar metallicity gradients; both the mass- and luminosity-weighted gradients are nearly constant or slightly negative for most  of the less massive RQEs, while both gradients are positive for most of the massive galaxies. Moreover, for the massive RQEs, $Z_{lw}$ is significantly lower than $Z_{mw}$, while for the less massive RQEs, $Z_{lw}$ is slightly higher than $Z_{mw}$ (see Fig. \ref{fig_Z}). 
The latter also has 
younger stellar populations outward, with metallicities higher than those of the dominant old populations, while for the massive RQEs, the younger populations tend to dominate in the central regions. 
Interestingly enough, the specific SFHs of the less massive RQEs show evidence of very weak inside-out quenching. 
Instead, for the massive RQEs, the SF in external regions seems to have quenched 
faster than in the inner regions on average (outside-in quenching). All of these results suggest that
RQEs that are more massive than $\sim10^{11}$ \msun\ differ on their 
radial stellar populations
from the less massive ones.

Determining why the RQE galaxies quenched so late and fast, whether they were early-type galaxies before that or if  the morphological transformation is what triggered their quenching, and whether they assemble by late mergers are discussed separately in the following for the RQEs that are less and more massive than $\sim 10^{11}$ \msun. 
Our discussion is based on the archaeological results presented above. By construction, from the archaeological approach, it is not possible to infer the dynamical assembly history of galaxies (mergers). However, some consequences of the dynamical assembly may be left imprinted on the global and radial properties of their stellar populations, thus helping us constrain scenarios that imply mergers.  
The scenarios discussed are highly speculative and are based on the reported generic trends of only a few objects, so they should be considered just as proposals that have yet to be confirmed with more detailed object-by-object studies.

$\bullet$ RQEs less massive than $10^{11}$ \msun:
A simple scenario, where the RQEs defined here are the tail of delayed evolution of CLEs is not enough to explain several of the differences mentioned above, particularly the different slopes of RQEs in their specific SFHs compared to CLEs of similar masses.
Based on our results, we can speculate that most of the RQEs suffered a relatively late, intermediate (1:3 to 1:8 mass ratio) merger, where at least one of the merging galaxies was star-forming and gas-rich (the less massive one, which has been accreted presumably by an earlier-type galaxy). 
This gas-rich late merger scenario (or the rejuvenation scenario in some cases, see below)
may explain why the specific SFH of these RQEs, which can correspond to at least two galaxies in the past, is relatively high and nearly constant until $\sim 2$ Gyr ago.

From Fig. \ref{fig_intSFH}, one sees that the mean specific SFHs of RQEs decrease relatively fast, similar to those of the CLEs; however, at some time, the mean specific SFHs of the former, instead of continuing to decrease, became roughly constant. This time can be indicative of the SF activity of the gas-rich secondary when the primary progenitor already entered in its quenching phase. The stellar mass fractions that formed at these times account, on average, for $\sim$60\% and 90\% of the final masses of the low- and intermediate-mass RQEs, respectively. Therefore, since then ($z<0.8$), the RQE main progenitors could have experienced accretion of a gas-rich satellite (or simply of gas), which comprises between 40\% and 10\% of their final mass, respectively. Before and during the late gas-rich event, the interaction enhances the SF (starburst), consuming 
the gas; the accreted gas can also activate a (weak) AGN. Indeed, gas-rich mergers have been suggested as a mechanism that triggers the shut-down of SF in quenched galaxies at $z \sim$ 0.8 \citep{WuP+2020}. If a disk still remains following the moment the system becomes strongly spheroid dominated, it likely becomes stable against disk instabilities \citep[morphological quenching;][]{Martig+2009}, which also quenches the SF. As a result, it is likely that a fraction of the young stellar populations forms from more enriched gas than the old populations belonging to the more massive, earlier-type progenitor, so their metallicities are expected to be higher than those of the older stellar populations (see lower panel of Fig. \ref{fig_Z}). From a preliminary analysis of fine morphological structures (Appendix \ref{S_Ap_FineStructure}), we find a high incidence of broad tides in the RQEs of masses above $\sim 10^{10}$ \msun; the broad tides (fans and plumes) are fossils of intermediate or major mergers that took place not too long ago \citep[e.g.,][and more references therein]{Mancillas+2019}. A lower fraction of the RQEs also have  
shells.

In addition to the gas-rich late merger scenario, for some low-mass (\ms\ $\lesssim 10^{10}$ \msun) RQE galaxies, a process of gas infalling from the cosmic web (typically of low metallicity) or gas that was acquired through internal stellar mass loss may have produced a ``rejuvenation'' of SF that stopped recently. 
Assuming that the BSF galaxies are well described by the ``rejuvenation'' of their SF (see below), and given the similarities between the RQE and BSF galaxies of the low-mass bin in the metallicity profiles and the radial specific SFHs, this scenario is plausible for some RQE galaxies. We also note that the incidence of fine structures strongly decreases for the low-mass RQEs (Appendix \ref{S_Ap_FineStructure}), although this could be partially due to effects of surface brightness sensitivity given the low luminosities of these galaxies.

$\bullet$ RQEs more massive than $10^{11}$ \msun:
For these galaxies\footnote{The MaNGA IDs (and Plate-IFU designs) of these galaxies are 1-152806 (manga-8936-6101), 1-44063 (manga-8718-6101), 1-594755 (manga-8612-1901), 1-604860 (manga-8714-3704), 1-633196 (manga-8485-12703), and 1-634238 (manga-7992-12701).}, or for some of them, the results discussed above can be consistent with the presence of a late ($z<$ 0.3--0.5),
intermediate, or major merger of a primary early-type progenitor with a secondary gas-poor galaxy but with younger populations on average than the primary.
Most of the accreted, smaller galaxy is expected to fall by dynamical friction to the center of the larger, older one. This scenario may explain why the stellar populations in the inner parts of the final RQE show evidence of younger ages, lower metallicities, higher sSFRs, and the later onset of quenching than in the outer parts. According to this scenario, most of the inner populations of the local RQE actually formed in the accreted satellite, which formed later and had more sustained SF activity than the primary early-type galaxy.
On the other hand, in contrast to the less massive RQEs, the merger for these RQEs could be dry because as the galaxies are more massive, they are poorer in gas content.  

The kinematics of the massive RQEs seems to support our gas-poor late merger scenario. For them,  the median values of ($\epsilon$,$\lambda_{R_e}$) are (0.19,0.35);
while for massive CLEs, these values are (0.14,0.14).
Thus, the median values of the massive RQEs lie well outside the region of non-regular rotators (see Fig. \ref{fig_rot}). Their values of $\lambda_{R_e}$ and galaxy mass fit within the class D described by \cite{Naab+2014}, which corresponds to galaxies that experienced, in addition to minor mergers, a recent ($z < 1$) collisionless major merger leading to a moderate rotation. 
The median structural parameters (S\'ersic index and concentration) tend to be similar between massive CLEs and RQEs (see Fig. \ref{fig_struct}), although with larger scatter for lower values, that is, more disky structures, in the case of the RQEs. 
From our preliminary morphological analysis of fine structures (Appendix \ref{S_Ap_FineStructure}), we find that 85\% of the massive RQEs present broad tides, which is possible evidence of a recent, major merger.

Finally, the massive RQEs are, in general, above the line that separates blue and red galaxies (see Fig. \ref{fig_gi_Ms_0.08}). The median $g-i$ colors also tend to be similar between massive CLEs and RQEs, although slightly redder for the former than for the latter (1.29 and 1.26, respectively). Therefore, they do not seem to currently belong to a green valley population of galaxies at these masses.
To confirm the generic (and highly speculative) scenarios proposed above for RQEs, or to favor other possibilities, we will study the IFS information of each RQE individually elsewhere, including the kinematics, their morphological features from deep optical images, and their environments.

%%%
\subsubsection{The particular case of the BSF galaxies}
\label{SD_BSF}
%%%

On average, BSFs are younger at all radii than RQEs and CLEs, especially when using $age_{lw}$. The BSFs also tend to have lower metallicities. The large differences between the ages or metallicities weighted by mass and those weighted by luminosity (Figs. \ref{fig_age}--\ref{fig_Z} and \ref{fig_ageMW_ageLW_grad}--\ref{fig_Z_grad}) show that the BSF ellipticals suffered very recent episodes of SF. 
The gradients in both $age_{mw}$ and $age_{lw}$ are nearly flat, but the gradients in $age_{lw}$ for low-mass BSFs tend to be positive. The gradients in $Z_{mw}$ and $Z_{lw}$ are flatter than those of the CLEs of the same mass, and they are even positive sometimes. We note that there are not very massive BSFs (they have \ms\ $< 10^{11}$  \msun{}). These results are consistent with the inferred mean CMTDs, which show that BSFs accumulated most of their stellar masses later than CLEs, regardless of the mass. The radial age$_{mw}$ profiles and CMTDs of BSFs show evidence of inside-out growth, suggesting they acquired their E morphology similar to the CLE galaxies. 
The main difference is when they accumulated most of their stellar masses.
The BSFs accumulated 90\% of their total mass late, between 2 and 4 Gyr ago on average. 
The integrated specific SFHs show that the BSFs had an early quenching phase similar to their CLEs counterparts, but then, while the CLEs suffered a long-term quenching, the BSFs have continued to form stars at relatively high rates (Fig. \ref{fig_intSFH}).
The departure between the BSFs and CLEs starts, on average, 4 (5) Gyr ago for low-mass (intermediate-mass) galaxies. 
At the moment of the departures in the sSFR, CLEs and BSFs had assembled more than 90\% and 80\% of their total mass, respectively. These results suggest that BSFs experienced accretion events at $z < 0.5$ that triggered SF and comprised around 20\% of the final mass in these E galaxies. Such late events are not frequent in CLEs.

Establishing whether the BSF galaxies are a sample of intermediate- and low-mass CLEs that are rejuvenated by recent gas accretion or gas-rich (wet) minor mergers  or, alternatively, whether they are post-starburst galaxies with a recent morphological transformation is of interest. As is whether BSF galaxies are located in particular environments. 

As mentioned above, our results suggest the BSF galaxies initially evolve similarly as the CLE galaxies of the same mass, but they formed a fraction of their stars later, and are still forming stars, following a rejuvenation. 
Regarding the environment, \cite{Lacerna+2016} did not find BSF Es in dense environments, such as the Coma supercluster. \cite{Lacerna+2018} found that isolated BSF Es are not located in dense large-scale environments compared with other isolated E galaxies using different proxies to define the environment within 5 Mpc. 
Therefore, the rejuvenation process of E galaxies seems to occur only in low or average density environments. 
However, these environments do not ensure that the E galaxies would be blue and star-forming in general; most of the isolated E galaxies in low-density environments actually show red colors with very low SFRs\footnote{Morphological quenching \citep{Martig+2009}, compaction \citep{Tacchella+2016}, type-Ia supernovae feedback, and AGN feedback  \citep{Yesuf+2014,Lacerna+2018} can be mechanisms to shut down SF in most of the intermediate- and low-mass E galaxies that live in average or low-density environments, where the environmental and halo quenching are not suitable.} \citep{Lacerna+2016,Lacerna+2018}. Finally, from the preliminary fine-structure morphological analysis presented in Appendix \ref{S_Ap_FineStructure}, we find that $\sim$30\% of the BSF Es present broad tides (as the BSFs are less massive, the broad tides are less frequent), that is, only less than a third of the BSFs show evidence of late mergers. The incidence of shells is slightly larger; shells are long-lived fine structures and are the product of mergers that could have taken place several gigayears ago \citep[e.g.,][]{Mancillas+2019}. 

We propose that BSF Es formed as E quiescent galaxies of similar masses, but they suffered late accretion of low-metallicity gas from the cosmic web or from a minor merger with a gas-rich galaxy that can sustain their SF for 4--5 Gyr at least.  In some cases, the renewed SF can even be due to gas acquired through internal stellar mass loss \citep{Shapiro+2010}. 
The rejuvenation acts before galaxies are quenched long-term.
The scenario of rejuvenation for BSF Es explains the following general results: i) their values for age$_{mw}$ are younger, on average, than those of the CLEs and much younger in the case of  age$_{lw}$, which is evidence of late bursts of star formation; ii) their
stellar metallicities are similar or slightly lower than those of the CLEs, but with a larger dispersion, which shows that a fraction of the stellar population in the BSFs formed, on average, from gas with less enrichment than the CLEs and with a large scatter in the level of enrichment;
iii) their specific SFHs are roughly continuous in the last 4--5 Gyr, that is, they present a sustained late activity of SF, though at earlier epochs they were quenching similar to their CLE counterparts; and iv) their $Z_{mw}$ and $Z_{lw}$ gradients are flatter than those of the CLEs, especially for $Z_{lw}$, which means their younger stellar populations were formed from gas with a roughly equal metallicity in the central and external parts of the galaxy. 
In the rejuvenation scenario by gas accretion from the cosmic web or minor gas-rich mergers, likely an inner disk forms within the E galaxy, and SF proceeds within it, probably in a low efficient regime due to  the spheroid that stabilizes the gaseous disk avoiding the efficient formation of molecular clouds. Interestingly enough, the BSF Es present roughly flat sSFR radial profiles (Fig. \ref{fig_sSFR_grad}), similar to what is observed in disk star-forming galaxies \citep[e.g.,][]{Sanchez+2018_AGN} but with lower amplitudes. These results are in line with those reported by \citet{Shapiro+2010}, who show that the SF in early-type galaxies is a scaled-down version of similar processes in more vigorously star-forming galaxies.

A key signature of the rejuvenation scenario is the presence of an inner star-forming disk within the BSF Es. From the morphological and kinematics analysis presented in Appendix \ref{S_Ap_FineStructure}, we find that almost all of the BSFs show a disky star-forming central structure, which is kinematically decoupled sometimes. 
The SF in inner disk components has been suggested as a mechanism that contributes to the rejuvenation of the stellar populations in the central regions, which may explain the outside-in formation history found in low-mass spheroidal dominated galaxies (bulge-to-total mass ratios > 0.5), in the EAGLE simulation \citep{Rosito+2019}. However, our results show that the inside-out formation is still present in E galaxies with central disks, and probably rejuvenated, such as the BSF  Es (and the other types of low-mass Es). Although their simulated galaxies are bulge dominated, they show very low S\'ersic indexes (mostly $n$ < 2), whereas our E galaxies, including the BSFs, show mostly $n$ > 2.5 (see Fig. \ref{fig_struct}).
For the MaNGA Es, the central disk structure is evident in BSFs ($\geq$93\%), partially in the RQEs ($\sim$40\%), and marginally in the CLEs (<20\%). 
Finally, we do not find very massive BSF galaxies. The rejuvenation process is likely quite inefficient in massive galaxies due to halo mass quenching \citep[e.g.,][and references therein]{ZuMandelbaum2016} and the effects of feedback from strong AGNs.

\section{Summary and conclusions}
\label{Sconclusions}
%%%%%%%%%%%%%%%%%%%%%%%%%%%%%%%%%
We have studied a sample of 343 E galaxies observed with the MaNGA survey (MPL-7). The morphological classification was done using the deep images of the DESI Legacy Imaging Surveys. We defined seven classes 
of E galaxies based on their integrated spectroscopically properties from MaNGA and their colors from SDSS photometry.
We paid attention to the CLE, RQE, and BSF types, which correspond to the classical "red and dead" (73\%),
recently quenched (10\%) and blue, star-forming (4\%) E galaxies, respectively.
 For them, we inferred the mass-weighted and luminosity-weighted stellar age and metallicity gradients, the sSFR radial profiles,
and we reconstructed their integrated and radial CMTDs and 
specific SFHs. 
The results are presented and analyzed in mass bins so that the effects due to volume incompleteness are minimized.
Our main results can be summarized as follows.

$\bullet$ 
The CLEs show mild negative 
luminosity-weighted age gradients ($-0.08_{-0.07}^{+0.06}$ dex/\re), nearly flat mass-weighted age gradients ($-0.03_{-0.05}^{+0.04}$ dex/\re), and mildly steep negative
gradients in both luminosity- and mass-weighted metallicities ($-0.11_{-0.07}^{+0.06}$ dex/\re{} and $-0.11_{-0.08}^{+0.07}$ dex/\re, respectively). 
The CLEs accumulated 90\% of their stellar masses
between 5 and 7 Gyr ago. We observe a "downsizing" trend in both stellar mass
growth  rates and SF quenching, that is to say the  most  massive  CLEs tend to form stars earlier and quench faster than the less massive CLE galaxies. 
In general, the CLEs quench relatively fast, roughly in half the cosmic time at which they attained their most active phase of SF as traced by their luminosity-weighted ages. 
The CLEs show a weak inside-out formation assembly: 
Their inner parts assembled 90\% of their mass 
$\sim$ 2 Gyr earlier than their external parts. This effect is more pronounced for massive (\ms > $10^{11}$ \msun) galaxies. 
The radial specific SFHs show that quenching is faster in the inner parts compared to the outer parts, especially for the low-mass CLEs (\ms < $10^{10.4}$ \msun). 

$\bullet$ The RQE galaxies are $\sim$ 4--5 Gyr younger than the CLEs using the light-weighted ages (by construction) and $\sim$ 3--4 Gyr using mass-weighted stellar ages. 
The median gradients in stellar age and metallicity of the RQEs are similar or flatter, but with much larger scatter, than those of the CLEs of the same mass, except the massive (\ms > $10^{11}$ \msun) RQEs that show positive median gradients 
in age$_{lw}$ (0.11$_{-0.37}^{+0.47}$ dex/\re{}), 
$Z_{lw}$ (0.09$_{-0.06}^{+0.11}$ dex/\re{}), and
$Z_{mw}$ (0.07$_{-0.08}^{+0.07}$ dex/\re{}).
The RQE galaxies reached 90\% of their total mass between 2 and 
6 Gyr ago without a clear trend with mass. 
On average, they quenched $1.2\pm0.9$ Gyr ago and very quickly, in less than 1/4th  the cosmic time at which they attained the most active SF phase. 
For the less massive RQE galaxies, their inner parts show some evidence of earlier mass assembly than the outer parts, while for the most massive RQEs, the assembly is nearly homogeneous or slightly outside-in in some cases. Furthermore, the less massive RQEs show evidence of very weak inside-out quenching. 
Instead, the massive RQEs tend to show a mild outside-in quenching.

$\bullet$ The BSF E galaxies are younger at all radii than the CLEs using both age$_{lw}$ and age$_{mw}$; the differences are much larger for the former case, showing that BSFs suffered late bursts of SF. Their age and metallicity profiles are, on average, flatter than those of the CLEs.
 While their CMTDs and specific SFHs at early epochs are similar to those of CLEs of the same mass; overall they formed a significant fraction of their stellar mass at later times. For them, 90\% of the total mass has been accumulated between 2 and 4 Gyr ago and they are still forming stars actively. Their specific SFHs have remained nearly constant since 4--5 Gyr ago.  We find that nearly all the BSFs have evidence of disky central structures. 
There are not massive BSF galaxies.

The results from our archaeological analysis of the CLE galaxies are consistent with a two-phase scenario, where these galaxies formed 
their (dominant in mass) central part early and nearly coeval, likely by gas-rich dissipative collapse, compaction, and a consequent intense burst of SF. As the SF is rapidly quenched by gas exhaustion (and probably, by AGN feedback), mergers (mostly dry) could have contributed to the assembly of the observed 
external parts of a significant fraction of CLEs.

We find that most of the properties of RQEs that are more massive than \ms$\sim 10^{11}$ \msun{} differ from those of the less massive RQEs.
For BSFs, we discussed the possibility that they formed early, similarly as the CLEs of comparable masses and then they suffered from a process of rejuvenation.
There are not very massive BSFs, suggesting that the rejuvenation process is quite inefficient in massive galaxies.

We stress that the generic evolutionary interpretations discussed here for RQEs and BSF ellipticals could be flawed by the low numbers in our samples or the large diversity of their evolutionary paths. A case-by-case spectroscopic, photometric, and environmental analysis is desirable to confirm or modify these interpretations. We will report on this analysis elsewhere.

%============================= 
\begin{acknowledgements}

We acknowledge the anonymous referee for his/her thoughtful revision of our paper.
V.A-R. acknowledges support from the projects CONACyT CB285721 and PAPIIT-UNAM IA104118.  S.F.S. acknowledges support from the projects Conacyt CB285080, FC-2016-01-1916 and PAPIIT IN100519.
This research made use of Montage. It is funded by the National Science Foundation under Grant Number ACI-1440620, and was previously funded by the National Aeronautics and Space Administration's Earth Science Technology Office, Computation Technologies Project, under Cooperative Agreement Number NCC5-626 between NASA and the California Institute of Technology.
This research also made use of \textsc{astropy}, a community-developed core \textsc{python} ({\tt http://www.python.org}) package for Astronomy \citep{astropy}; 
\textsc{matplotlib} \citep{Hunter:2007}; \textsc{numpy} \citep{:/content/aip/journal/cise/13/2/10.1109/MCSE.2011.37};  
\textsc{pandas} \citep{mckinney2010};
and \textsc{scipy} \citep{citescipy}. 

Funding for the Sloan Digital Sky Survey IV has been provided by the Alfred P. Sloan Foundation, the U.S. Department of Energy Office of Science, and the Participating Institutions. SDSS-IV acknowledges
support and resources from the Center for High-Performance Computing at
the University of Utah. The SDSS web site is www.sdss.org.

SDSS-IV is managed by the Astrophysical Research Consortium for the 
Participating Institutions of the SDSS Collaboration including the 
Brazilian Participation Group, the Carnegie Institution for Science, 
Carnegie Mellon University, the Chilean Participation Group, the French Participation Group, Harvard-Smithsonian Center for Astrophysics, 
Instituto de Astrof\'isica de Canarias, The Johns Hopkins University, Kavli Institute for the Physics and Mathematics of the Universe (IPMU) / 
University of Tokyo, the Korean Participation Group, Lawrence Berkeley National Laboratory, 
Leibniz Institut f\"ur Astrophysik Potsdam (AIP),  
Max-Planck-Institut f\"ur Astronomie (MPIA Heidelberg), 
Max-Planck-Institut f\"ur Astrophysik (MPA Garching), 
Max-Planck-Institut f\"ur Extraterrestrische Physik (MPE), 
National Astronomical Observatories of China, New Mexico State University, 
New York University, University of Notre Dame, 
Observat\'ario Nacional / MCTI, The Ohio State University, 
Pennsylvania State University, Shanghai Astronomical Observatory, 
United Kingdom Participation Group,
Universidad Nacional Aut\'onoma de M\'exico, University of Arizona, 
University of Colorado Boulder, University of Oxford, University of Portsmouth, 
University of Utah, University of Virginia, University of Washington, University of Wisconsin, 
Vanderbilt University, and Yale University.

The Legacy Surveys consist of three individual and complementary projects: the Dark Energy Camera Legacy Survey (DECaLS; NOAO Proposal ID 2014B-0404; PIs: David Schlegel and Arjun Dey), the Beijing-Arizona Sky Survey (BASS; NOAO Proposal ID 2015A-0801; PIs: Zhou Xu and Xiaohui Fan), and the Mayall z-band Legacy Survey (MzLS; NOAO Proposal ID 2016A-0453; PI: Arjun Dey). DECaLS, BASS and MzLS together include data obtained, respectively, at the Blanco telescope, Cerro Tololo Inter-American Observatory, National Optical Astronomy Observatory (NOAO); the Bok telescope, Steward Observatory, University of Arizona; and the Mayall telescope, Kitt Peak National Observatory, NOAO. The Legacy Surveys project is honored to be permitted to conduct astronomical research on Iolkam Du'ag (Kitt Peak), a mountain with particular significance to the Tohono O'odham Nation.

NOAO is operated by the Association of Universities for Research in Astronomy (AURA) under a cooperative agreement with the National Science Foundation.

This project used data obtained with the Dark Energy Camera (DECam), which was constructed by the Dark Energy Survey (DES) collaboration. Funding for the DES Projects has been provided by the U.S. Department of Energy, the U.S. National Science Foundation, the Ministry of Science and Education of Spain, the Science and Technology Facilities Council of the United Kingdom, the Higher Education Funding Council for England, the National Center for Supercomputing Applications at the University of Illinois at Urbana-Champaign, the Kavli Institute of Cosmological Physics at the University of Chicago, Center for Cosmology and Astro-Particle Physics at the Ohio State University, the Mitchell Institute for Fundamental Physics and Astronomy at Texas A\&M University, Financiadora de Estudos e Projetos, Fundacao Carlos Chagas Filho de Amparo, Financiadora de Estudos e Projetos, Fundacao Carlos Chagas Filho de Amparo a Pesquisa do Estado do Rio de Janeiro, Conselho Nacional de Desenvolvimento Cientifico e Tecnologico and the Ministerio da Ciencia, Tecnologia e Inovacao, the Deutsche Forschungsgemeinschaft and the Collaborating Institutions in the Dark Energy Survey. The Collaborating Institutions are Argonne National Laboratory, the University of California at Santa Cruz, the University of Cambridge, Centro de Investigaciones Energeticas, Medioambientales y Tecnologicas-Madrid, the University of Chicago, University College London, the DES-Brazil Consortium, the University of Edinburgh, the Eidgenossische Technische Hochschule (ETH) Zurich, Fermi National Accelerator Laboratory, the University of Illinois at Urbana-Champaign, the Institut de Ciencies de l'Espai (IEEC/CSIC), the Institut de Fisica d'Altes Energies, Lawrence Berkeley National Laboratory, the Ludwig-Maximilians Universitat Munchen and the associated Excellence Cluster Universe, the University of Michigan, the National Optical Astronomy Observatory, the University of Nottingham, the Ohio State University, the University of Pennsylvania, the University of Portsmouth, SLAC National Accelerator Laboratory, Stanford University, the University of Sussex, and Texas A\&M University.

BASS is a key project of the Telescope Access Program (TAP), which has been funded by the National Astronomical Observatories of China, the Chinese Academy of Sciences (the Strategic Priority Research Program "The Emergence of Cosmological Structures" Grant XDB09000000), and the Special Fund for Astronomy from the Ministry of Finance. The BASS is also supported by the External Cooperation Program of Chinese Academy of Sciences (Grant 114A11KYSB20160057), and Chinese National Natural Science Foundation (Grant  11433005).

The Legacy Survey team makes use of data products from the Near-Earth Object Wide-field Infrared Survey Explorer (NEOWISE), which is a project of the Jet Propulsion Laboratory/California Institute of Technology. NEOWISE is funded by the National Aeronautics and Space Administration.

The Legacy Surveys imaging of the DESI footprint is supported by the Director, Office of Science, Office of High Energy Physics of the U.S. Department of Energy under Contract No. DE-AC02-05CH1123, by the National Energy Research Scientific Computing Center, a DOE Office of Science User Facility under the same contract; and by the U.S. National Science Foundation, Division of Astronomical Sciences under Contract No. AST-0950945 to NOAO.

\end{acknowledgements}

\bibliographystyle{aa}
\bibliography{references}

%============================= 
\appendix

%%%%%%
\section{Individual age profiles}
\label{S_Ap_compGoddard2017}
%%%%%%

%%%fig LW-MW 
\begin{figure*}
\includegraphics[width=6.3cm]{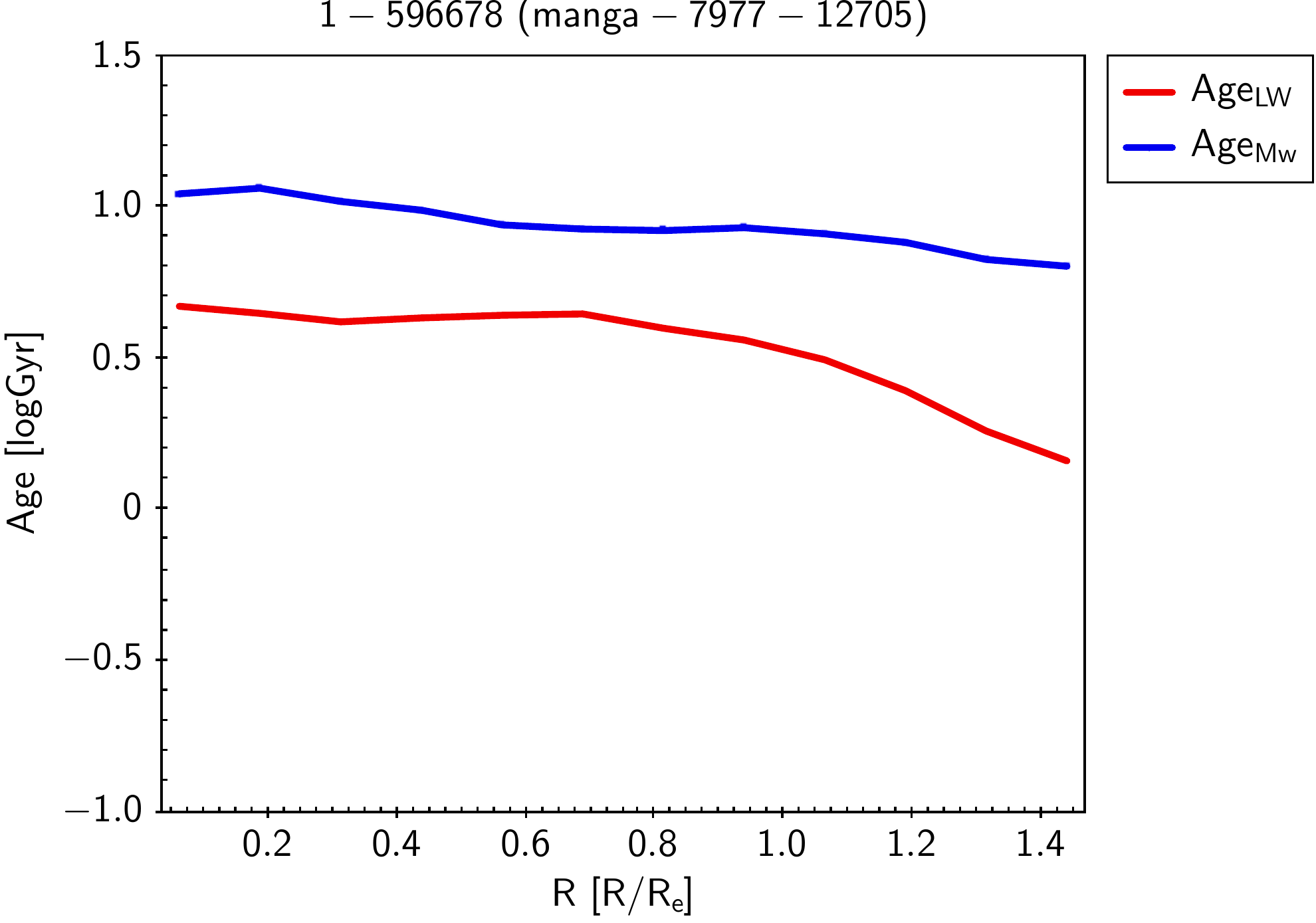}
\includegraphics[width=6.3cm]{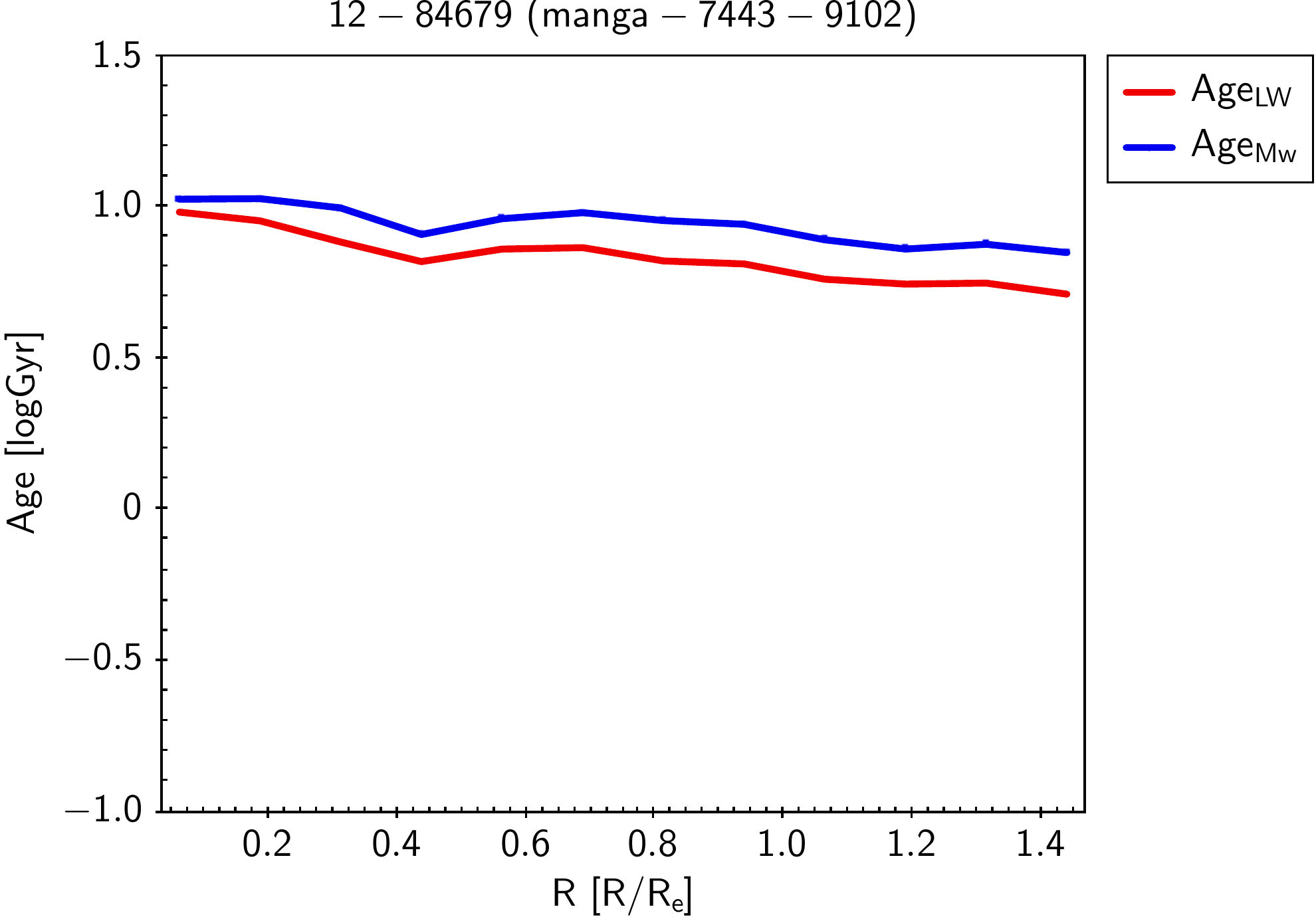}
\includegraphics[width=6.3cm]{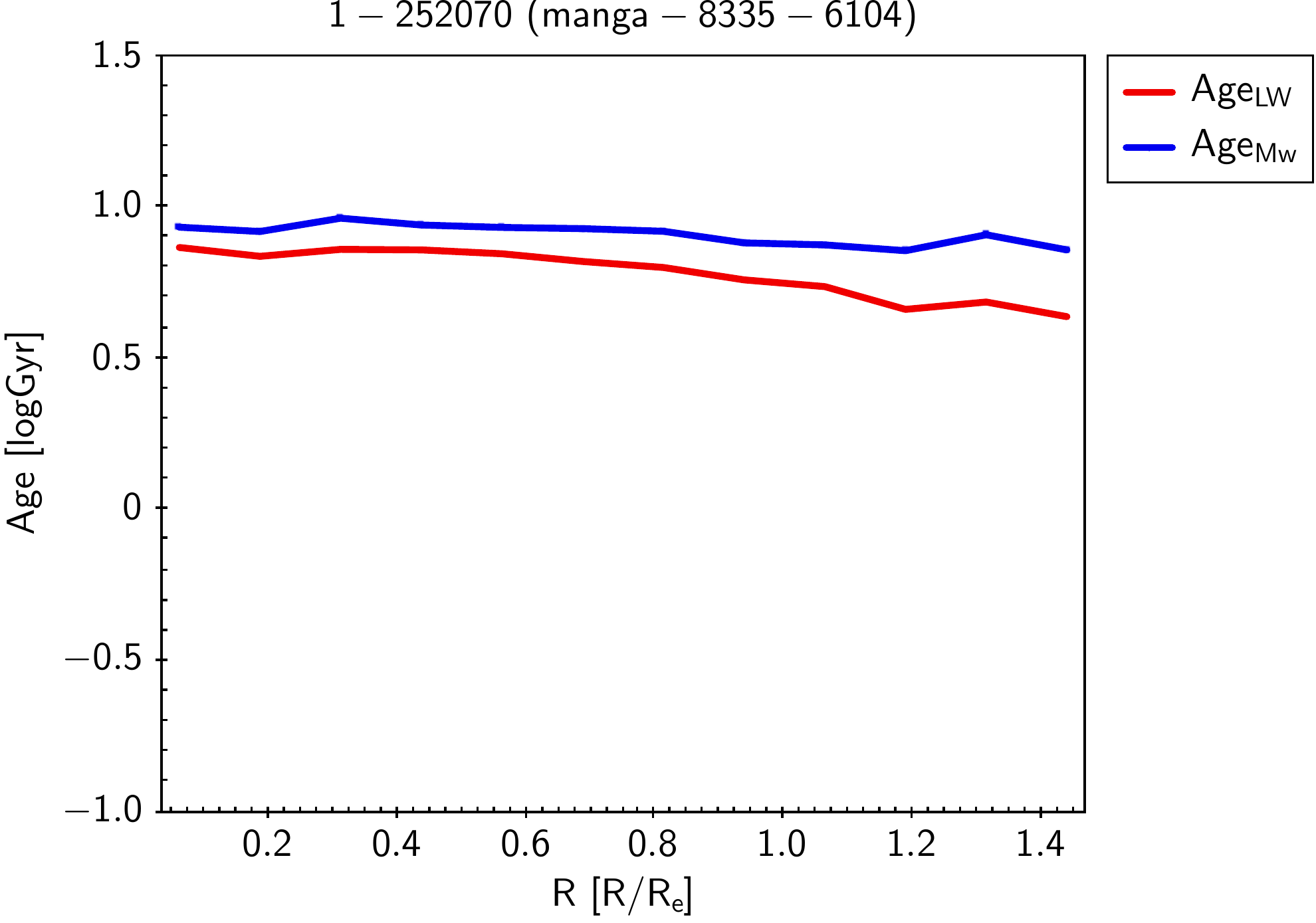}
\includegraphics[width=6.3cm]{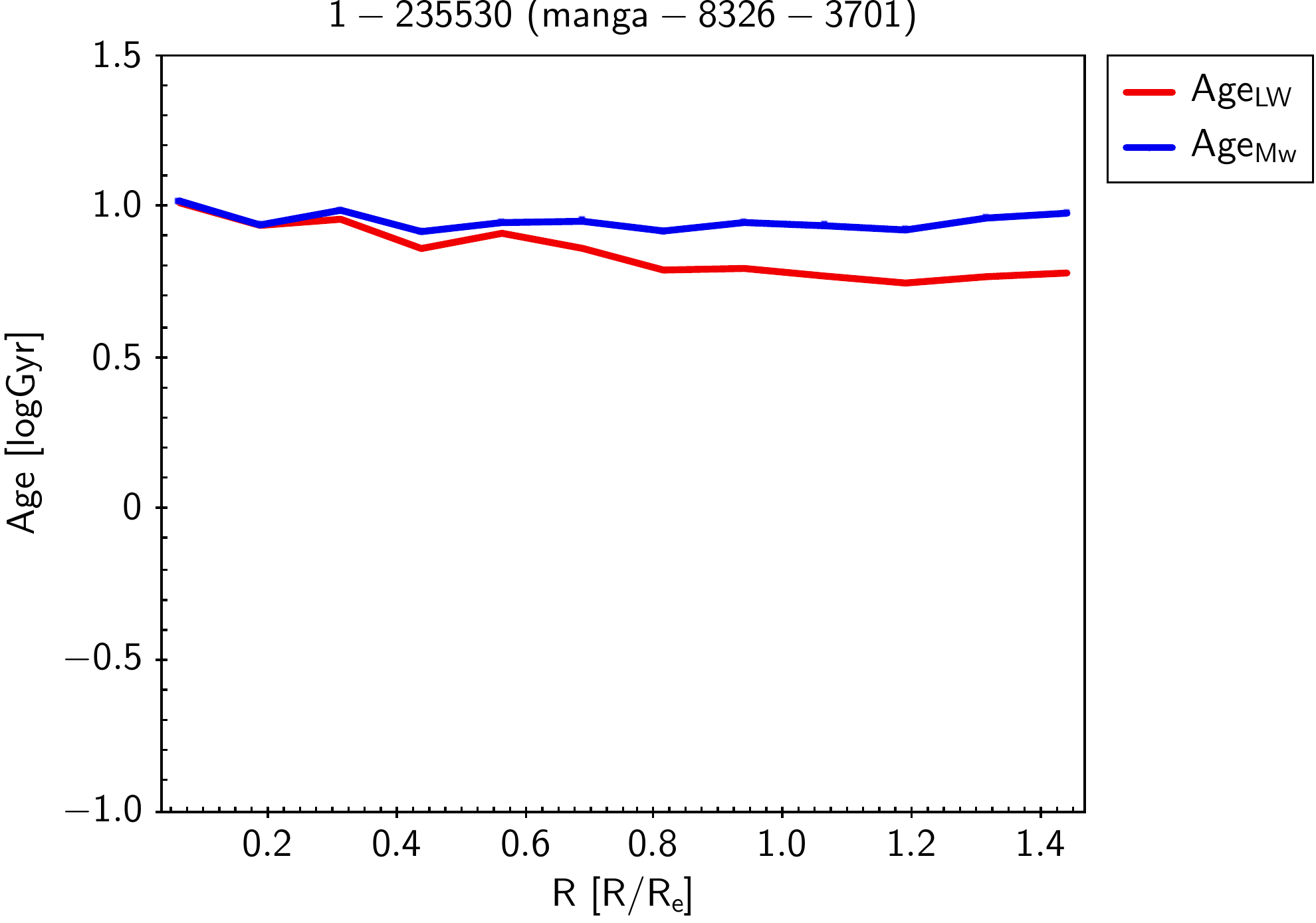}
\includegraphics[width=6.3cm]{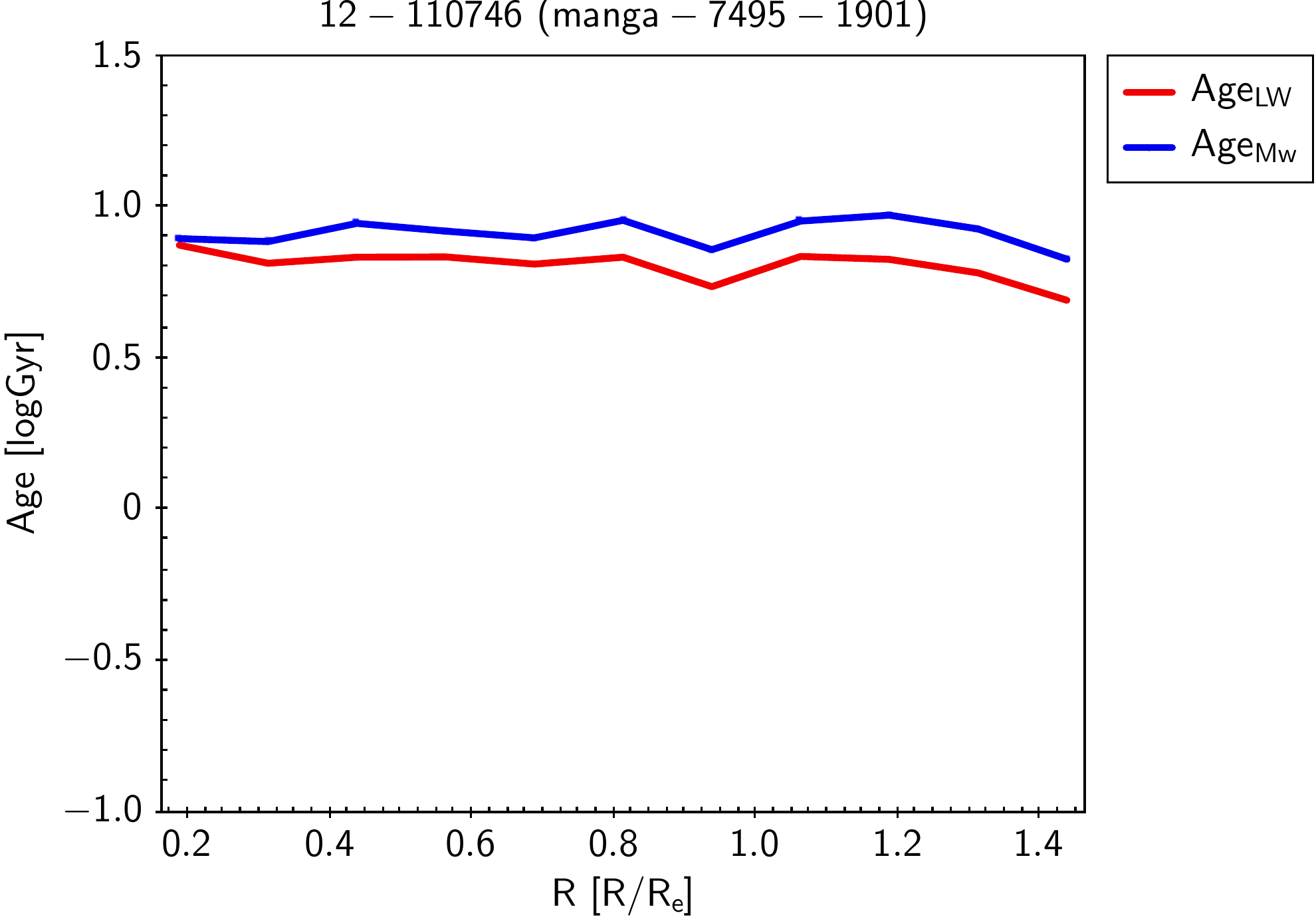}
\caption{Our mass- and light-weighted age profiles (blue and red lines, respectively) for the galaxies in Fig. 7 of \cite{Goddard+2017_466mass}. The MaNGA ID and the Plate-IFU design are shown above each panel.
}
\label{figApp_LW_MW}
\end{figure*}

In Sect. \ref{Sdiscussion}, we discuss that there is a diversity in the radial trends reported in the literature for the stellar age profiles.  
Different softwares, stellar population models, and methods to derive the stellar age gradients may explain the diversity of results. 
Here, we do not pretend to make an exhaustive comparison of the methods, but we show our results for the same five galaxies in Fig. 7 of \cite{Goddard+2017_466mass} as examples of radial profiles for individual galaxies, rather than comparing average trends. Fig. \ref{figApp_LW_MW} shows our mass- and light-weighted age profiles for galaxies observed in MaNGA with a different number of fibers. 
The light-weighted age profiles (red lines) tend to be mildly negative or flat, which is consistent with the results of \cite{Goddard+2017_466mass}. However, our mass-weighted age profiles are also mildly negative or flat, which is opposite to the positive age gradients reported by \cite{Goddard+2017_466mass}.

\begin{figure*}
\includegraphics[width=6.cm]{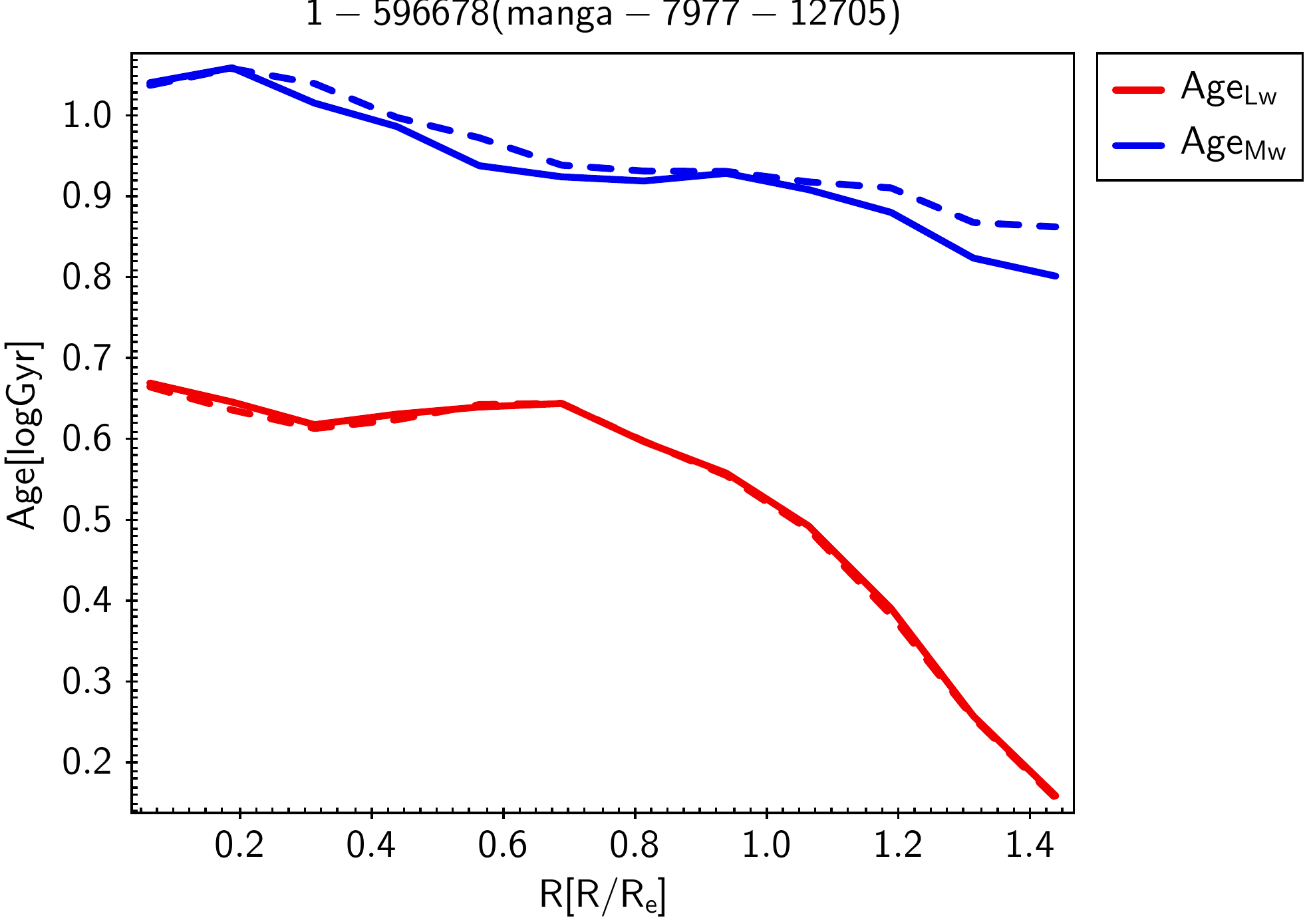}
\includegraphics[width=6.3cm]{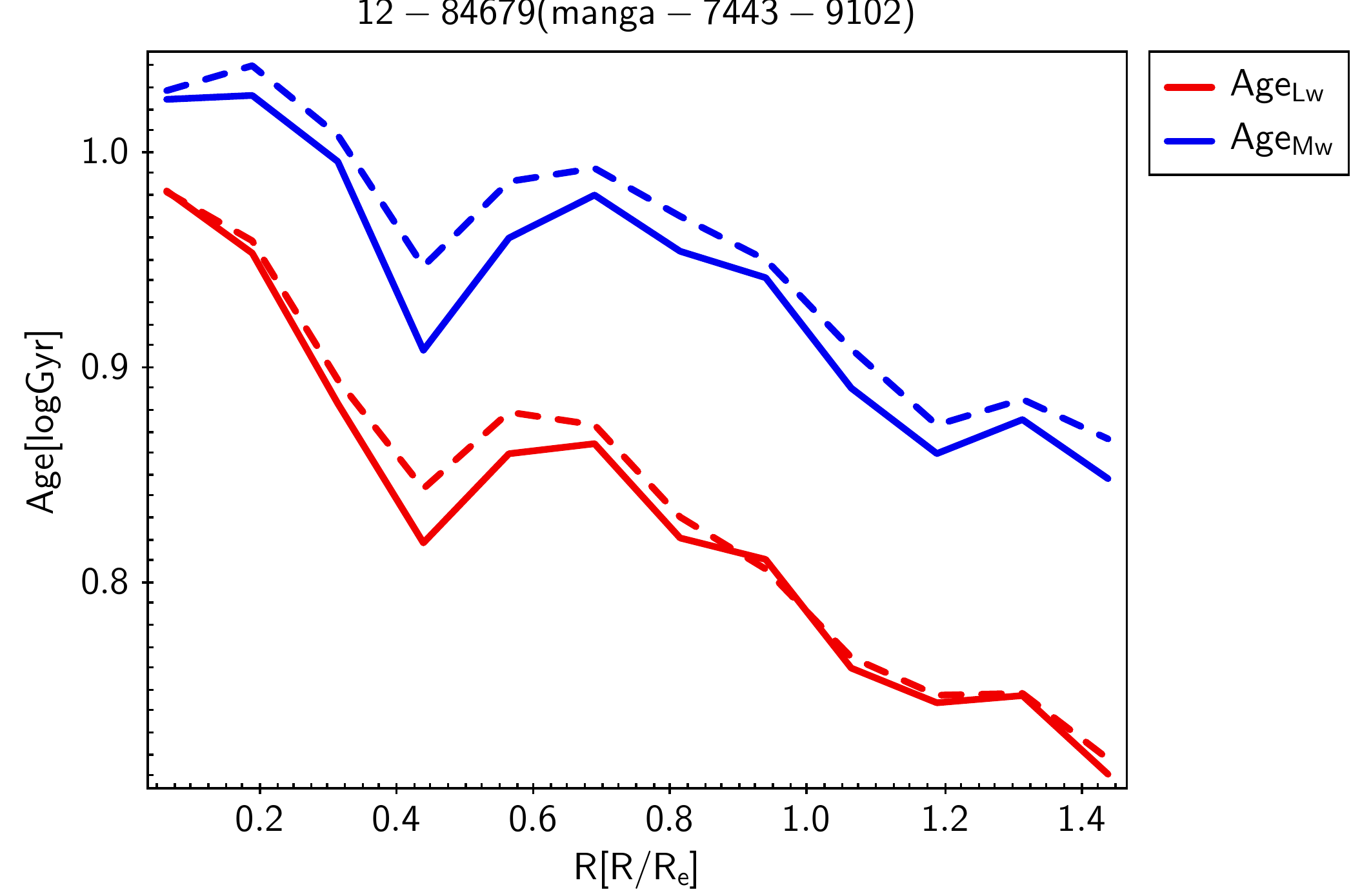}
\includegraphics[width=6.3cm]{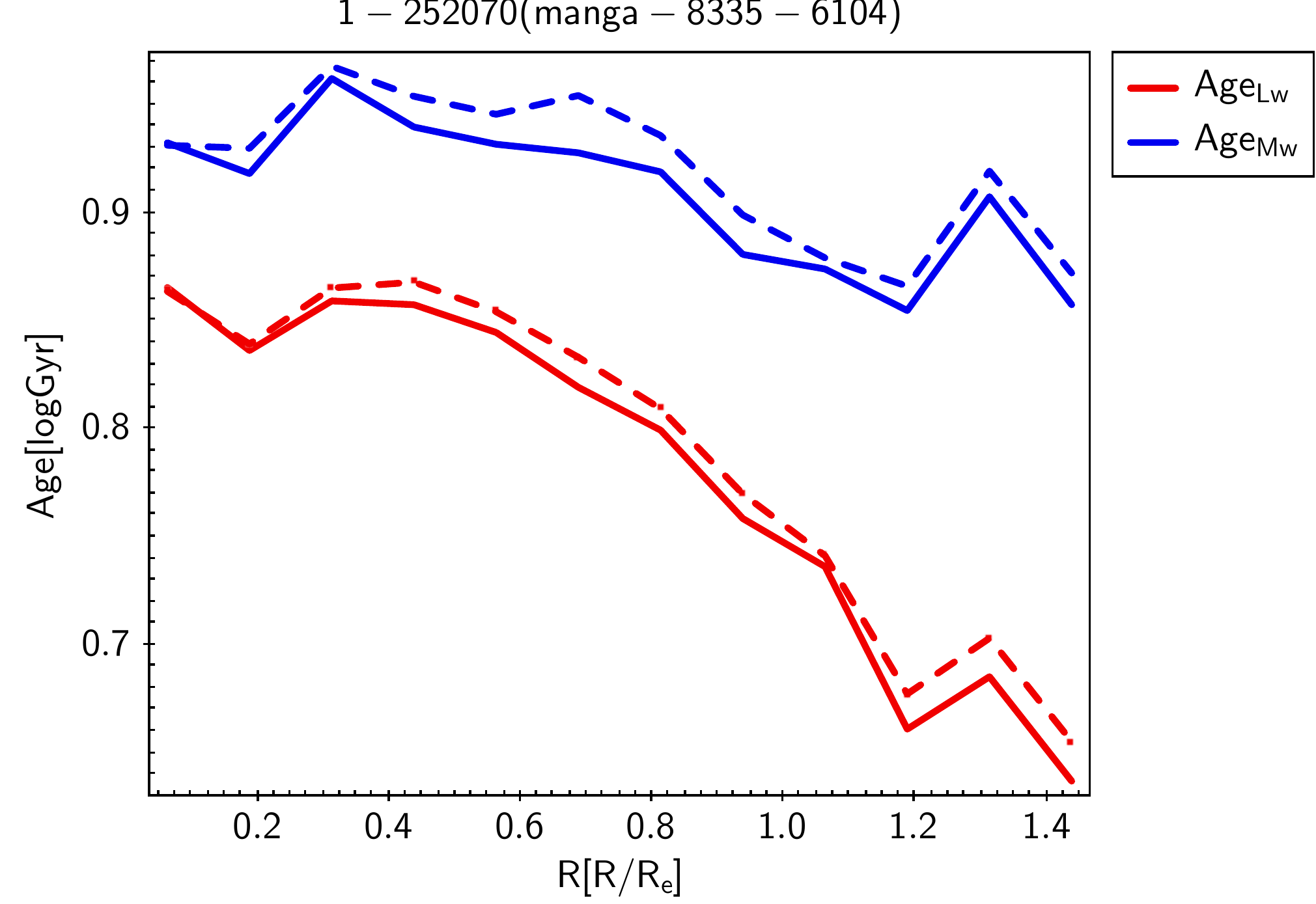}

\includegraphics[width=6.cm]{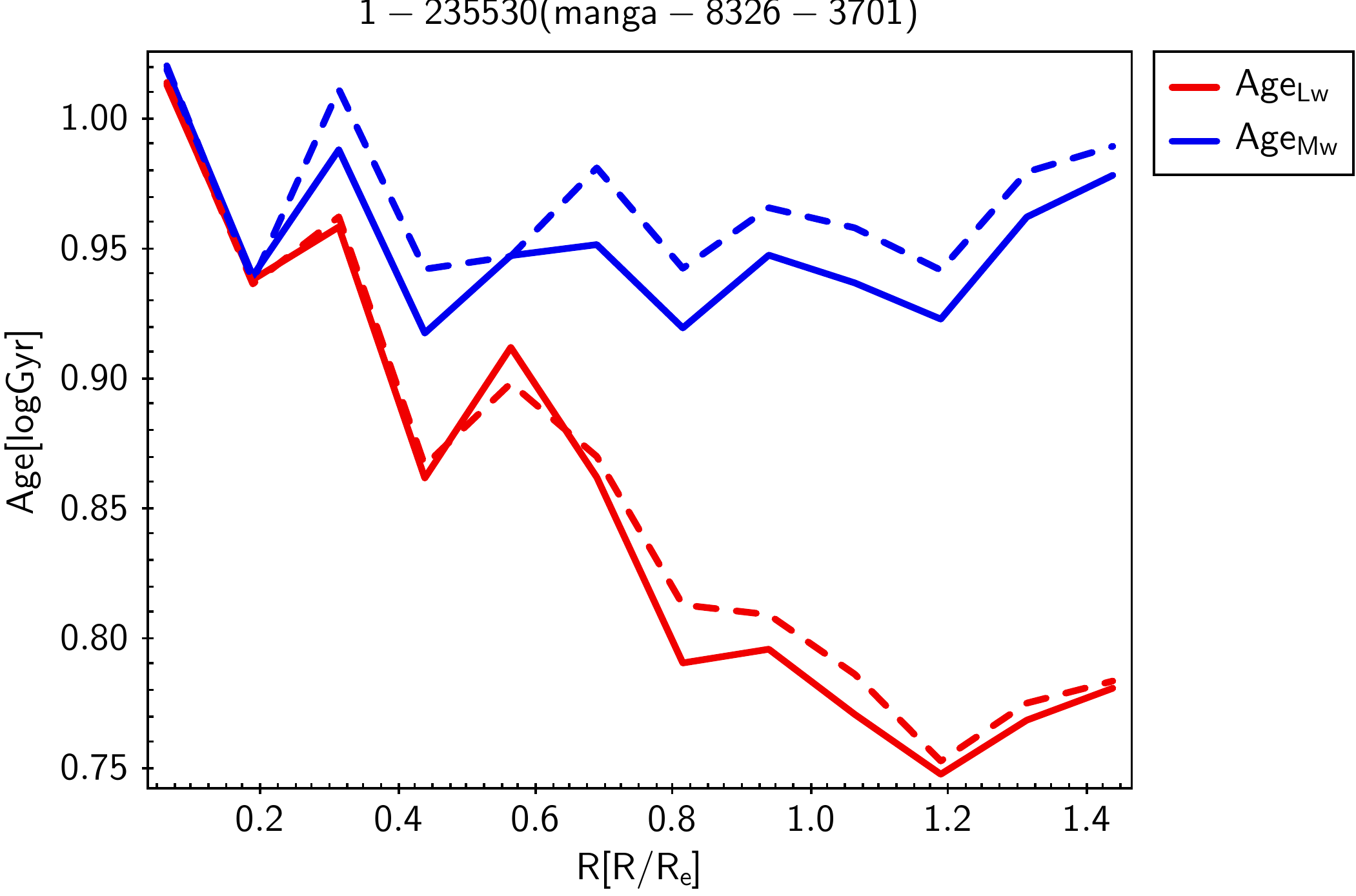}
\includegraphics[width=6.cm]{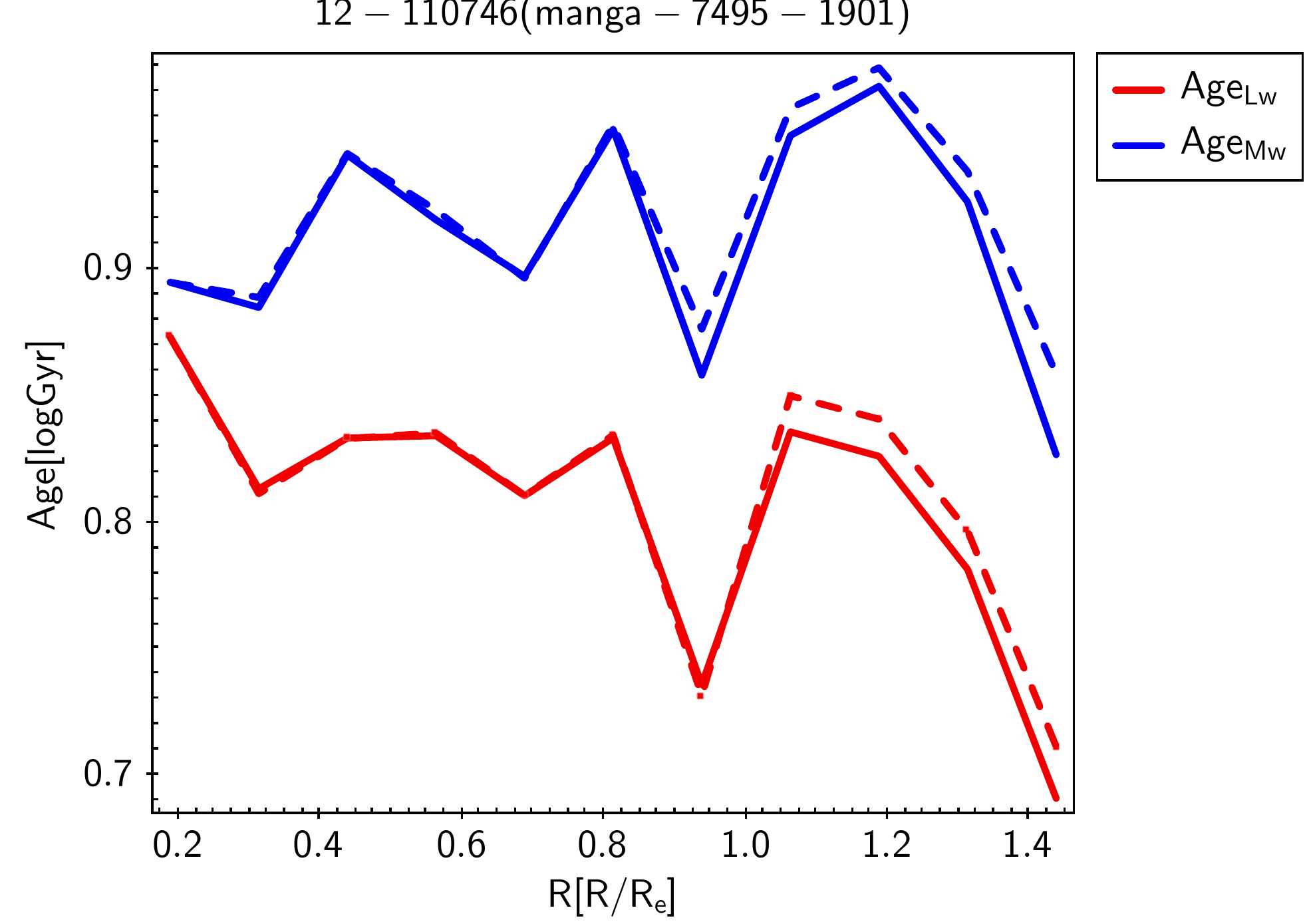}
\caption{Our mass- and light-weighted age profiles (blue and red lines, respectively) for the galaxies in Fig. 7 of \cite{Goddard+2017_466mass}. The dashed lines are the same as in Fig. \ref{figApp_LW_MW}. The solid lines are the profiles without weighting by mass or light within each radial bin.
The MaNGA ID and the Plate-IFU design are shown above each panel.
}
\label{figApp_LW_MW_with_wo}
\end{figure*}

\begin{figure*}
\centering
\includegraphics[width=5.5cm]{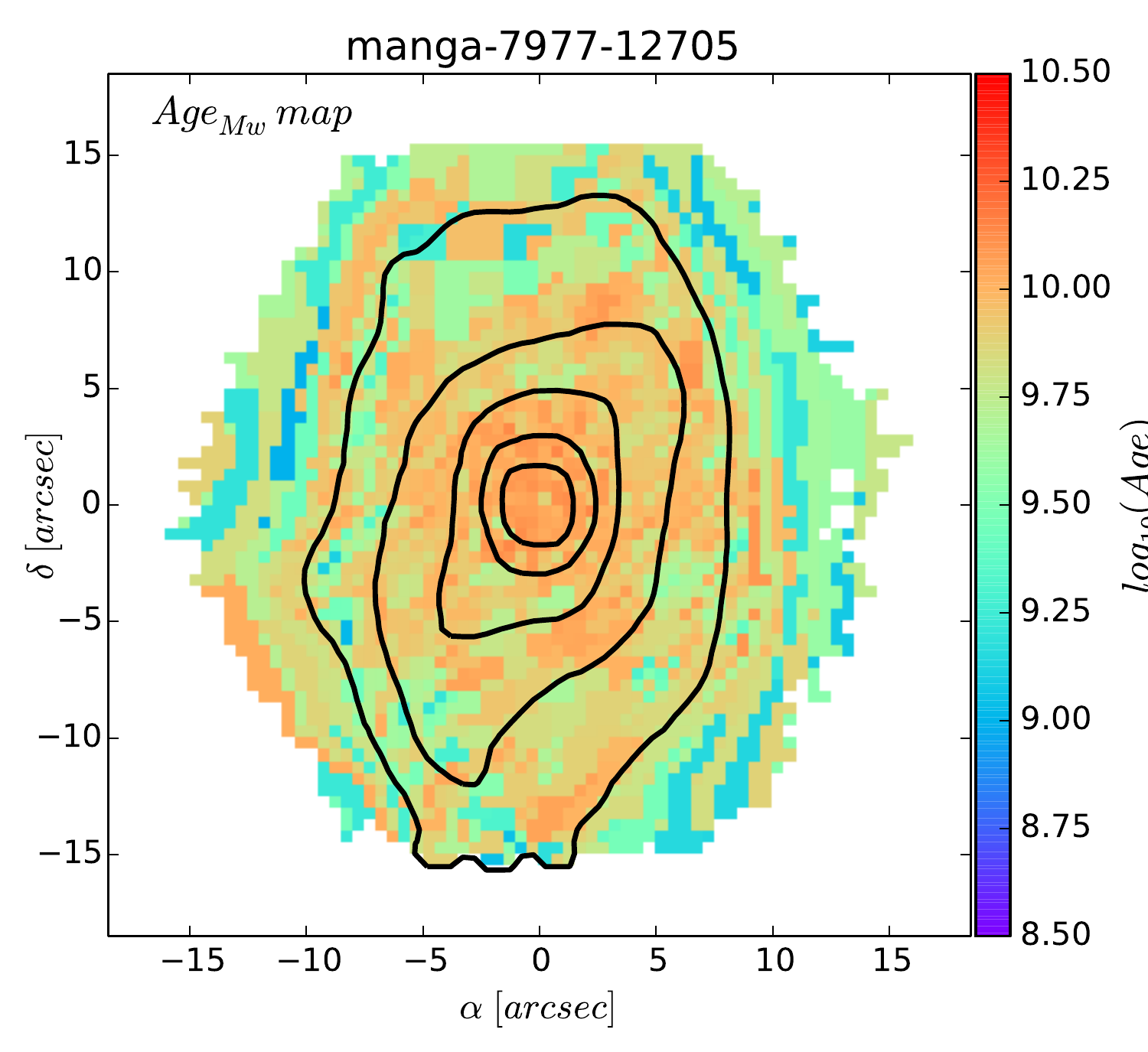}
\includegraphics[width=5.5cm]{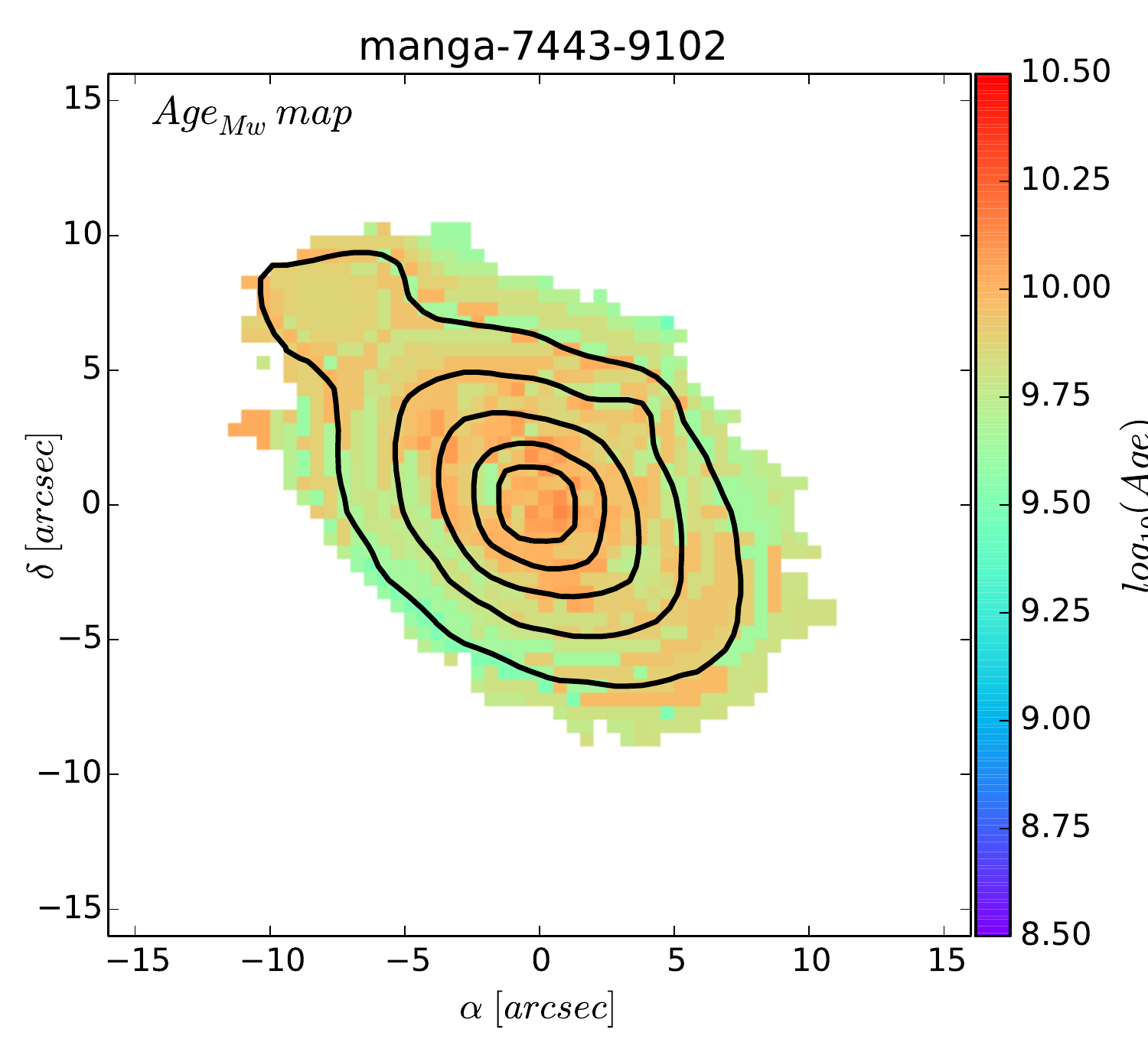}
\includegraphics[width=5.5cm]{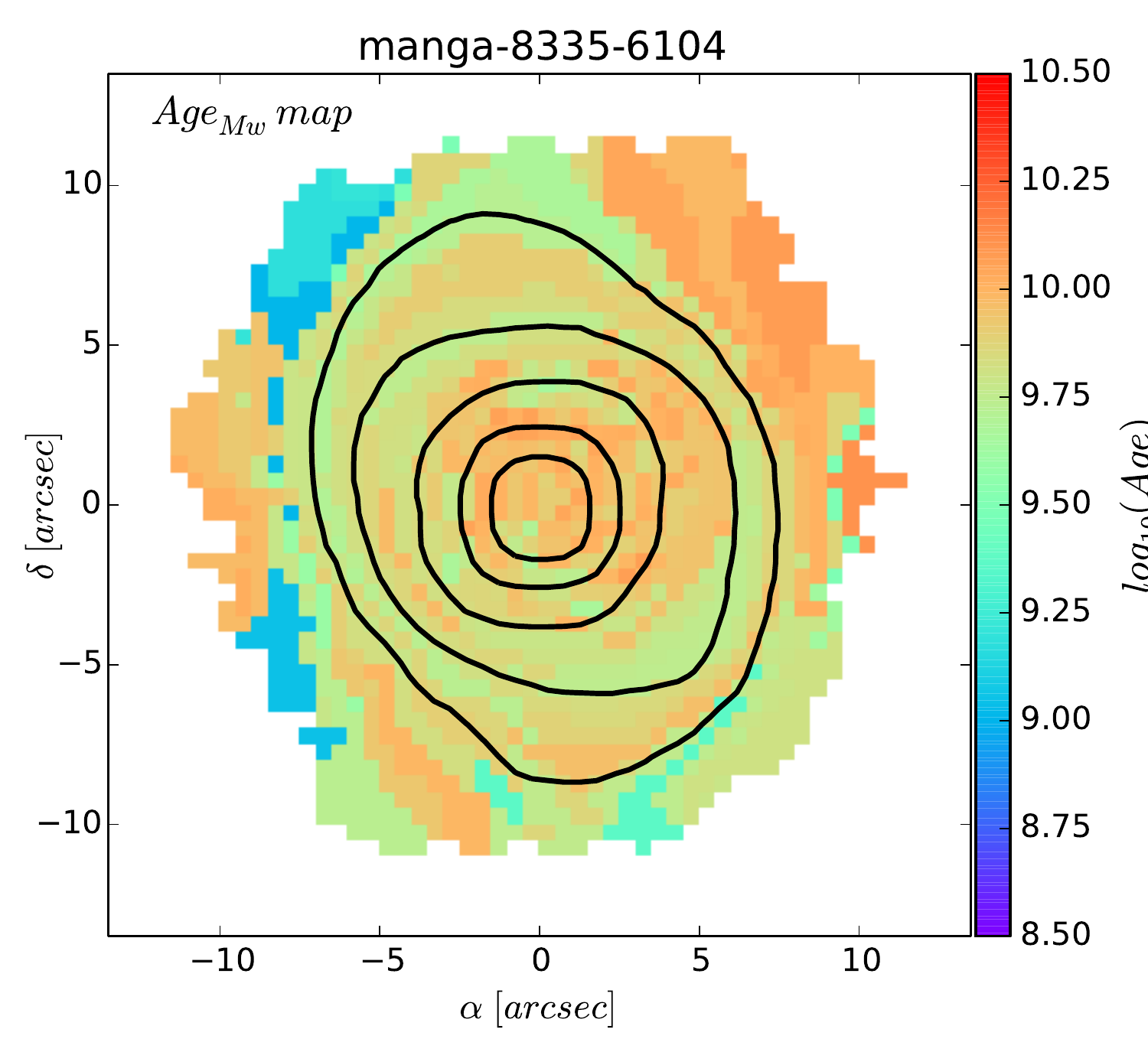}
\includegraphics[width=5.5cm]{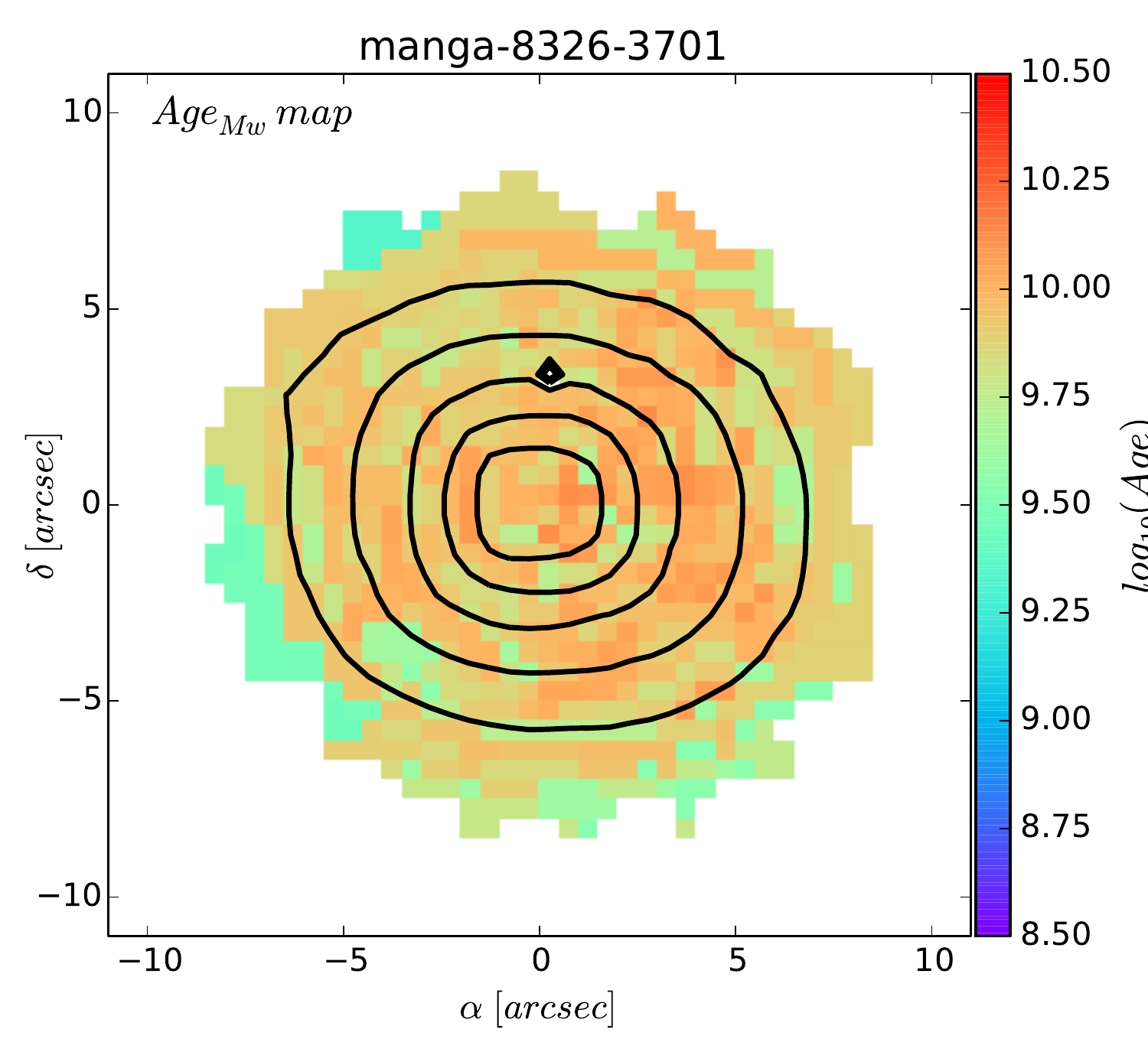}
\includegraphics[width=5.5cm]{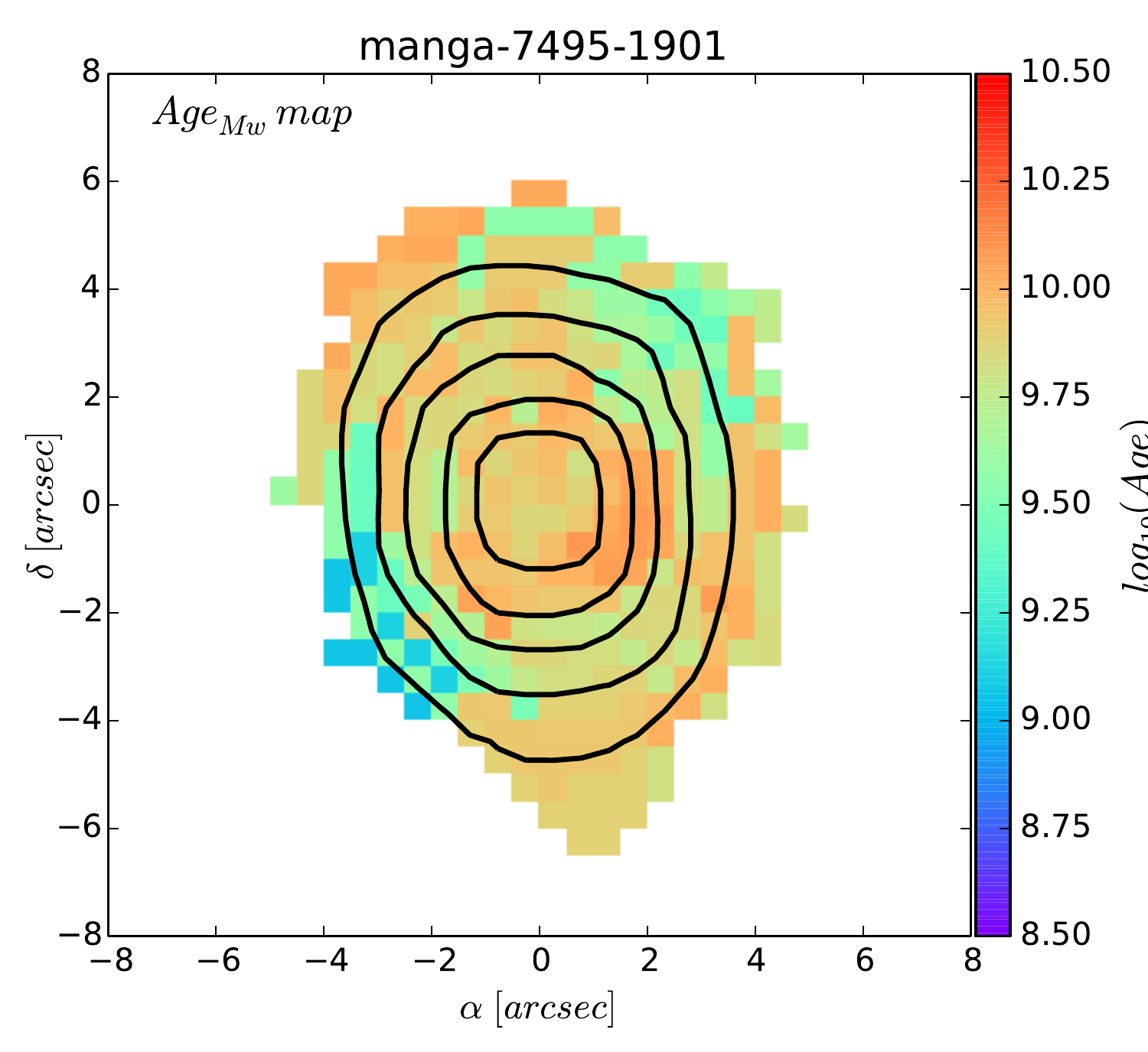}
\caption{Our mass-weighted age maps for the galaxies in Fig. 7 of \cite{Goddard+2017_466mass}. The Plate-IFU design is shown above each panel.
}
\label{figApp_MWmaps}
\end{figure*}

A possible explanation for the different trends is that we weight by the stellar mass (or light) 2D maps within each radial bin. Fig. \ref{figApp_LW_MW_with_wo} shows the radial profiles without this weighting (solid lines). We still obtain mildly negative or flat gradients regardless of the weighting approach. Furthermore, we do not see evident positive gradients in the mass-weighted age maps (see Fig. \ref{figApp_MWmaps}), that is, the galaxy outskirts do not seem to be older than the internal parts.

%%%fig LW-MW with and w/o weighting

There are other differences between \cite{Goddard+2017_466mass} and our methods, which may explain why we obtain opposite trends with the mass-weighted ages. They used inputs from the MaNGA data analysis pipeline \citep[DAP,][]{Westfall+2019} when running their spectral fitting code \verb|FIREFLY| \citep{Wilkinson+2017} with a minimum S/N = 5 in the $r$-band, whereas we used \verb|Pipe3D| with a threshold S/N = 50 for the segmentation and integration within the $V$ band. 
The most significant difference between \verb|FIREFLY| and our results (\verb|Pipe3D|) is that \verb|FIREFLY| arithmetically averages its light and mass weighed ages, while  \verb|Pipe3D| uses a logarithmic average 
(see Eqs. \ref{eq_ageMW}--\ref{eq_ageLW}), so the weight of young stars goes to zero in \verb|FIREFLY|.
Furthermore, there are many differences in the methodology to obtain the SSP decomposition: different stellar libraries, different extinction laws, and different S/N definitions in order to indicate the most important differences. 
An in depth exploratory comparison of the methods will be presented in a future paper.

%%%fig MW maps

%%%%%%
\section{Radial profiles with more than five bins}
\label{S_Ap_12bins}
%%%%%%

\begin{figure*}
\includegraphics[width=15.9cm]{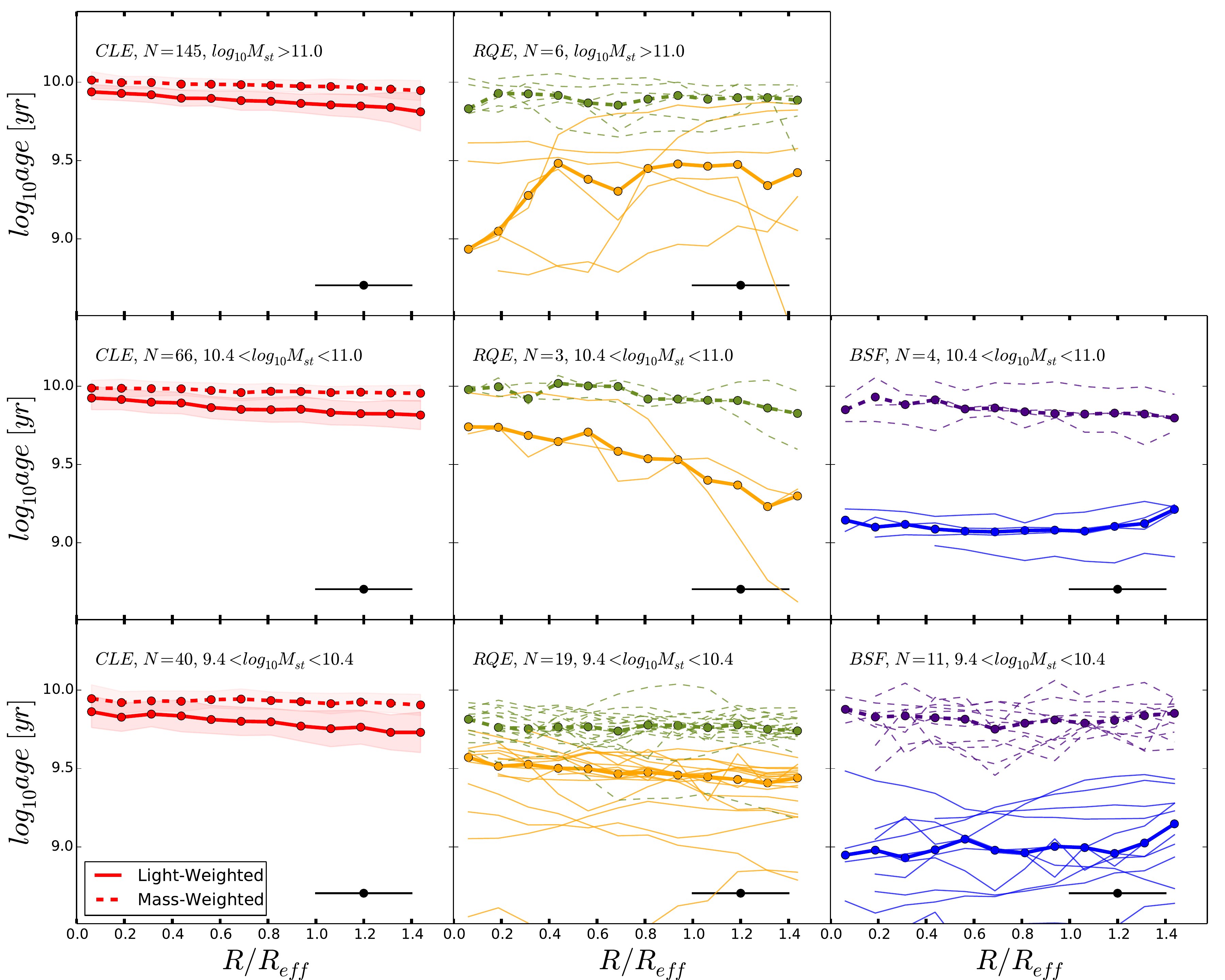}
\caption{Similar to Fig. \ref{fig_ageMW_ageLW_grad}, 
but using 12 radial bins.
}
\label{fig_ageMW_ageLW_grad12}
\end{figure*}

In Sect. \ref{S4} we show both individual and ``stacked'' radial profiles using only five radial bins in units of $R_{eff}$ given the typical PSF size of MaNGA observations (see the horizontal error bars in panels of Figs. \ref{fig_ageMW_ageLW_grad}--\ref{fig_sSFR_grad}).
\citet{Zibetti+2020} suggest the presence of a ``U-shape'' in the inner regions of the light-weighted stellar age profiles of E galaxies, with a minimum around 0.4 \re, which is not present in our results.
Here, we also calculated the radial profiles using twelve radial bins for each galaxy to see if the shape of the profiles shows some dependence on the radial binning. Fig \ref{fig_ageMW_ageLW_grad12} corresponds to the mass-weighted and luminosity-weighted age profiles using twelve radial bins. 
Although we increased the number of bins, our analysis always considers the typical PSF size (horizontal error bar).
The shapes of the profiles are similar to those shown in Fig. \ref{fig_ageMW_ageLW_grad}. We obtain similar results for the metallicity and sSFR profiles. Therefore, our trends do not change with a smaller radial bin. We do not observe an evident ``U-shape'' in the inner region of the age profiles of Es, which is consistent with others works \citep[e.g.,][]{GonzalezDelgado+2015,ZhengZheng+2017},
although there is a tiny dip in the light-weighted age profiles at $\sim$ 0.2 \re{}, only for low-mass Es. 
As discussed above, different softwares, stellar population models, and methods to derive the stellar age gradients may explain the diversity of results in the stellar age profiles. 

%%%fig age profiles (12 radial bins) 

%%%%%%%
\section{Fine-structure and kinematic morphological features}
\label{S_Ap_FineStructure}
%%%%%%

This section presents preliminary results after a uniform visual search for faint substructures in the BSF, RQE, and CLE galaxies using moderately-deep imaging from the DESI Legacy Imaging Surveys.  
We also inspected  the central regions of CLE, RQE, and BSF galaxies to identify dusty non-regular features that may provide a hint as to the presence of central disks. Additionally, from the DAP products available in Marvin\footnote{ https://www.sdss.org/dr15/manga/marvin/} \citep{Cherinka+2019}, we extracted the projected stellar rotation velocity, the ionized gas rotation velocity, and the EW(H$\alpha$) corrected for underlying stellar continuum absorption.

The BSF galaxies show evidence of dusty, non-regular (disk-like) features in their central regions in practically all of the cases 
(93\%). This fraction amounts to 69\%
in RQEs and decreases drastically to 21\% in CLEs. 
The spatial distribution of EW(H$\alpha$) is very different between BSFs, RQEs, and CLEs. All (100\%) BSF galaxies show a 
spatially dominant H$\alpha$ concentration in the central region of these galaxies, while in RQEs and CLEs, that happens with a relatively low frequency of 18\% and 13\%, respectively. 
The velocity maps show evidence of disk rotation in nearly all the BSF galaxies (93\%),
and this fraction decreases to 31\% for the RQE galaxies. 
In summary, the fraction of BSF galaxies with evidence of central disks is $\geq$ 93\%, while this fraction is $\sim$40\%, on average, for the RQE galaxies.

We have taken advantage of the post-processing catalog by the DESI collaboration \citep{Dey+2019} to generate a census of fine morphological  structures at a surface brightness limit of 27.5 mag arcsec$^{-2}$. We visually categorize these structures as broad tides, shells, and filaments or tidal streams. Here, we report the incidence of broad tidal structures (fans and plumes), which emerge from the main body of the galaxy and shells. The incidence of filaments is very low in the three E galaxy samples.  
Broad tides mostly result from major or intermediate dry mergers and they do not survive for a long time \citep[e.g.,][]{vanDokkum2005, Mancillas+2019}. Thus, if they are observed, then the merger was relatively recent. The shells are commonly found around E galaxies \citep[e.g.,][]{MalinCarter1983, SchweizerSeitzer1992} and are the result of intermediate mergers that could have happened a long time ago, given that they can survive several gigayears if more recent mergers do not destroy them \citep[e.g.,][]{Mancillas+2019}.

We report the incidence of broad tides and shells for each 
stellar mass bin employed in this paper: $\log$(\ms/\msun)$\ge11.0$, $10.4\le\log$(\ms/\msun)$<11.0$, and 
$9.4\le\log$(\ms/\msun)$<10.4$ correspond to the high-, intermediate-, and low-mass bins, respectively. 
For CLEs, the incidence of broad tides in these three mass bins is 65\%, 43\%, and 30\%,
respectively, while the respective incidences of shells are 24\%, 31\%, and 5\%.
For RQEs, the incidence of broad tides in the three mass bins is 85\%, 66\%, and 11\%,
respectively, while the respective incidences of shells are 0\%, 33\%, and 5\%.
For BSFs, there are no objects that are more massive than $10^{11}$ \msun. The overall incidence of broad tides in this galaxy sample is of 30\% (50\% and 23\% in the intermediate- and low-mass bins,
respectively), while the overall incidence of shells is 48\% (50\% and 44\%, respectively).
The reported fractions must be considered with caution, especially for the low-mass bin, in which mergers or interactions with other low-mass or dwarf galaxies produce features of very low-surface brightness. The visibility of fine structure features depends on the sample's surface brightness limit and the galaxy orientation \citep{Mancillas+2019}.

\end{document}